\documentclass[preprint2]{aastex}

\shorttitle{HD209458b}
\shortauthors{Morello G.}

\usepackage{amsmath,nicefrac}
\usepackage{enumitem}
\usepackage{multirow}
\usepackage[flushleft]{threeparttable}

\begin{document}


\title{A SEA BASS on the exoplanet HD209458b}


\author{G. Morello}
\affil{Irfu, CEA, Universit\'e Paris-Saclay, 91191 Gif-sur-Yvette, France}
\affil{Department of Physics \& Astronomy, University College London, Gower Street, WC1E 6BT, UK}
\email{giuseppe.morello@cea.fr}



\begin{abstract}
We present here the first application of Stellar and Exoplanetary Atmospheres Bayesian Analysis Simultaneous Spectroscopy (SEA BASS) on real datasets.  SEA BASS is a scheme that enables the simultaneous derivation of four-coefficient stellar limb-darkening profiles, transit depths, and orbital parameters from exoplanetary transits at multiple wavelengths. It relies on the wavelength-independence of the system geometry and on the reduced limb-darkening effect in the infrared. This approach has been introduced by Morello et al. (2017) (without the SEA BASS acronym), who discuss several tests on synthetic datasets.
Here, we (1) improve on the original algorithm by using multiple \textit{Spitzer}/InfraRed Array Camera passbands and a more effective set of geometric parameters, (2) demonstrate its ability with \textit{Hubble Space Telescope}/Space Telescope Imaging Spectrograph datasets, by (3) measuring the HD209458 stellar limb-darkening profile over multiple passbands in the 290--570 nm range with sufficient precision to rule out some theoretical models that have been adopted previously in theliterature, and (4) simultaneously extracting the transmission spectrum of the exoplanet atmosphere. The higher photometric precision of the next-generation instruments, such as those onboard the \textit{James Webb Space Telescope}, will enable modeling the star-planet systems with unprecedented detail, and increase the importance of SEA BASS for avoiding the potential biases introduced by inaccurate stellar limb-darkening models.
\end{abstract}


\keywords{planets and satellites: individual (HD209458b) - planets and satellites: atmospheres - planets and satellites: fundamental parameters - stars: atmospheres - techniques: photometric - techniques: spectroscopic}



\section{INTRODUCTION}
Over the past 15 years, transit spectroscopy has enabled the detection of atomic, ionic, and molecular species in the atmospheres of exoplanets (e.g., \citealp{charbonneau02, vidal-madjar03, vidal-madjar04, barman07, tinetti07}), as well as the characterization of their thermal and cloud profiles (e.g., \citealp{sing16, tsiaras17}). The diagnostic parameter is the squared planet-to-star radii ratio, $p^2 = (R_p/R_*)^2$, the so-called ``transit depth''. The optically thick area of a transiting exoplanet is wavelength-dependent due to the differential opacity of its gaseous envelope. The expected spectral modulations are of the order of 100 parts per million (ppm) relative to the stellar flux for the hot Jupiters \citep{brown01}, and smaller for other classes of planets.

\begin{table}[!t]
\begin{center}
\begin{threeparttable}
\caption{HD209458 system parameters \label{tab1}}
\begin{tabular}{lc}
\tableline\tableline
\multicolumn{2}{c}{Stellar parameters} \\
\tableline
$T_{\mbox{eff}}$ (K) \tnote{a} & 6065$\pm$50 \\
$\log{g_*}$ (cgs) \tnote{a} & 4.361$\pm$0.008 \\
$[\mbox{Fe/H}]$ (dex) \tnote{a} & 0.00$\pm$0.05 \\
$M_*$ ($M_{\odot}$) \tnote{a} & 1.119$\pm$0.033 \\
$R_*$ ($R_{\odot}$) \tnote{a} & 1.155$\pm$0.016 \\
\tableline
\multicolumn{2}{c}{Planetary parameters} \\
\tableline
$M_p$ ($M_{Jup}$) \tnote{a} & 0.685$\pm$0.015 \\
$R_p$ ($R_{Jup}$) \tnote{a} & 1.359$\pm$0.019 \\
$a$ (AU) \tnote{a} & 0.04707$\pm$0.00047 \\
\tableline
\multicolumn{2}{c}{Transit parameters} \\
\tableline
$R_p/R_*$ \tnote{a} & 0.12086$\pm$0.00010 \\
$a/R_*$ \tnote{a} & 8.76$pm$0.04 \\
$i$ (deg) \tnote{a} & 86.71$\pm$0.05 \\
$P$ (days) \tnote{b} & 3.52474859$\pm$0.00000038 \\
$E.T.$ (HJD) \tnote{b} & 2452826.628521$\pm$0.000087 \\
\tableline
\end{tabular}
\begin{tablenotes}
\item[a] \cite{torres08}
\item[b] \cite{knutson07}
\end{tablenotes}
\end{threeparttable}
\end{center}
\end{table}

The shape of a transit light-curve also depends on the stellar limb-darkening effect, i.e., the wavelength-dependent radial decrease in specific intensity, coupled with the star-planet geometry and timing. Accurate knowledge of the stellar intensity profile is necessary to infer the correct transit depth from observations, especially at the precision level required to study the exoplanetary atmospheres. In most cases, a set of limb-darkening coefficients is computed either from specific stellar-atmosphere models or by interpolation on a precalculated grid. In some cases, for the highest quality datasets, empirical limb-darkening coefficients are estimated from the light-curve fits by assuming a linear \citep{schwarzschild1906} or quadratic \citep{kopal50} law:
\begin{eqnarray}
\frac{ I_{\lambda}( \mu ) }{ I_{\lambda}( 1 ) } = 1 - u (1-\mu) \\
\frac{ I_{\lambda}( \mu ) }{ I_{\lambda}( 1 ) } = 1 - u_1 (1-\mu) - u_2 (1-\mu)^2
\end{eqnarray}
Here $\mu=\cos{\theta}$, where $\theta$ is the angle between the line-of-sight and the normal to the stellar surface. Several authors have found that the linear and quadratic limb-darkening laws can be inadequate to approximate the stellar intensity profile. \cite{morello17} showed that such inadequate parameterizations can provide a perfect match to the observed transit light-curves while significantly biasing the transit depth and other parameter estimates. \cite{claret00} proposed a four-coefficient law, hereinafter ``claret-4'', which enables a more accurate approximation to the stellar model-atmosphere intensity profiles:
\begin{equation}
\frac{ I_{\lambda}( \mu ) }{ I_{\lambda}( 1 ) } = 1 - \sum_{n=1}^{4} a_n \left ( 1 - \mu^{ \nicefrac{n}{2}} \right )
\label{eqn:law_claret-4}
\end{equation}
The claret-4 coefficients are not commonly fitted to the transit light-curves, as the strong parameter degeneracies are likely to hamper the convergence of the fit.

\cite{morello17} suggest an optimal strategy to break some degeneracies by adopting informative priors on the orbital parameters, hence enabling the convergence of model-fits with free claret-4 coefficients. Their proposal was to measure the ``geometric'' parameters, $a/R_*$ and $i$, from infrared transit observations, then implement the results as informative priors when model-fitting at shorter wavelengths. This approach is motivated by the smaller limb-darkening effect in the infrared, which mitigates the potential biases that are due to inaccurate models, and the negligible wavelength-dependence of $a/R_*$ and $i$. Its effectiveness has been demonstrated over a set of simulated observations.

Here, we apply this technique (now named Stellar and Exoplanetary Atmospheres Bayesian Analysis Simultaneous Spectroscopy (SEA BASS)), for the first time to real datasets, taken with the InfraRed Array Camera (IRAC) onboard \textit{Spitzer} and the Space Telescope Imaging Spectrograph (STIS) onboard the Hubble Space Telescope (\textit{HST}).
We focus on the HD209458 system, which consists of a hot Jupiter orbiting a solar analog star. Stellar and planetary parameters available in the literature are summarized in Table~\ref{tab1}. HD209458b was the first transiting exoplanet to be discovered \citep{charbonneau00, henry00}, and also the first one on which a chemical element has been detected (Na, \citealp{charbonneau02}). HD209458 is currently one of the most frequently studied exoplanetary systems, thanks to the relatively large amplitude of the transit signal and the apparent brightness ($V=$7.6) of its parent star.

\begin{table*}[!t]
\begin{center}
\caption{\textit{Spitzer}/IRAC datasets analysed for this study (infrared observations). \label{tab2}}
\begin{tabular}{cccccccc}
\tableline\tableline
Channel & Prog. ID & AOR\tablenotemark{a} & UT Date & Mode\tablenotemark{b} & Type\tablenotemark{c} & Mission\tablenotemark{d} & Pip. \tablenotemark{e} \\
\tableline
Ch1 & 461 & 24740096 & 2007 Dec 31 & full, 0.4 & Transit & Cold & 18.25.0 \\
(3.6 $\mu$m) & 461 & 27928832 & 2008 Jul 19 & full, 0.4 & Transit & Cold & 18.25.0 \\
 & 60021 & 41628928 & 2011 Jan 14 & sub, 0.1 & Full-phase & Warm & 19.2.0 \\
 & 60021 & 50798336 & 2014 Feb 15 & sub, 0.1 & Full-phase & Warm & 19.2.0 \\
Ch2 & 461 & 27928576 & 2008 Jul 22 & full, 0.4 & Transit & Cold & 18.25.0 \\
(4.5 $\mu$m) & 60021 & 38704$_{\mbox{128}}^{\mbox{384}}$\tablenotemark{f} & 2010 Jan 19 & sub, 0.4 & Full-phase & Warm & 19.2.0 \\
Ch3 & 461 & 24740096 & 2007 Dec 31 & full, 2.0 & Transit & Cold & 18.25.0 \\
(5.8 $\mu$m) & 461 & 27928832 & 2008 Jul 19 & full, 2.0 & Transit & Cold & 18.25.0 \\
Ch4 & 40280 & 24649984 & 2007 Dec 24 & sub, 0.4 & Half-phase & Cold & 18.25.0 \\
(8.0 $\mu$m) & 461 & 27928576 & 2008 Jul 22 & full, 2.0 & Transit & Cold & 18.25.0 \\
\tableline
\end{tabular}
\tablenotetext{a}{Astronomical Observation Request.}
\tablenotetext{b}{Readout mode and frame time in seconds.}
\tablenotetext{c}{Observations are either restricted to a short time interval that contains the transit, or include the half or full orbital period.}
\tablenotetext{d}{Observations taken before the cryogen exhaustion on 2009 May 15 (Cold \textit{Spitzer} mission), or after it (Warm \textit{Spitzer} mission). }
\tablenotetext{d}{Pipeline version of the Basic Calibrated Data.}
\tablenotetext{f}{This transit is split over two AORs.}
\end{center}
\end{table*}

\subsection{Structure of the paper}
Section~\ref{sec:observations} reports the observations analyzed in this paper. Section~\ref{sec:analysis} describes the state-of-the-art data-detrending pipelines adopted for the different instruments, which are not novel to this work, except for the ``test of aperture sizes'' (Section~\ref{ssec:test_aperture}). Sections~\ref{ssec:results_irac1e2} and \ref{ssec:results_irac3e4} report the results obtained for the infrared observations; more technical details are discussed in Appendices~\ref{app:background}--\ref{app:ramps}. Section~\ref{ssec:irac_overview} compares our parameter results over multiple epochs and passbands, and with previous results reported in the literature. Section~\ref{ssec:IRAC_priors} discusses different ways of combining the geometric parameter estimates into a  Bayesian prior to adopt in the analyses of visible light-curves. Section~\ref{sec:results_STIS} reports the results obtained for the visible observations. Appendix~\ref{app:ld_laws} discusses the impact of using two-coefficient limb-darkening laws instead of the most accurate claret-4.

\section{OBSERVATIONS}
\label{sec:observations}

\subsection{Infrared photometry}
\label{ssec:obs_irac}

We reanalyzed all the transits of HD209458b observed with \textit{Spitzer}/IRAC across the four passbands centered at 3.6, 4.5, 5.8, and 8.0 $\mu$m, i.e., the IRAC channels 1--4, respectively. Observational and detector information for the individual light-curves and their identifiers is given in Table~\ref{tab2}.

In full-array mode, frames of 256$\times$256 pixels are recorded with a cadence of 8.4 s (frame time + delay time).  Except for the 2007 December 24 observation, a detector switch occurred after each full frame, in order to sample the same transit event almost simultaneously at 3.6 and 5.8 $\mu$m or 4.5 and 8.0 $\mu$m. The telescope needed frequent repointings to keep the centroid within one pixel. This observing strategy is not available for the warm \textit{Spitzer} mission, as the two longest wavelength channels are not operative. The adjectives ``cold'' and ``warm'' associated with \textit{Spitzer} refer to the datasets taken before and after the cryogen exhaustion on 2009 May 15, respectively.

In sub-array mode, 64 frames from an area of 32$\times$32 pixels are taken in succession. The delay time between consecutive groups of 64 frames is 2 s. No repointings are needed in this mode to keep the centroid within one pixel over a few hours.

\subsection{Visible spectroscopy}
\label{ssec:obs_stis}
We also reanalyzed two transits of HD209458b that were observed spectroscopically with the \textit{HST}/STIS G430L grating (wavelength range 290--570~nm) as part of the GO-9447 program. Each visit consists of five consecutive \textit{HST} orbits (duration $\sim$95 minutes), including 50-minute gaps due to Earth occultations. Frames are 64$\times$1024 detector images with an integration time of 22~s and a reset time of 20~s. The dispersion direction is almost parallel to the longest side of the detector, and it is stable within the same pixel column during each visit. Observational and detector information for the individual light-curves and their identifiers is given in Table~\ref{tab3}.

\begin{table}[!t]
\begin{center}
\caption{\textit{HST}/STIS datasets analyzed for this study (visible observations). \label{tab3}}
\begin{tabular}{cccc}
\tableline\tableline
Grating & Program & Rootnames & UT Date \\
\tableline
\multirow{5}{*}{G430L}  & \multirow{5}{*}{GO-9447} & o6n301010 & \multirow{5}{*}{2003 May 3}\\
&  &  o6n3a10d0 & \\
&  &  o6n3a10c0 & \\
&  &  o6n3a10b0 & \\
&  &  o6n3a10a0 & \\
\tableline
\multirow{5}{*}{G430L}  & \multirow{5}{*}{GO-9447} & o6n303010 & \multirow{5}{*}{2003 Jun 25}\\
&  &  o6n3a30d0 & \\
&  &  o6n3a30c0 & \\
&  &  o6n3a30b0 & \\
&  &  o6n3a30a0 & \\
\tableline
\tableline
\end{tabular}
\end{center}
\end{table}

\section{DATA ANALYSIS}
\label{sec:analysis}

In this section we describe the specific data-detrending pipelines adopted for each instrument, from the extraction of raw light-curves to the final parameter fitting.

The model-fitting procedure consists of two steps, which are common to all of our data analyses:
\begin{enumerate}
\item least-squares minimization between the reference light-curve and the parametric model;
\item Markov Chain Monte Carlo (MCMC) sampling of the parameters posterior distribution.
\end{enumerate}
Least-squares parameter estimates are used as starting parameters for the chains. Then we run an Adaptive Metropolis algorithm with delayed rejection \citep{haario06} for 1,000,000 iterations, with a burn-in of 100,000 iterations. 
The transit parameters that are not free in the fit are fixed to the values reported in Table~\ref{tab1}. For the \textit{Spitzer}/IRAC light-curves, we adopt the passband-integrated limb-darkening coefficients provided by \cite{hayek12}, here reported in Table~\ref{tab4}.

\begin{table}[!t]
\begin{center}
\begin{threeparttable}
\caption{Claret-4 limb-darkening coefficients for the HD209458 star in the \textit{Spitzer}/IRAC passbands. \label{tab4}}
\begin{tabular}{ccccc}
\tableline\tableline
Channel  & $a_1$ & $a_2$ & $a_3$ & $a_4$ \\
\tableline
Ch1, 3.6 $\mu$m & 0.5564 & -0.5462 & 0.4315 & -0.1368 \\
Ch2, 4.5 $\mu$m & 0.4614 & -0.4277 & 0.3362 & -0.1074 \\
Ch3, 5.8 $\mu$m & 0.4531 & -0.5119 & 0.4335 & -0.1431 \\
Ch4, 8.0 $\mu$m & 0.4354 & -0.6067 & 0.5421 & -0.1816 \\
\tableline
\end{tabular}
\begin{tablenotes}
\item Using 3D \texttt{Stagger Code} models; taken from \cite{hayek12}.
\end{tablenotes}
\end{threeparttable}
\end{center}
\end{table}

\subsection{Detrending \textit{Spitzer}/IRAC data}
We downloaded the Basic Calibrated Data (BCD) from the \textit{Spitzer} Heritage Archive \citep{wu10}. BCD are flat-fielded and flux-calibrated frames \citep{fazio04, irachand}. Each observation consists of a time series of frames taken with regular cadences (see Section~\ref{ssec:obs_irac} for timing details).

We take square arrays of pixels with the stellar centroid at their centers as fixed photometric apertures during each observation, following the same procedure as adopted by, e.g., \cite{morello15, morello16}, and \cite{ingalls16}. The centroid position varies within the same pixel during each observation.

For the full-phase and half-phase observations (see Table~\ref{tab2}), we restrict the data analysis to an interval $|\phi | < 0.05$, where $\phi$ denotes the time from mid-transit in units of the orbital period, commonly referred to as orbital phase. This cut allows an out-of-transit baseline longer than 6~hr, $\lesssim$1.5 times longer than those associated with transit-only observations.

We flag and correct outliers in the flux time series. Outliers are those points that deviate from the median of the four nearest temporal neighbors by more than five times the noise level, estimated as the median standard deviation of the sets of five consecutive points. Outliers are replaced with the mean value of the points immediately before and after, or, in the case of two consecutive outliers, with a linear interpolation. In order to identify more than two consecutive outliers, \O$_i$ ($i=1 \dots n$), we considered the light-curves binned in time (a binning factor of up to 128 for the sub-array mode, and with different starting points), such that a single outlier, $\overline{O}$, in a binned time series corresponds to the arithmetic mean of the consecutive outliers in the unbinned version. $\overline{O}$ is then replaced with the mean value of the points immediately before and after, $(\overline{A}+\overline{B})/2$ (in the binned version). This operation is equivalent to shifting $\overline{O}$ by $(\overline{A}+\overline{B})/2 - \overline{O}$. As we wish to correct the
unbinned light-curve, all the consecutive outliers, O$_i$ ($i=1 \dots n$), are then shifted by $(\overline{A}+\overline{B})/2 - \overline{O}$. The percentage of outliers is always smaller than 0.4$\%$.

\subsubsection{IRAC channels 1 and 2}
The two IRAC channels with the shortest wavelengths (3.6 and 4.5 $\mu$m) both use InSb detectors \citep{fazio04}. The most significant instrumental effects arise from fluctuations of the centroid position coupled with a non-uniform inter- and intra-pixel response \citep{fazio04, morello15}. The flux measurements are correlated to the centroid of the Point Spread Function (PSF) on the detector, which varies over multiple timescales \citep{grillmair12, hora14, krick16}. In particular, the most prominent effect is a sawtooth modulation associated with the periodic on-off cycling of a battery heater within the spacecraft bus, although its amplitude depends on the solar angle during the observations.

We apply the pixel-Independent Component Analysis (ICA) technique to remove the instrumental systematics in the observations at 3.6 and 4.5 $\mu$m \citep{morello14, morello15, morello15b}. The key step is the ICA of the pixel time series within the chosen photometric aperture, i.e., a linear transformation of pixel time series set into maximally independent components  \citep{hyvarinen00}. \cite{morello16} proposed a wavelet pixel-ICA algorithm, in which the pixel time series undergo Discrete Wavelet Transform (DWT) before the ICA separation. The so-called wavelet pixel-ICA algorithm has proven to perform as well as or better than other state-of-the-art pipelines to detrend \textit{Spitzer}/IRAC data \citep{ingalls16}. The $n$-level DWT preliminarily separates the high-frequency components while down-sampling by a factor of $2^n$. This trade-off was worthwhile in detrending observations with low signal-to-noise-ratio (S/N$\sim$0.4\footnote{Here S/N is defined as the amplitude ratio between the transit/eclipse signal and the high-frequency noise.}) such as those analyzed by \cite{morello16}. No preference between the two variants of the pixel-ICA algorithm has been found for the observations analyzed here (S/N$\sim$3--5) hence we selected the simpler one (no wavelet).

One of the independent components has the morphology of a transit; the others are referred to as non-transit components, and may correspond to different source signals, either astrophysical or instrumental in nature, and noise. 
The raw light-curves are model-fitted as linear combinations of the non-transit components and a transit model, as in \cite{morello16}. We set uniform priors for the MCMC. The output chains (approximately gaussian) are used to estimate the best parameter values (means) and lower limit error bars (standard deviations, $\sigma_{\mbox{\footnotesize par}, 0}$). 

We also take into account the uncertainty associated with the independent components \citep{morello16}:
\begin{equation}
\label{eqn:sigmaica}
\sigma_{ICA}^{2} = \sum_j o_j^2 ISR_j
\end{equation}
where ISR is the so-called Interference-to-Signal-Ratio matrix \citep{tichavsky08, waldmann12}, and $o_j$ are the coefficients of non-transit components. The ICA variance, $\sigma_{ICA}^{2}$, is added to the MCMC-sampled likelihood variance, $\sigma_0^2$, approximately equal to the variance of the residuals. The final parameter error bars are rescaled as
\begin{equation}
\sigma_{par} = \sigma_{par,0} \sqrt{ \frac{\sigma_0^2+\sigma_{ICA}^2}{\sigma_0^2} } .
\end{equation}

\subsubsection{IRAC channels 3 and 4}
\label{ssec:irac3e4_detrending}
The two longest wavelength IRAC channels (5.8 and 8.0 $\mu$m) use Si:As detectors \citep{fazio04}. The most significant instrumental effects are the so-called ``detector ramps'' \citep{harrington07, knutson07, machalek08}.

Assuming a charge-trapping model, \cite{agol10} derived an exponential formula to model the increase in flux per pixel over time observed at 8.0~$\mu$m:
\begin{equation}
\label{eqn:ramp_exp}
F(t) = a_0 - a_1 e^{-t/\tau_1}
\end{equation}
They recommend the use of a double exponential formula to approximate the ramp on the photometric flux (e.g., integrated over an aperture):
\begin{equation}
\label{eqn:ramp_exp2}
F(t) = a_0 - a_1 e^{-t/\tau_1} - a_2 e^{-t/\tau_2}
\end{equation}
The double exponential formula has the asymptotic behaviour predicted by the charge-trapping model, $\tau_1$ and $\tau_2$ are characteristic timescales associated with pixels with different illumination levels, and in the analyses of \cite{agol10}, led to a smaller Bayesian Information Criterion (BIC, \citealp{schwarz78}).

The ramp observed at 5.8~$\mu$m has the opposite behavior in our datasets, i.e., the photometric flux decreases over time. While this behavior may not be explained by the same charge-trapping model \citep{seager10}, Equations~\ref{eqn:ramp_exp} and \ref{eqn:ramp_exp2} can also be used to approximate the ramp at 5.8 $\mu$m.

A long list of alternative ramp parameterizations has appeared in the literature for both channels (e.g., \citealp{charbonneau08, knutson08, knutson09, desert09, beaulieu10}), all of which are linear combinations of three basic forms: exponential, logarithmic, and powers of time. We tested the different functions found in the literature, and compare their results.

Another frequently adopted procedure is to discard the initial part of the ramp (e.g., \citealp{charbonneau08, knutson08, beaulieu10}), before any data detrending. We tested different lengths for the trimmed interval of each dataset: from 0 to 91 minutes, with a 3.5-minute step.

The raw light-curves are model-fitted as the product of transit and ramp models.

\subsubsection{Test of aperture sizes}
\label{ssec:test_aperture}
In the study of exoplanetary transits, the relevant information is the fraction of stellar light that is occulted by the planet over time, while the absolute calibrated flux is not required. However, an additive constant on the flux measurements does change their ratios, therefore biasing the inferred stellar and planetary parameters (see, e.g., \citealp{kipping10} for the case of additive flux from the nightside exoplanet emission, \citealp{ballerini12, csizmadia13} for the case of stellar spots).

Here we derive an analogous formula for the transit depth affected by background. 
Let us consider the simplest transit model, i.e., a dark circle (the planet) that passes in front of another circle (the star) with uniform brightness \citep{seager03}. Before and after transit, the detector will record the unperturbed stellar flux: $F_{\mbox{out}} = \alpha F_*$. Here $F_*$ denotes the stellar flux, the $\alpha$ factor accounts for the fact that typically, the user does not calculate the absolute source flux, but rather a detector number that also depends on user choices (e.g., size of the photometic aperture, PSF fitting strategy). During the full transit (when the planet circle lies inside the stellar circle), the decrease in flux is proportional to the planet-to-star area ratio: $\Delta F = F_{\mbox{\footnotesize out}} - F_{\mbox{\footnotesize in}} = (R_p/R_*)^2 \alpha F_*$. Here $F_{\mbox{\footnotesize in}}$ denotes the in-transit flux, and $R_p$ and $R_*$ the planetary and stellar radii, respectively. The transit depth is defined as follows:
\begin{equation}
\label{eqn:transit_depth}
p^2 = \frac{\Delta F}{F_{\mbox{\footnotesize out}}} = \left ( \frac{R_p}{R_*} \right )^2
\end{equation}
Now let us assume that the measured flux includes some background, $B$, constant over time. Note that $B$ will affect the absolute flux values, but not the difference $\Delta F$. Therefore, the observed transit depth becomes
\begin{equation}
\label{eqn:transit_depth_background}
p^2(B) = \frac{\Delta F}{F_{\mbox{\footnotesize out}}} = \frac{\Delta F}{\alpha F_* + B} = \frac{p^2}{1+\frac{B}{\alpha F_*}}.
\end{equation}
In other words, if the background is neglected, the inferred planet-to-star area ratio will be biased by a factor $[1+B/(\alpha F_*)]^{-1}$. This relation holds in the more general case with stellar limb-darkening, although the more complex light-curve morphology requires a different definition of the transit depth parameter \citep{mandel02}.

Here we propose the following ``test of aperture sizes'' to measure the background to be subtracted from the data. The purpose of this test is not finding the best aperture to optimize the S/N on the result, but rather using the different apertures to measure the impact of background on the inferred transit depth, and correcting for it. We assume a uniform background per pixel, $b$, within a region that includes the source star. The background associated with a given aperture is proportional to the number of pixels enclosed, $n$, eventually including fractional pixels: $B=nb$.  By injecting this relation in Equation~\ref{eqn:transit_depth_background}:
\begin{equation}
\label{eqn:transit_depth_aperture}
p^2(n, b) = \frac{p^2}{1+\frac{nb}{\alpha F_*}}
\end{equation}
By measuring the transit depth using different aperture sizes, the background per pixel, $b$, can be estimated as the best-fit parameter to Equation~\ref{eqn:transit_depth_aperture}.
Note that in the limit of low background ($nb \ll \alpha F_*$) and large apertures (constant $\alpha$), by using the first-order Taylor expansion for the denominator, Equation~\ref{eqn:transit_depth_aperture} becomes linear in $n$:
\begin{equation}
\label{eqn:transit_depth_aperture_linear}
p^2(n, b) \approx p^2 \left ( 1 - \frac{b}{\alpha F_*} n \right )
\end{equation}

Traditional approaches to estimating the background consist of measuring the mean/median flux integrated on a ring around the star, or on a dark sky region (that does not contain the source). Possible sources of error with these methods are contamination from the tails of the source PSF or inhomogeneities between different sky regions. The test of aperture sizes minimizes these issues, as the background is measured from the variation in transit depth using multiple apertures containing the source. A negative background may be due to inaccurate bias/dark subtraction. The test automatically corrects for these potential offsets. The case of a background star underneath the source PSF is more subtle, as it undermines the assumption of locally uniform background in Equations~\ref{eqn:transit_depth_aperture}--\ref{eqn:transit_depth_aperture_linear}.

\subsection{Detrending \textit{HST}/STIS data}
We downloaded the flat-fielded science images (extension: flt.fits, \citealp{stishand}) from the Mikulski Archive for Space Telescopes (MAST, https://archive.stsci.edu). We computed white light-curves and spectral light-curves for five passbands uniformly spaced in wavelength as in \cite{knutson07}, allowing a direct comparison of the results. We tested rectangular apertures centered along the spectral trace with the cross-dispersion width ranging from 3 to 49 pixels. The available \textit{HST} wavelength solution was adopted to delimit the apertures in the dispersion direction. 

The notable instrumental effects are short- and long-term ramps (see Figure~\ref{fig10}). The short-term ramp is a flux variation correlated with the spacecraft orbital phase, which is highly repeatable between orbits within a visit, except for the first orbit, as a result of thermal settling of the telescope at its new pointing \citep{knutson07}. \textit{HST} observations of exoplanetary transits are purposefully scheduled to have two pre-transit orbits, and discard the first one \citep{brown01b, charbonneau02, knutson07, sing11, berta12, vidal-madjar13, fischer16}. The long-term ramp approximates a linear trend in the mean flux value of each orbit. We adopt the \texttt{divide-oot} method to correct for these systematics \citep{berta12}. The two in-transit orbits are divided by the time-weighted mean of the two out-of-transit orbits. We combine the observations into a unique series ordered by the exoplanet orbital phase, using the ephemerides reported in Table~\ref{tab1}, but allowing independent phase shifts for the two visits. This approach, also followed by \cite{knutson07}, is useful for obtaining good sampling of the full transit. Finally, the \texttt{divide-oot}-corrected, joint time series are MCMC-fitted using a transit model with transit depth, orbital parameters, phase shifts, and four limb-darkening coefficients as free parameters. Additionally, we tested different Bayesian priors on the orbital parameters (this is explained in Section~\ref{ssec:IRAC_priors}), stellar limb-darkening coefficients reported in the literature, and different limb-darkening laws (see Appendix~\ref{app:ld_laws}).

\begin{figure*}[!t]
\epsscale{1.90}
\plotone{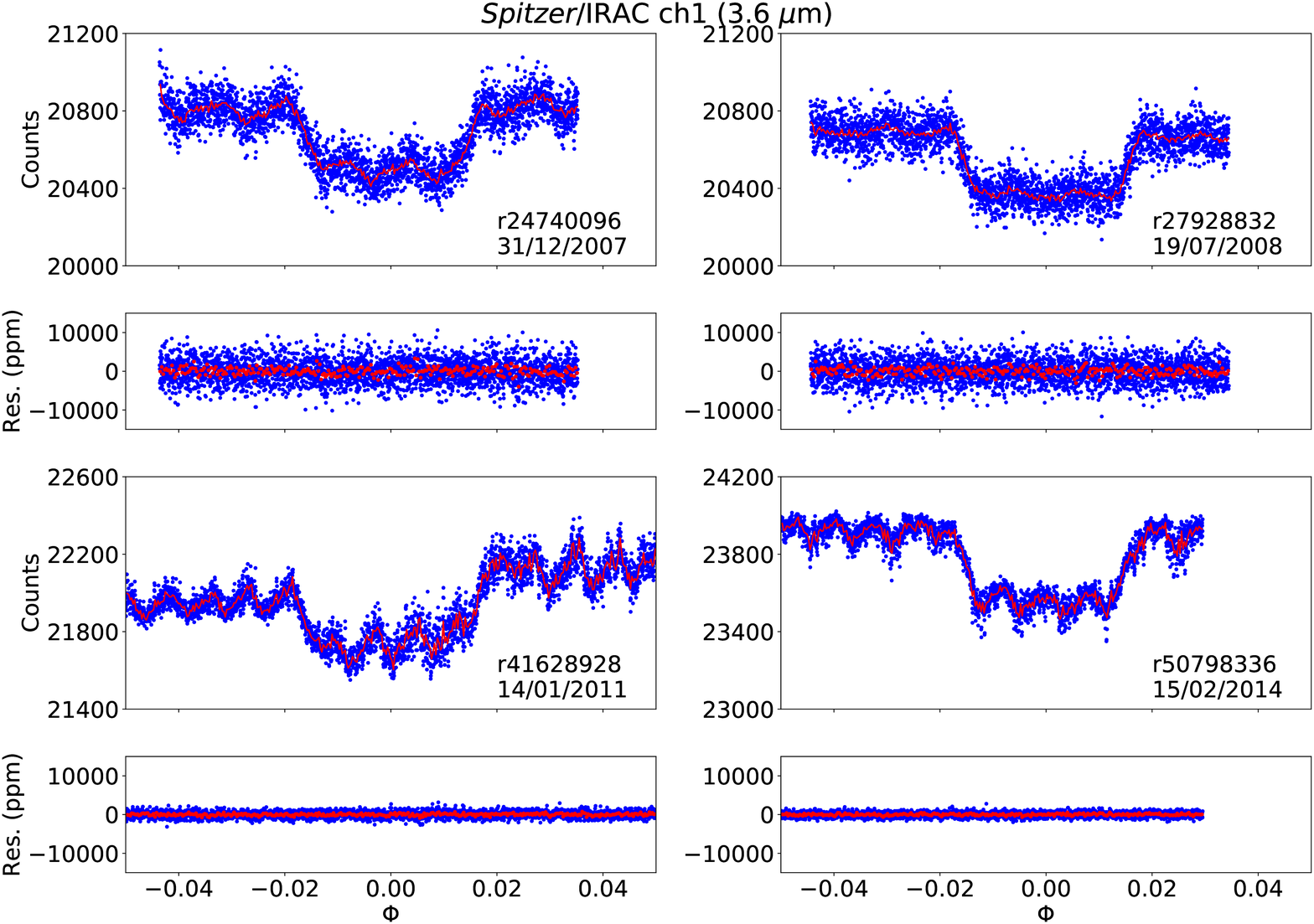}
\plotone{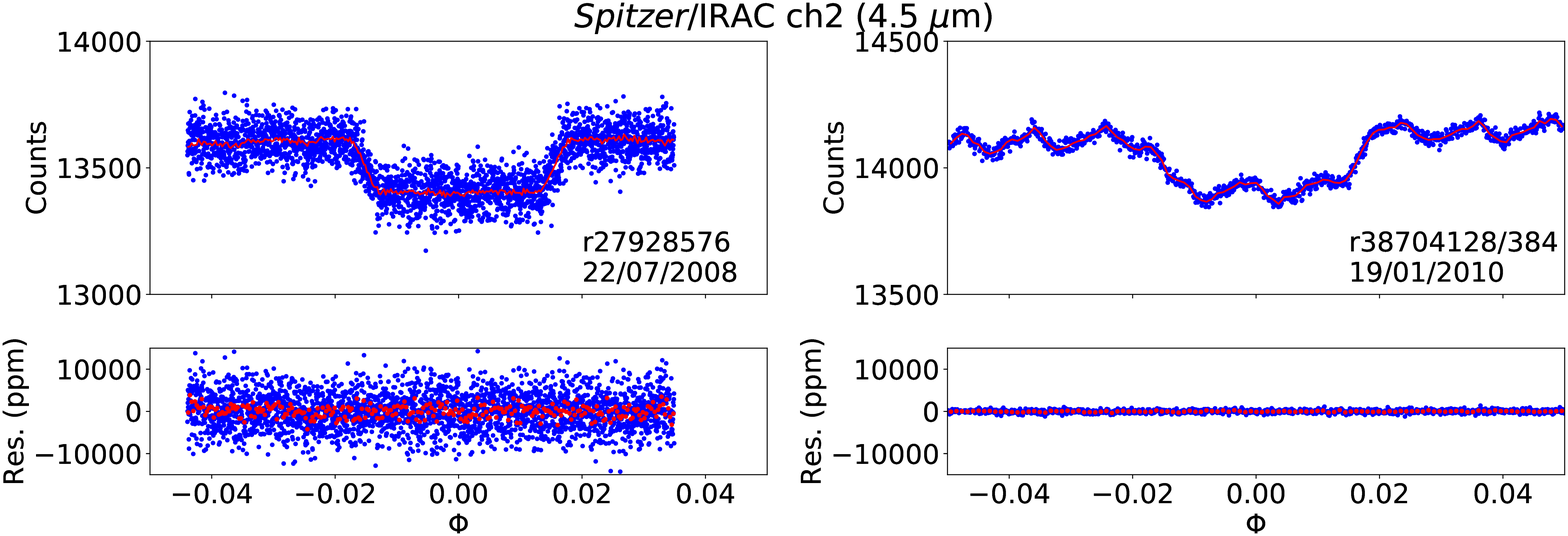}
\caption{Top panels: raw light-curves (blue dots) obtained for the \textit{Spitzer}/IRAC observations at 3.6 and 4.5 $\mu$m, and relevant best-fit models (red lines). The light-curve models are binned by a factor of 10 to improve their visualization. Bottom panels: residuals from the above light-curves and models (blue points), binned by a factor of 10 (red points).
\label{fig1}}
\end{figure*}

\begin{figure*}[!t]
\epsscale{1.0}
\begin{minipage}{0.5 \textwidth}
\plotone{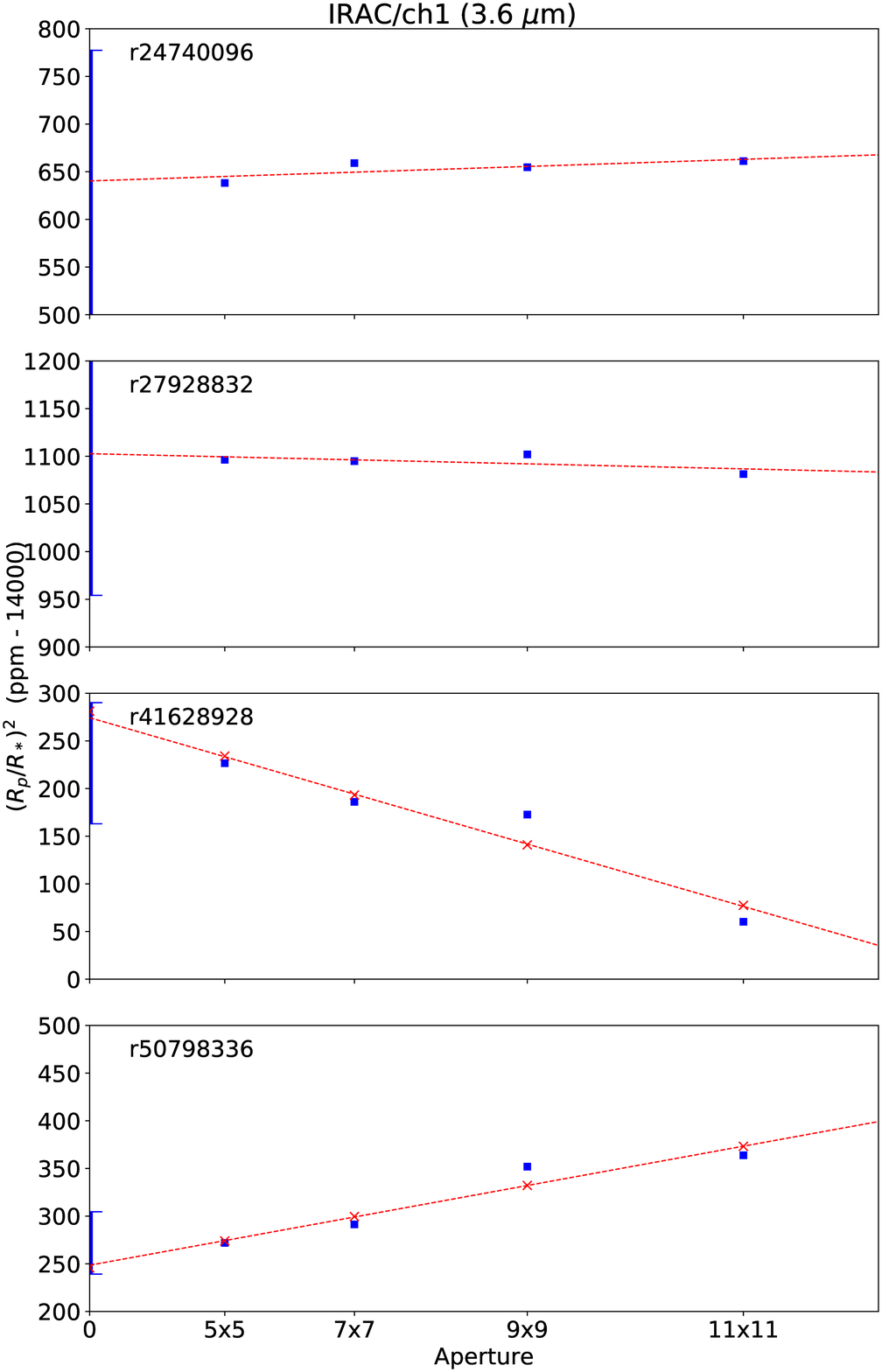}
\end{minipage}
\begin{minipage}{0.5 \textwidth}
\vspace{-6.1cm}
\plotone{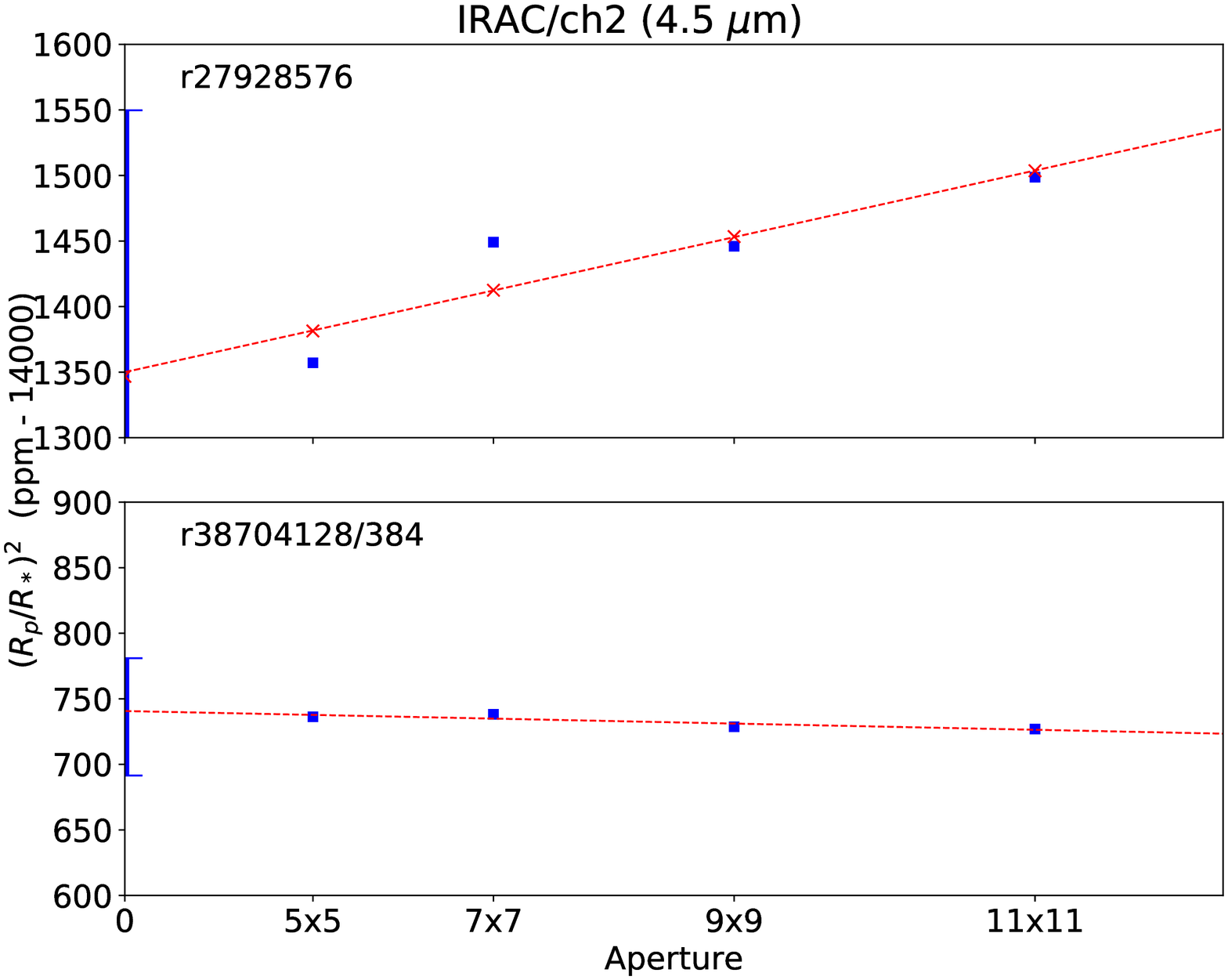}
\end{minipage}
\caption{Best-fit transit depths (blue squares) vs aperture size for the \textit{Spitzer}/IRAC observations at 3.6 and 4.5 $\mu$m, all system parameters fixed except for the transit depth. The reported error bars are for the 5$\times$5 array (other arrays lead to larger error bars). Best $p^2 (n, b)$ model (red crosses, see Equation~\ref{eqn:transit_depth_aperture}), and approximate linear model (red line). 
\label{fig2}}
\end{figure*}

\begin{figure*}[!t]
\epsscale{0.98}
\plotone{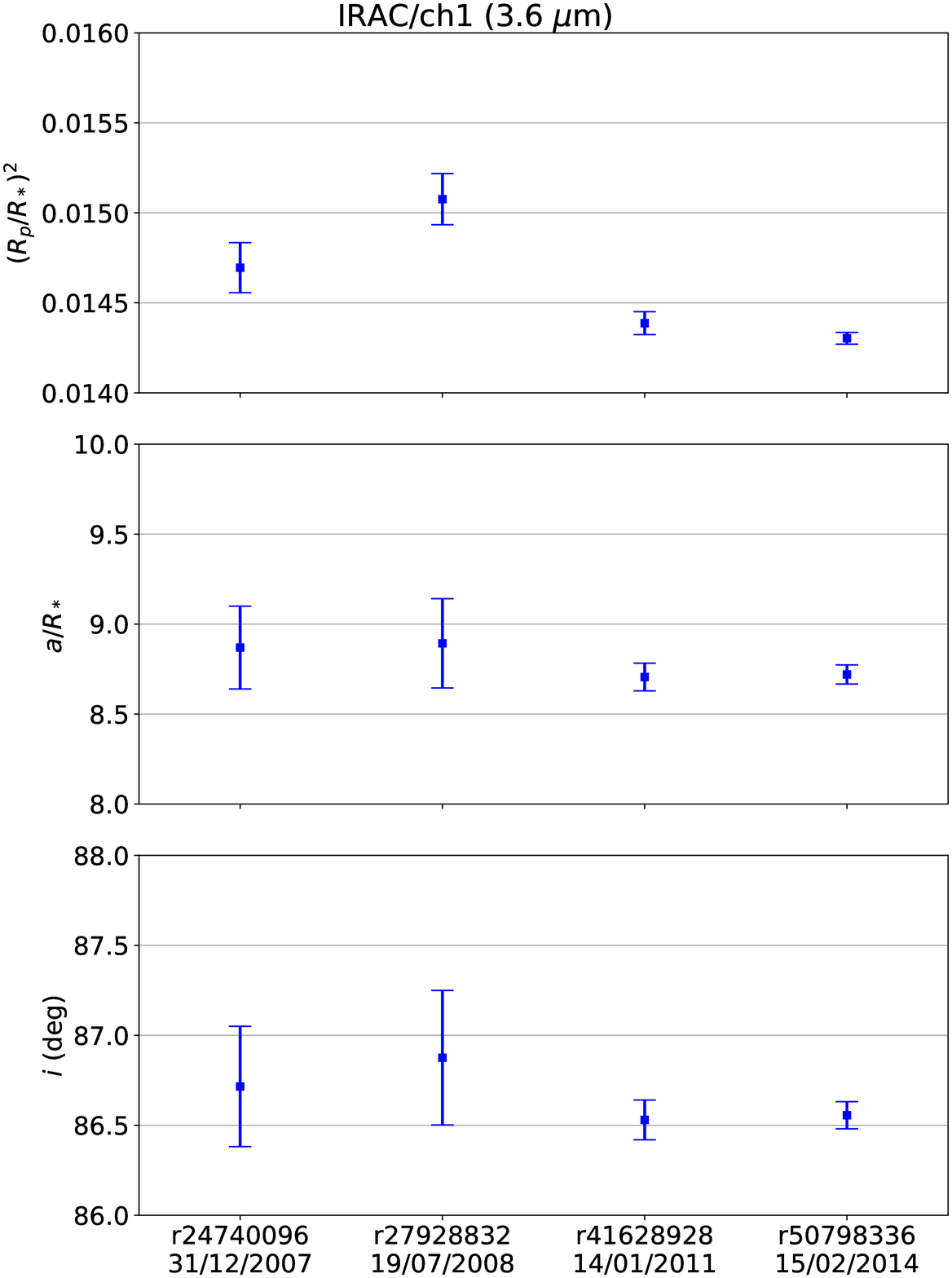}
\plotone{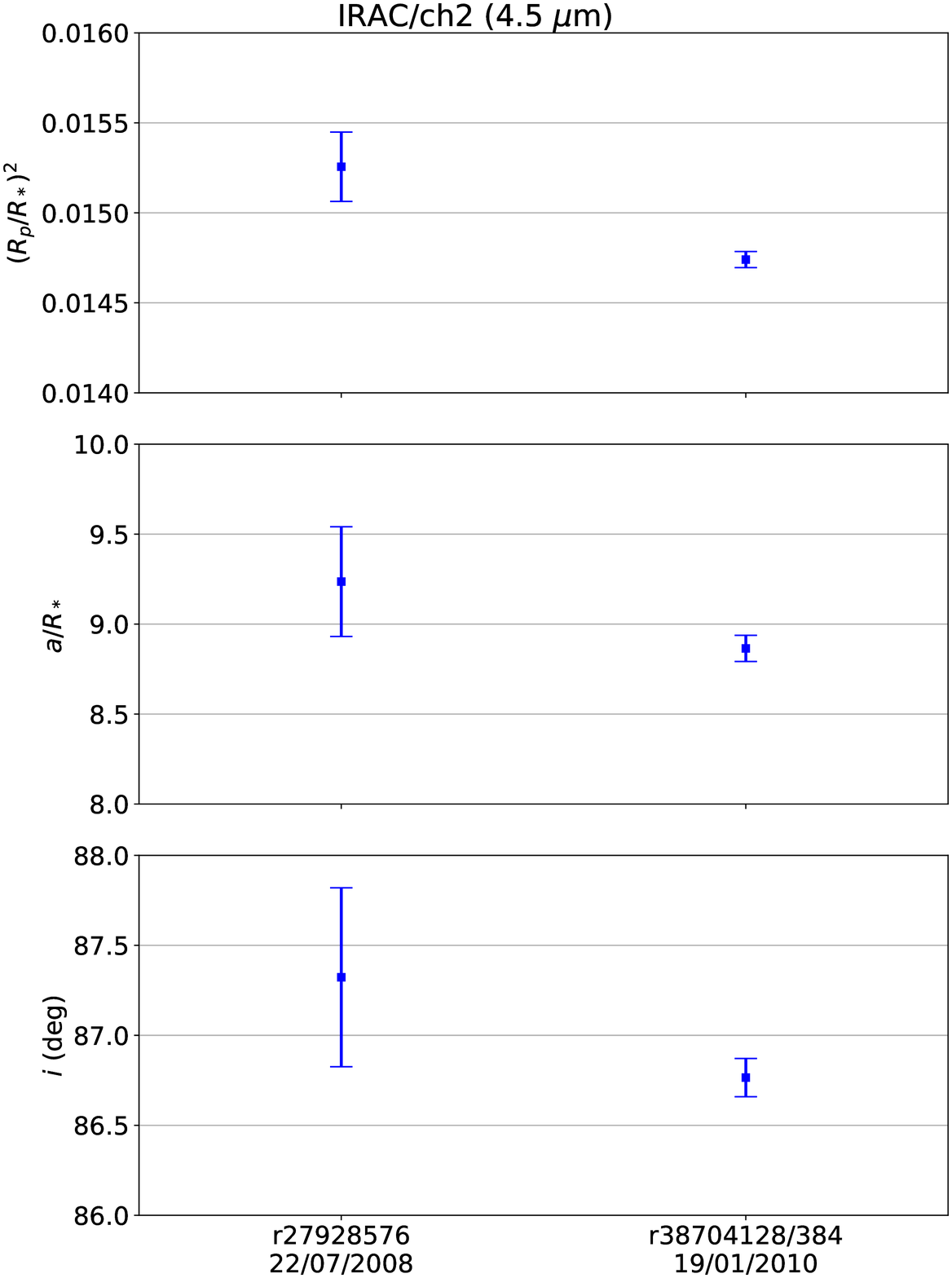}
\caption{Final parameter results for individual \textit{Spitzer}/IRAC observations at 3.6 and 4.5 $\mu$m. The earlier observations have larger uncertainties because of a less optimal observing strategy (see Sections~\ref{ssec:obs_irac} and \ref{ssec:results_irac1e2}).
\label{fig3}}
\end{figure*}

\begin{figure*}[!ht]
\epsscale{2.0}
\plotone{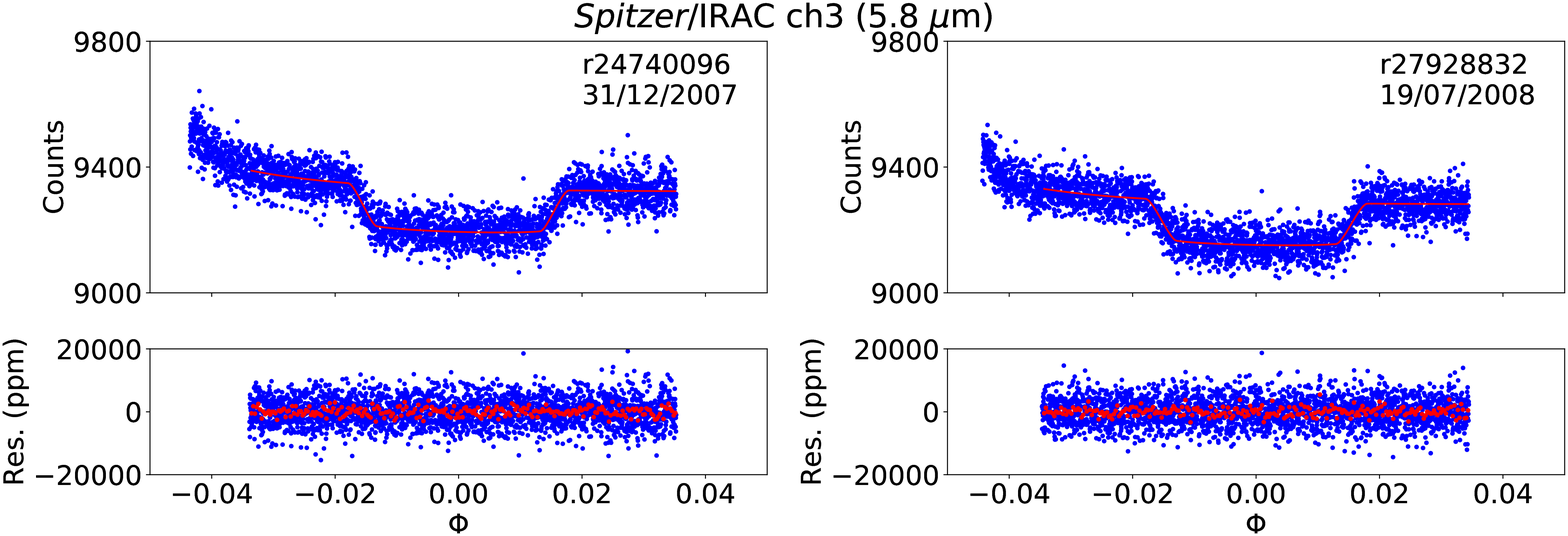}
\plotone{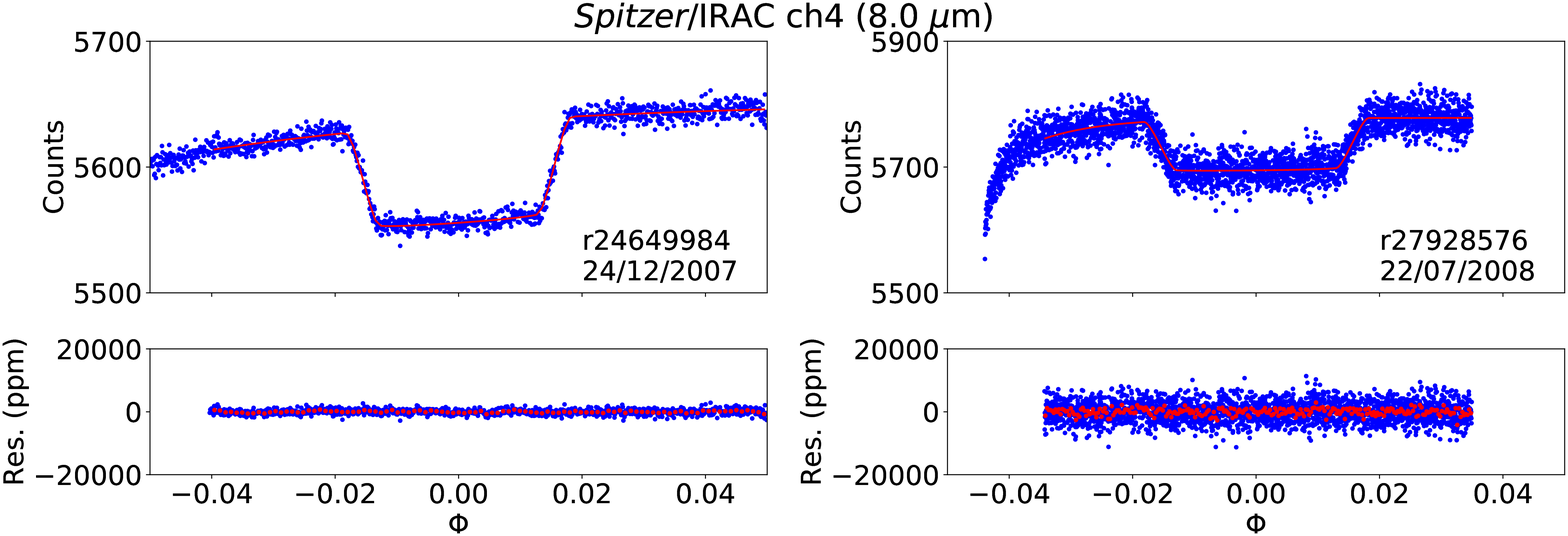}
\caption{Top panels: Raw light-curves (blue dots) obtained for the \textit{Spitzer}/IRAC observations at 5.8 and 8.0 $\mu$m, and relevant best-fit models (red lines). Bottom panels: Residuals from the above light-curves and models (blue points), binned by a factor of 10 (red points).
\label{fig4}}
\end{figure*}

\begin{figure*}[!ht]
\epsscale{1.0}
\begin{minipage}{0.5 \textwidth}
\plotone{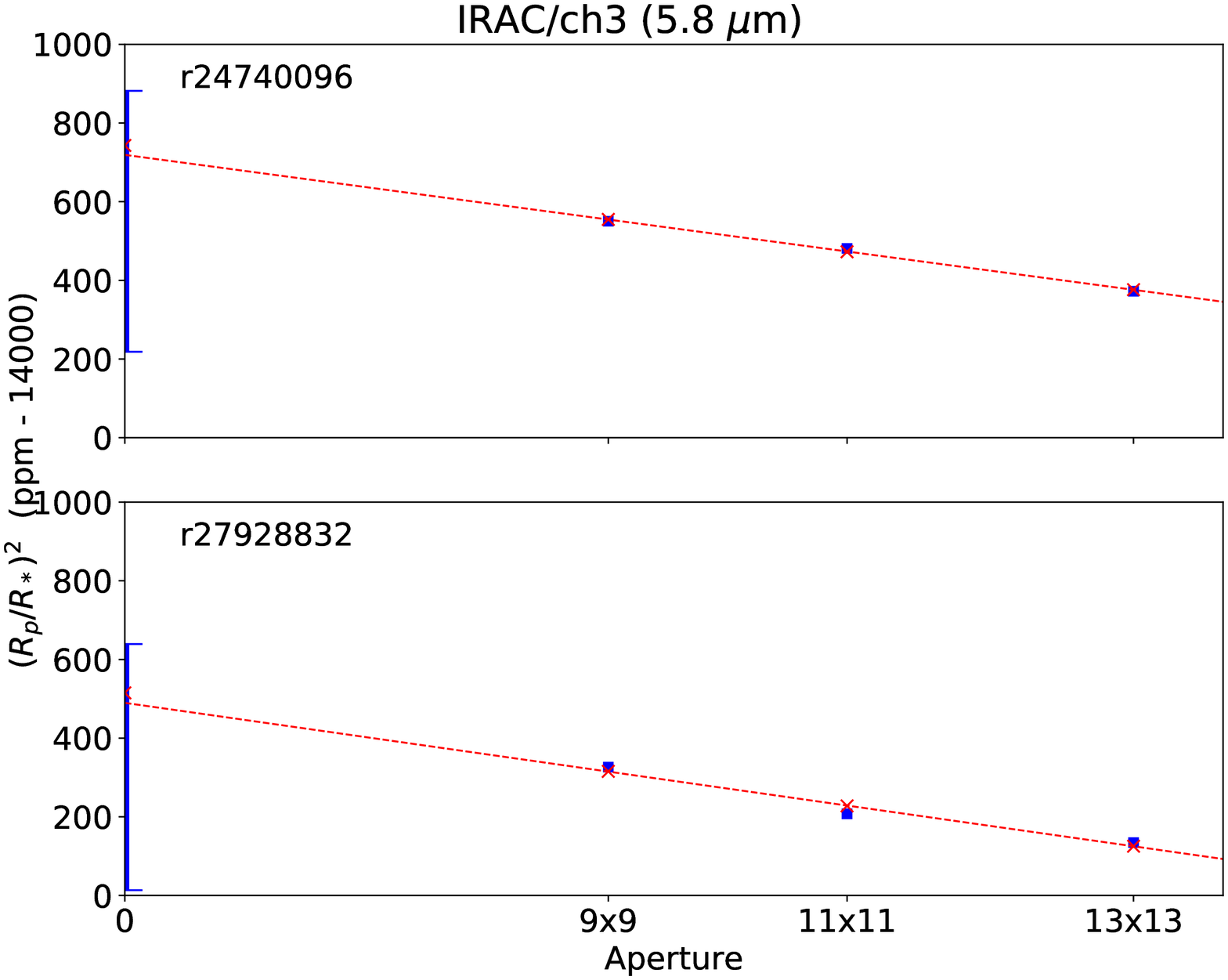}
\end{minipage}
\begin{minipage}{0.5 \textwidth}
\plotone{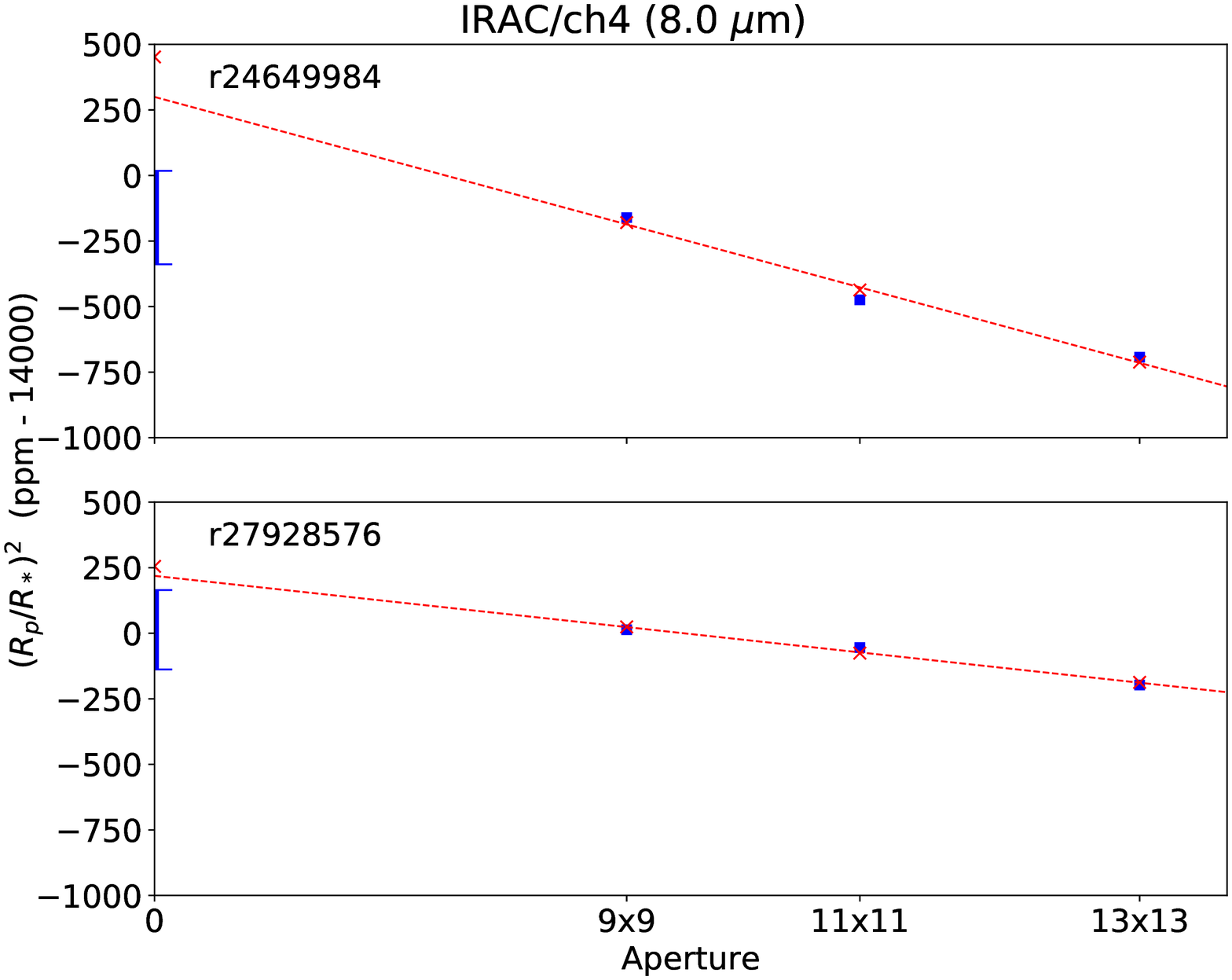}
\end{minipage}
\caption{Best-fit transit depths (blue squares) vs. aperture for the \textit{Spitzer}/IRAC observations at 5.8 and 8.0 $\mu$m, all parameters fixed except for the transit depth and the normalization factor. The reported error bars are for the 9$\times$9 array. Best $p^2 (n, b)$ model (red crosses, see Equation~\ref{eqn:transit_depth_aperture}), and approximate linear model (red line). 
\label{fig5}}
\end{figure*}

\begin{figure*}[!ht]
\epsscale{0.98}
\plotone{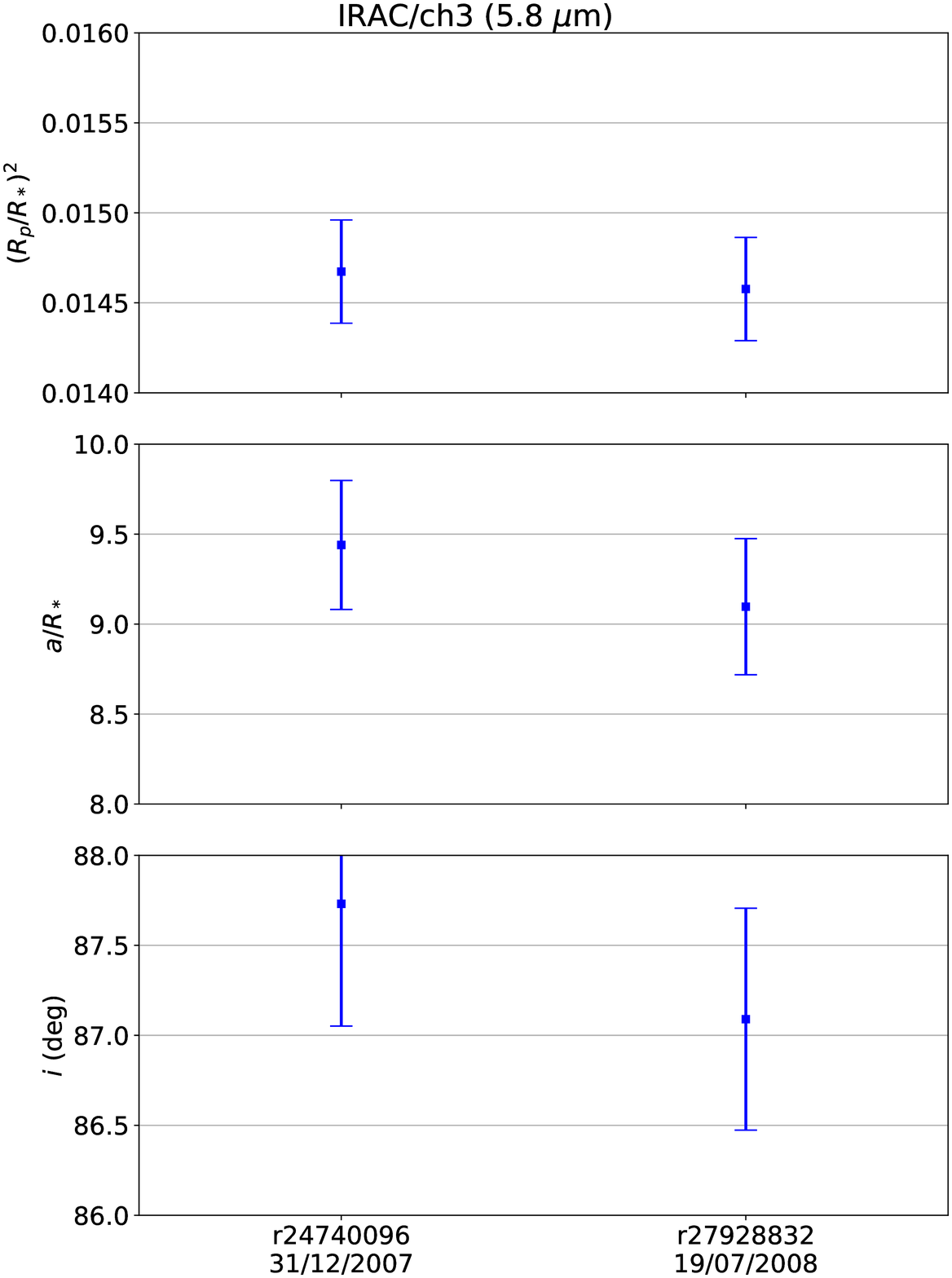}
\plotone{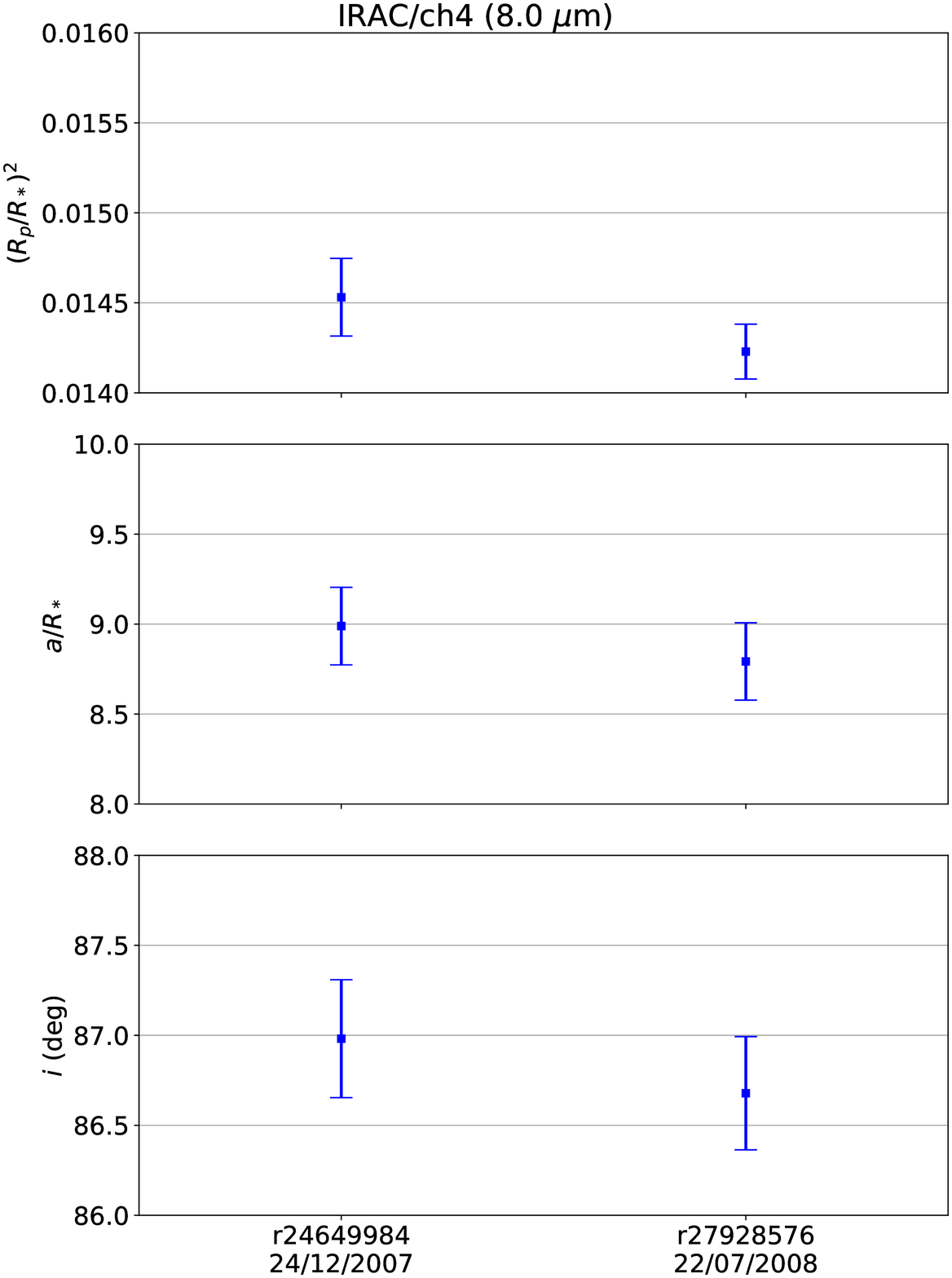}
\caption{Final parameter results for individual \textit{Spitzer}/IRAC observations at 5.8 and 8.0 $\mu$m.
\label{fig6}}
\end{figure*}

\begin{figure*}[!ht]
\epsscale{1.50}
\plotone{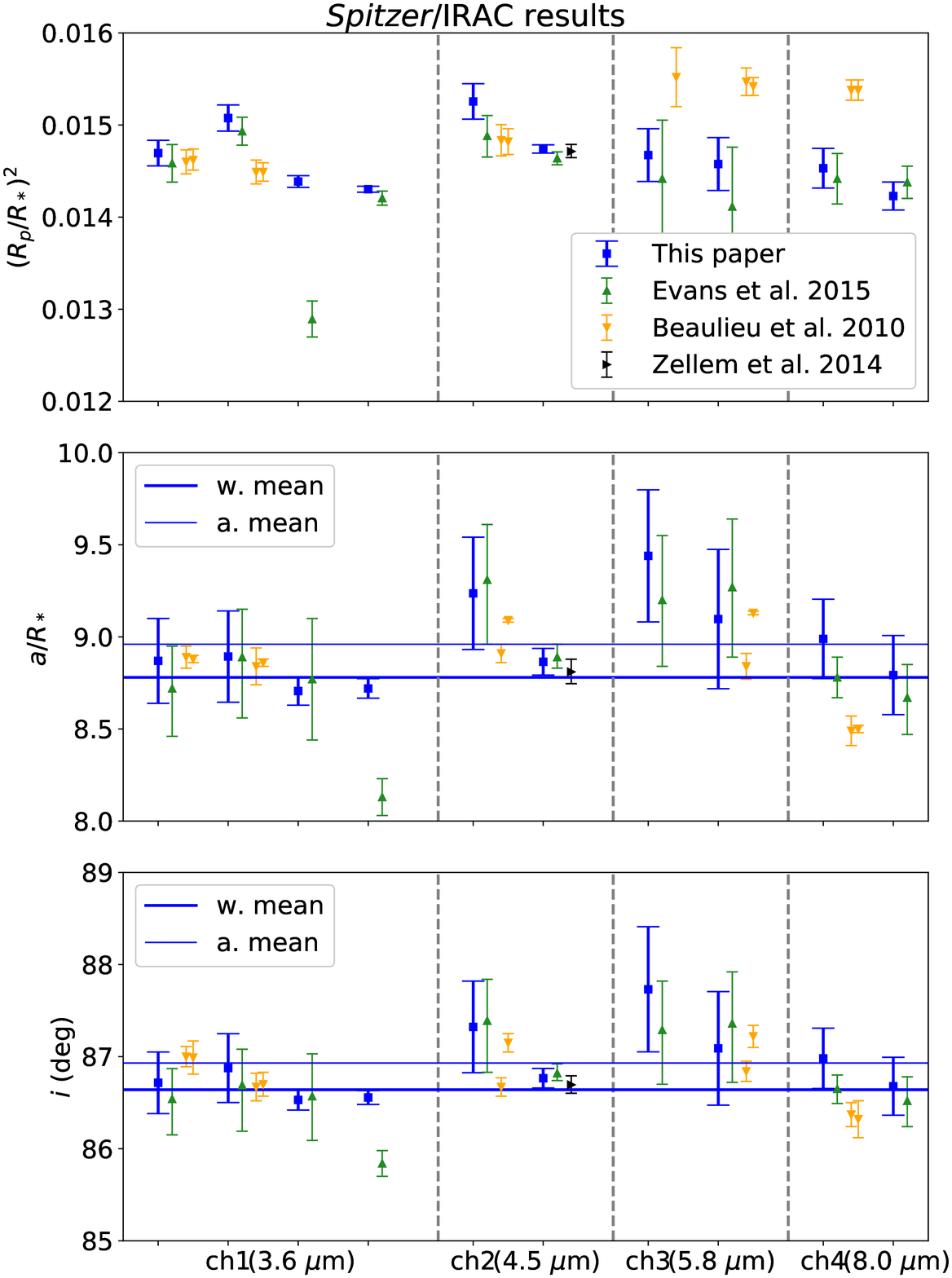}
\caption{Comparison between the \textit{Spitzer}/IRAC parameter results obtained in this paper (blue squares) with the previously published ones by \cite{evans15} (green upward-pointing triangles), \cite{beaulieu10} (yellow downward-pointing triangles), and \cite{zellem14} (black right-pointing triangles).
\label{fig7}}
\end{figure*}

\begin{figure*}[!ht]
\epsscale{1.5}
\plotone{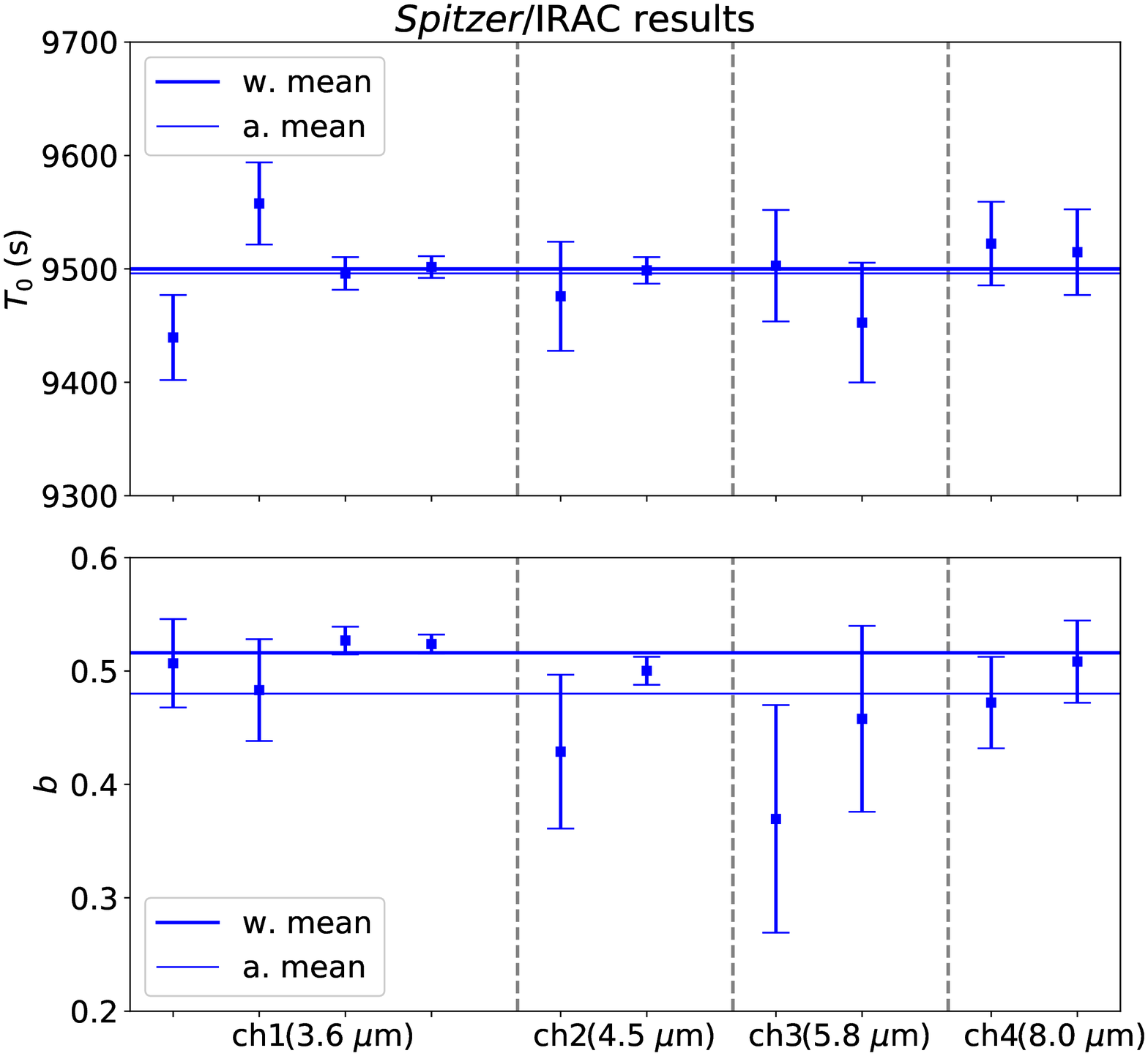}
\caption{\textit{Spitzer}/IRAC derived results for the transit duration and impact parameter, and their means.
\label{fig8}}
\end{figure*}

\section{INFRARED RESULTS}

\subsection{\textit{Spitzer}/IRAC channels 1 and 2 - 3.6 and 4.5 $\mu$m}
\label{ssec:results_irac1e2}

Figure~\ref{fig1} shows the raw light-curves, the relevant best-fit models, and residuals for the observations taken with the \textit{Spitzer}/IRAC channels 1 and 2 (3.6 and 4.5~$\mu$m, respectively).
Figure~\ref{fig2} shows the results of the test of aperture sizes (Section~\ref{ssec:test_aperture}) using four arrays: 5$\times$5, 7$\times$7, 9$\times$9, and 11$\times$11 pixels. For the two cold \textit{Spitzer} observations in channel 1 (left, top two panels) and the warm \textit{Spitzer} observation in channel 2 (right, bottom panel; see also Table~\ref{tab2}), the measured transit depths dependence on the aperture size is significantly below the uncertainty measurements. The best-fit values fall within intervals of 23 ppm (cold, channel 1) and 11 ppm (warm, channel 2). For comparison, the final error bars in transit depth for these observations are $\sim$140 ppm (cold, channel 1) and 44 ppm (warm, channel 2), i.e., they are at least four times larger than the relevant variations with the aperture size. The other three observations show clear trends in transit depth versus aperture size, even if the projected transit depths, with the background removed, are consistent with the uncorrected results within 1~$\sigma$. In two cases the transit depth increases with the aperture size, revealing a negative background, most likely due to imperfect dark subtraction.

Figure~\ref{fig3} and Table~\ref{tab5} report our final results, after background subtraction, for the transit depth and the orbital parameters. We selected the 5$\times$5 array (both as photometric aperture and set of pixel light-curves for ICA), as it gives the lowest residuals and smallest uncertainties in the ICA components (Equation~\ref{eqn:sigmaica}), in agreement with our previous studies \citep{morello14, morello16}. 

The parameters estimated from warm \textit{Spitzer} observations have 2--5 times smaller error bars than those obtained from cold \textit{Spitzer} observations, thanks to an optimized observing strategy with higher efficiency and no repointings (see Section~\ref{ssec:obs_irac}). The warm transit depths appear to be systematically smaller than the cold ones. The discrepancies between the different couples of measurements at the same wavelength are 2--5 $\sigma$, or 300--800 ppm between their central values. The other transit parameters also appear to have systematic differences between cold and warm observations, but they are consistent within 1 $\sigma$.

In Appendix~\ref{app:offset}, we show that when we inject appropriate offsets, i.e., additive constants, into the original light-curves in order to force the transit depths to assume identical values at all epochs, the corresponding best-fit orbital parameters also tend to be closer to each other. This behavior suggests that the discrepancy between cold and warm measurements is, most likely, caused by the imperfect correction of the different detector nonlinearities. If this is the case, the offset will be reduced with a careful reanalysis of the available data in collaboration with the \textit{Spitzer} team. The effect should be smaller for targets fainter than HD209458. Further studies on this hypothesis are beyond the scope of the current paper.

\subsection{\textit{Spitzer}/IRAC channels 3 and 4 - 5.8 and 8.0 $\mu$m}
\label{ssec:results_irac3e4}

Figure~\ref{fig4} shows the raw light-curves, best-fit models, and residuals for the observations taken with the \textit{Spitzer}/IRAC channels 3 and 4.
For these channels we selected larger apertures: 9$\times$9, 11$\times$11, and 13$\times$13 pixels. The reason is that the PSFs are larger for these channels, so that smaller (and static) apertures are more sensitive to the pointing variations. The experiment with the use of static apertures is to avoid the issues related to the determination of centroid coordinates (e.g., \citealp{stevenson10, lust14}). Here we focus on the results obtained by fitting a single exponential ramp with the first 49 minutes discarded from each dataset. The results obtained using different ramp parameterizations and discarded intervals are discussed in Appendix~\ref{app:ramps}. We anticipate that the best-fit orbital parameters are insensitive to the choice of ramp parameterization and discarded interval.

Figure~\ref{fig5} shows the tests of the aperture sizes for the observations taken with the \textit{Spitzer}/IRAC channels 3 and 4. The tests show stronger trends in transit depth versus aperture size than those observed at the shorter wavelengths, and these trends decrease with the aperture size, revealing a positive background (see Section~\ref{ssec:test_aperture}). The greater impact of background at the longer wavelengths is expected because of the higher zodiacal flux and the lower stellar flux \citep{krick12}. At 5.8 $\mu$m, the projected transit depths with the background subtracted are consistent with the results without subtraction within 1 $\sigma$, because the error bars are also larger. At 8.0 $\mu$m, the background subtraction changes the best-fit transit depth by 3.4 and 1.6 $\sigma$ for the two observations.

Figure~\ref{fig6} and Table~\ref{tab5} report our final results for the transit depth and the orbital parameters. We selected the 9$\times$9 array, as it gives slightly smaller residuals and error bars in most cases. The results obtained using larger arrays are almost identical (see Appendix~\ref{app:background}). All parameter estimates from observations at different epochs are mutually consistent to within 1 $\sigma$.

\begin{table*}[!ht]
\begin{center}
\caption{Final parameter results for individual \textit{Spitzer}/IRAC observations and calculated means. \label{tab5}}
\begin{tabular}{ccccccc}
\tableline\tableline
Channel & UT date & $(R_p/R_*)^2$ & $a/R_*$ & $i$ & $b$ & $T_0$ \\
($\mu$m) &  & percent  &  & (deg) &  & (s) \\
\tableline
3.6 & 2007 Dec 31 & 1.470$\pm$0.014 & 8.87$\pm$0.23 & 86.7$\pm$0.3 & 0.51$\pm$0.04 & 9439$\pm$38 \\
 & 2008 Jul 19 & 1.508$\pm$0.014 & 8.89$\pm$0.25 & 86.9$\pm$0.4 & 0.48$\pm$0.04 & 9558$\pm$36 \\
 & 2011 Jan 14 & 1.439$\pm$0.006 & 8.71$\pm$0.08 & 86.53$\pm$0.11 & 0.527$\pm$0.012 & 9496$\pm$14 \\
 & 2014 Feb 15 & 1.430$\pm$0.003 & 8.72$\pm$0.05 & 86.56$\pm$0.08 & 0.524$\pm$0.008 & 9502$\pm$10 \\
4.5 & 2008 Jul 22 & 1.526$\pm$0.019 & 9.2$\pm$0.3 & 87.3$\pm$0.5 & 0.43$\pm$0.07 & 9476$\pm$48 \\
 & 2010 Jan 19 & 1.474$\pm$0.004 & 8.86$\pm$0.07 & 86.77$\pm$0.11 & 0.500$\pm$0.012 &  9499$\pm$12 \\
5.8 & 2007 Dec 31 & 1.467$\pm$0.029 & 9.4$\pm$0.4 & 87.7$\pm$0.7 & 0.37$\pm$0.10 & 9503$\pm$49 \\
 & 2008 Jul 19 & 1.458$\pm$0.029 & 9.1$\pm$0.4 & 87.1$\pm$0.6 & 0.46$\pm$0.08 & 9453$\pm$53 \\
8.0 & 2007 Dec 24 & 1.453$\pm$0.022 & 8.99$\pm$0.22 & 87.0$\pm$0.3 & 0.47$\pm$0.04 & 9522$\pm$37 \\
 & 2008 Jul 22 & 1.423$\pm$0.015 & 8.79$\pm$0.21 & 86.7$\pm$0.3 & 0.51$\pm$0.04 & 9515$\pm$38 \\
\tableline
a. mean &  &  & 8.96$\pm$0.24 & 86.93$\pm$0.38 & 0.48$\pm$0.05 & 9496$\pm$32 \\
w. mean &  &  & 8.78$\pm$0.035 & 86.64$\pm$0.05 & 0.516$\pm$0.006 & 9500$\pm$6 \\
$\chi_0^2$ &  &  & 1.24 & 1.13 & 1.05 & 0.76 \\
\tableline
\end{tabular}
\end{center}
\end{table*}

\begin{figure*}[!ht]
\epsscale{0.98}
\plotone{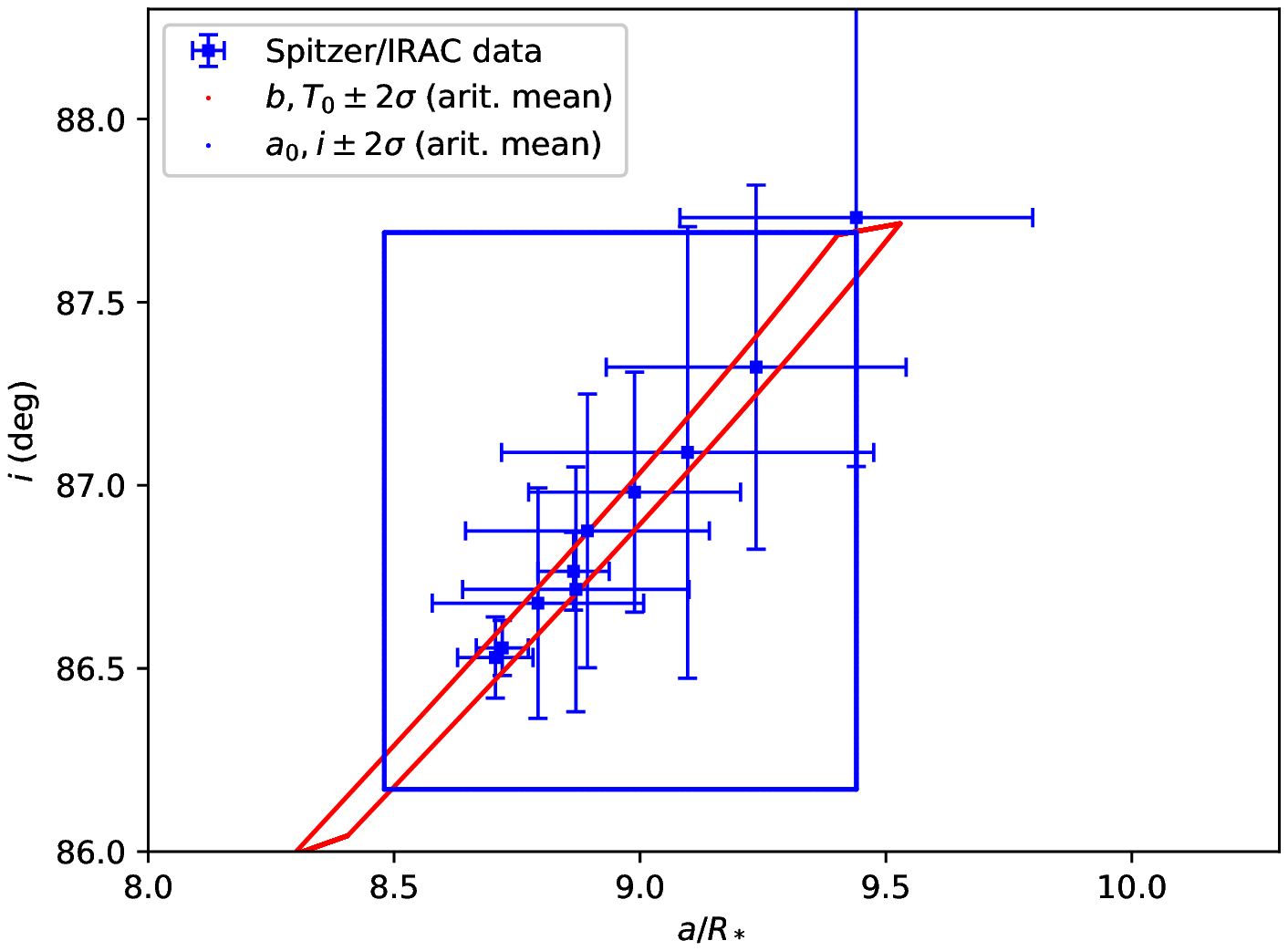}
\plotone{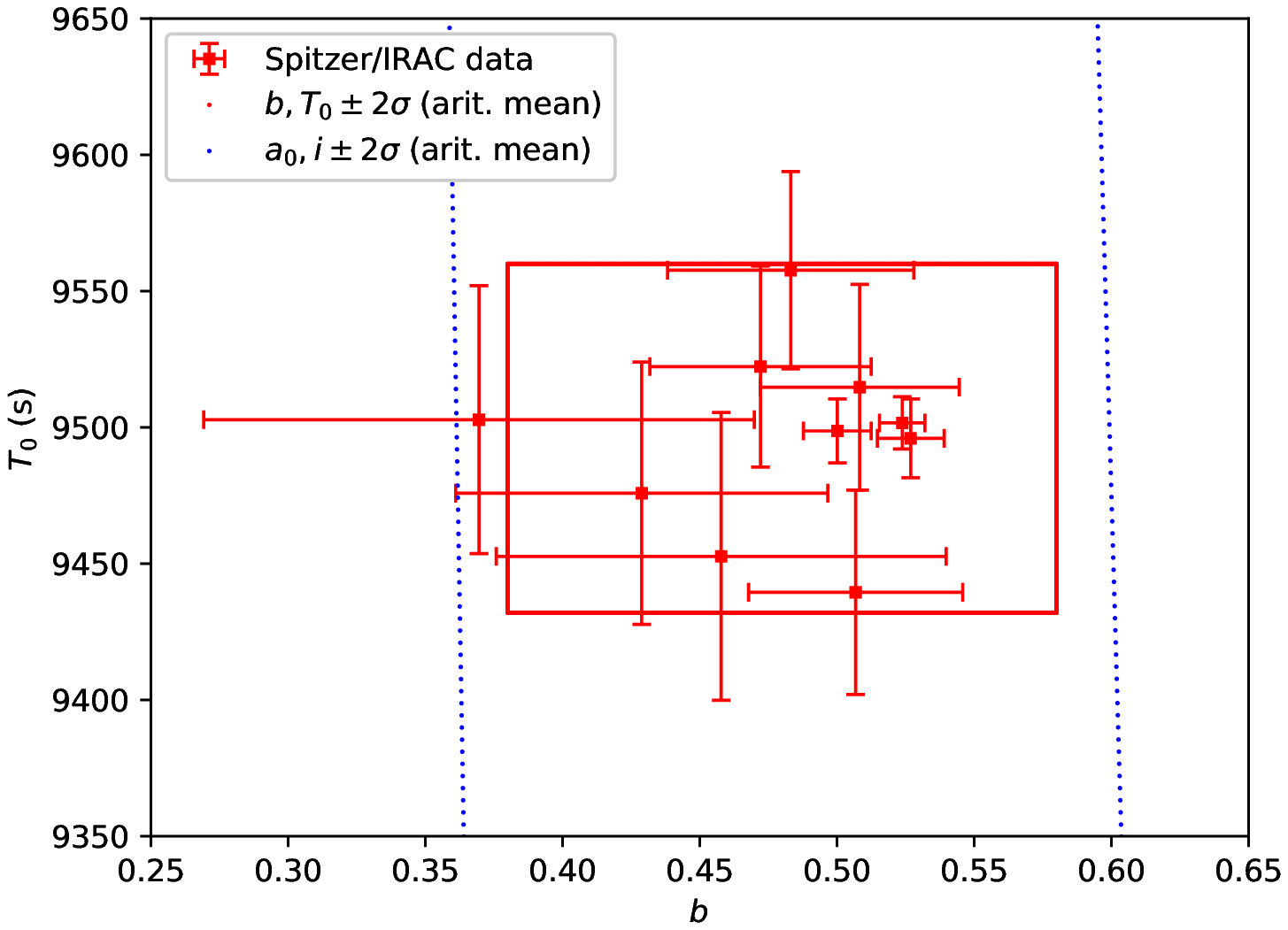}
\caption{Left panel: Scatter plot of $a/R_*$ vs. $i$ for the 10 \textit{Spitzer}/IRAC observations. The blue line encloses the 2~$\sigma$ confidence region defined by the arithmetic means on $a/R_*$ and $i$ (see Table~\ref{tab5}). The red line encloses the 2~$\sigma$ confidence region defined by the arithmetic means on $b$ and $T_0$. Right panel: Scatter plot of $b$ vs. $T_0$, with corresponding confidence regions.
\label{fig9}}
\end{figure*}

\subsection{Overview of \textit{Spitzer}/IRAC results}
\label{ssec:irac_overview}

Figure~\ref{fig7} summarizes the results obtained for the four passbands, and compares them with those reported in the previous literature. The orbital parameters estimated from different epochs and passbands are consistent within 1 $\sigma$, except in two cases, in which they are consistent within 2 $\sigma$.

Our results are very similar to those reported by \cite{evans15}, i.e., the differences are generally much smaller than 1 $\sigma$. Two exceptions are the two warm datasets at 3.6 $\mu$m, which are outliers in their analysis. Our results are also consistent within 1 $\sigma$ with those reported by \cite{beaulieu10} for the orbital parameters, except at 8.0 $\mu$m, for which their reported values are inconsistent with those obtained in the other passbands. They also obtained larger transit depths at 5.8 and 8.0 $\mu$m, most likely due to different ramp parameterizations (see Appendix~\ref{app:ramps}). It appears that the significantly smaller error bars reported by \cite{beaulieu10} do not fully account for the degeneracies between $a/R_*$ and $i$. In fact, the results that they obtained with two different fitting algorithms are inconsistent in some cases, as is clearly shown in Figure~\ref{fig7} (but the transit depths assume identical values). \cite{zellem14} analysed the 4.5 $\mu$m full phase-curve, obtaining transit parameters consistent with ours within 0.4 $\sigma$. 

\subsection{The \textit{Spitzer}/IRAC prior information}
\label{ssec:IRAC_priors}
Following the SEA BASS procedure as developed by \cite{morello17}, we derive the information on the orbital parameters from the infrared measurements in the form of gaussian probability distributions, which may be used as Bayesian priors in subsequent analyses at shorter wavelengths.

The most common way to combine $n$ multiple measurements of the same parameter, $x_i \pm \sigma_i$, into a unique estimate, $\bar x \pm \bar \sigma$, is by computing the weighted mean:
\begin{eqnarray}
\bar x = \frac{ \sum_{i=1}^{n} w_i x_i }{ \sum_{i=1}^{n} w_i }, \quad
\bar \sigma = \sqrt{ \frac{1}{ \sum_{i=1}^{n} w_i } }, 
\end{eqnarray}
where the weights are $w_i = 1/ \sigma_i^2$. The underlying assumption when taking the weighted mean is that there are no systematic errors \citep{taylor96}. A more conservative estimate is obtained by computing the arithmetic mean and standard deviation. Unlike the weighted mean error, the standard deviation is sensitive to differential systematic errors across the measurements, and does not scale as $n$ increases.
We calculated both the weighted and arithmetic means of $a/R_*$ and $i$ over the 10 \textit{Spitzer}/IRAC measurements; the results are reported in Table~\ref{tab5} and Figure~\ref{fig7}. 

It is known that $a/R_*$ and $i$ are strongly correlated parameters (e.g., \citealp{carter08}). The information on their correlation is lost if we report only their mean values and relevant uncertainties, regardless of how they are calculated. A simple way to retain this information without increasing the mathematical complexity is to consider an equivalent set of less strongly correlated parameters to express in the standard form ($\bar x \pm \bar \sigma$). A possible set consists of the impact parameter, $b$, and the ``central'' transit duration, $T_0$ \citep{seager03}:
\begin{eqnarray}
b = \frac{a}{R_*} \cos{i} \\
T_0 = \frac{P}{\pi} \arcsin{ \left ( \frac{\sqrt{1-b^2}}{\frac{a}{R_*}\sin{i}} \right ) }
\end{eqnarray}
The impact parameter is the sky-projected planet-star separation in units of
the stellar radius at the instant of inferior conjunction; the central transit duration is the length of the interval during which the planet center covers the star, i.e., the projected separation is not larger than the stellar radius.

Figure~\ref{fig8} and Table~\ref{tab5} report the individual estimates, arithmetic and weighted means of $b$ and $T_0$. These parameters are less widely scattered around their weighted means than $a/R_*$ and $i$, as measured by the smaller reduced chi-squared values, $\chi_0^2$.

Figure~\ref{fig9} shows the scatter plots for the two sets of parameters $(a/R_*, \ i)$ and $(b, \ T_0)$. In the former set, the correlation is striking, while in the latter, it is not evident by eye. The overplotted regions determined using the new set of parameter priors ($b$ and $T_0$) are also clearly more representative of the data distribution than the analogous definitions with $a/R_*$ and $i$.

\section{VISIBLE RESULTS}
\label{sec:results_STIS}
\begin{figure*}[!ht]
\epsscale{0.98}
\plotone{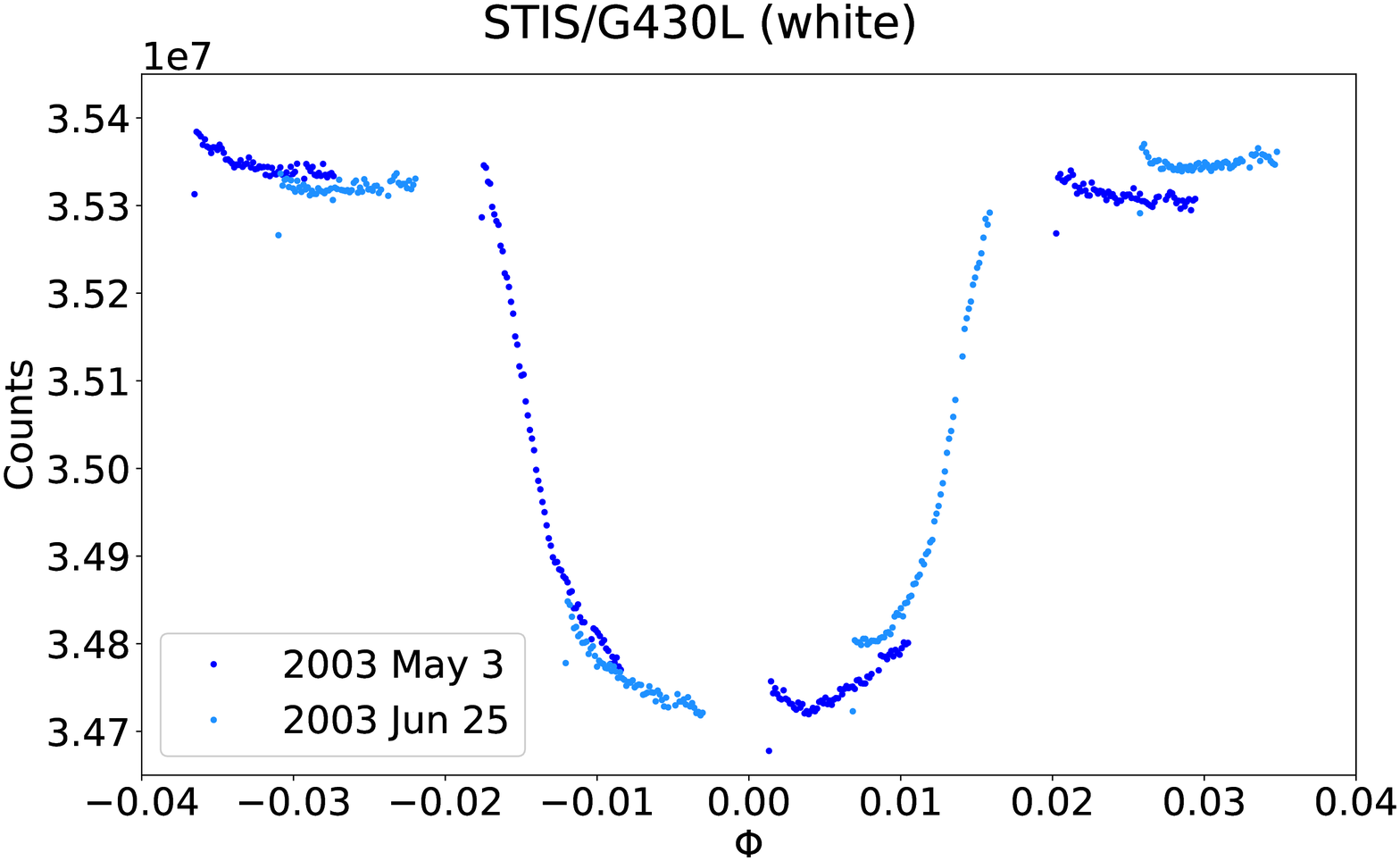}
\plotone{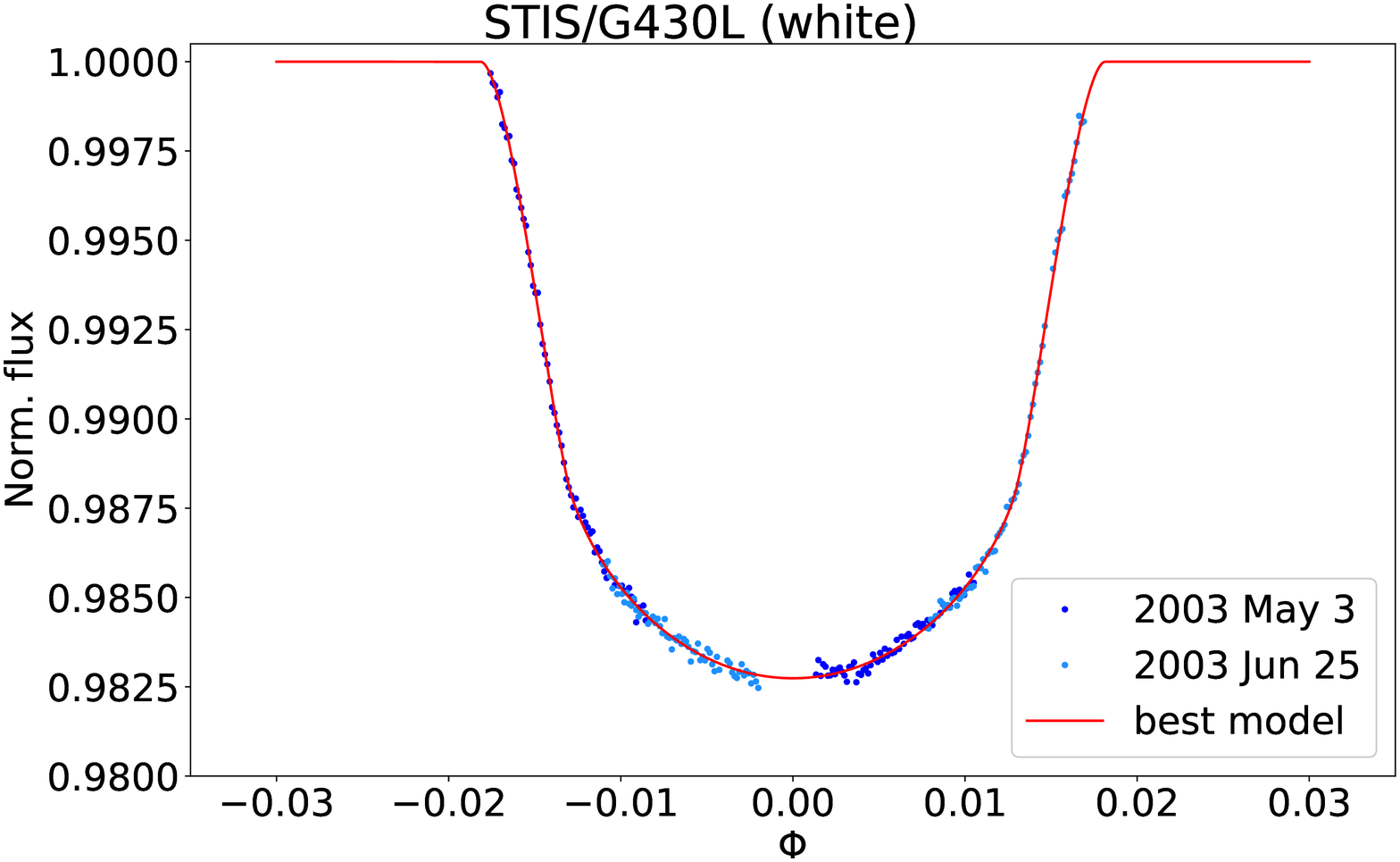}
\caption{Left panel: Raw white light-curves for the \textit{HST}/STIS G430L observations (two shades of blue). Right panel: \texttt{Divide-oot} detrended light-curves and best-fit transit model (red). The detrended out-of-transit data are not represented, as they are identical to 1 by definition of the \texttt{divide-oot} method \citep{berta12}.
\label{fig10}}
\end{figure*}

\begin{figure}[!ht]
\epsscale{0.98}
\plotone{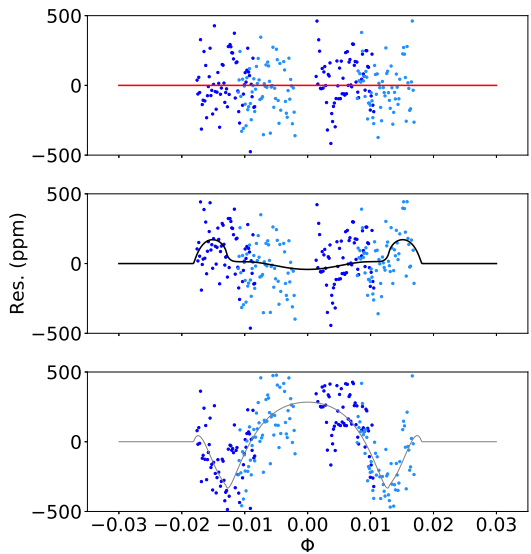}
\caption{Residuals between the detrended \textit{HST} light-curves and the different best-fit transit models. Top panel: With free limb-darkening coefficients. Middle panel: Using fixed limb-darkening coefficients, 3D model reported by \cite{hayek12}. Bottom panel: Using fixed limb-darkening coefficients, 1D model reported by \cite{hayek12}. The black and gray lines are the difference between the transit models obtained with fixed (3D and 1D respectively) and free limb-darkening coefficients.
\label{fig11}}
\end{figure}

\subsection{\textit{HST}/STIS G430L - white light-curve}
\begin{figure*}[!ht]
\epsscale{2.0}
\plotone{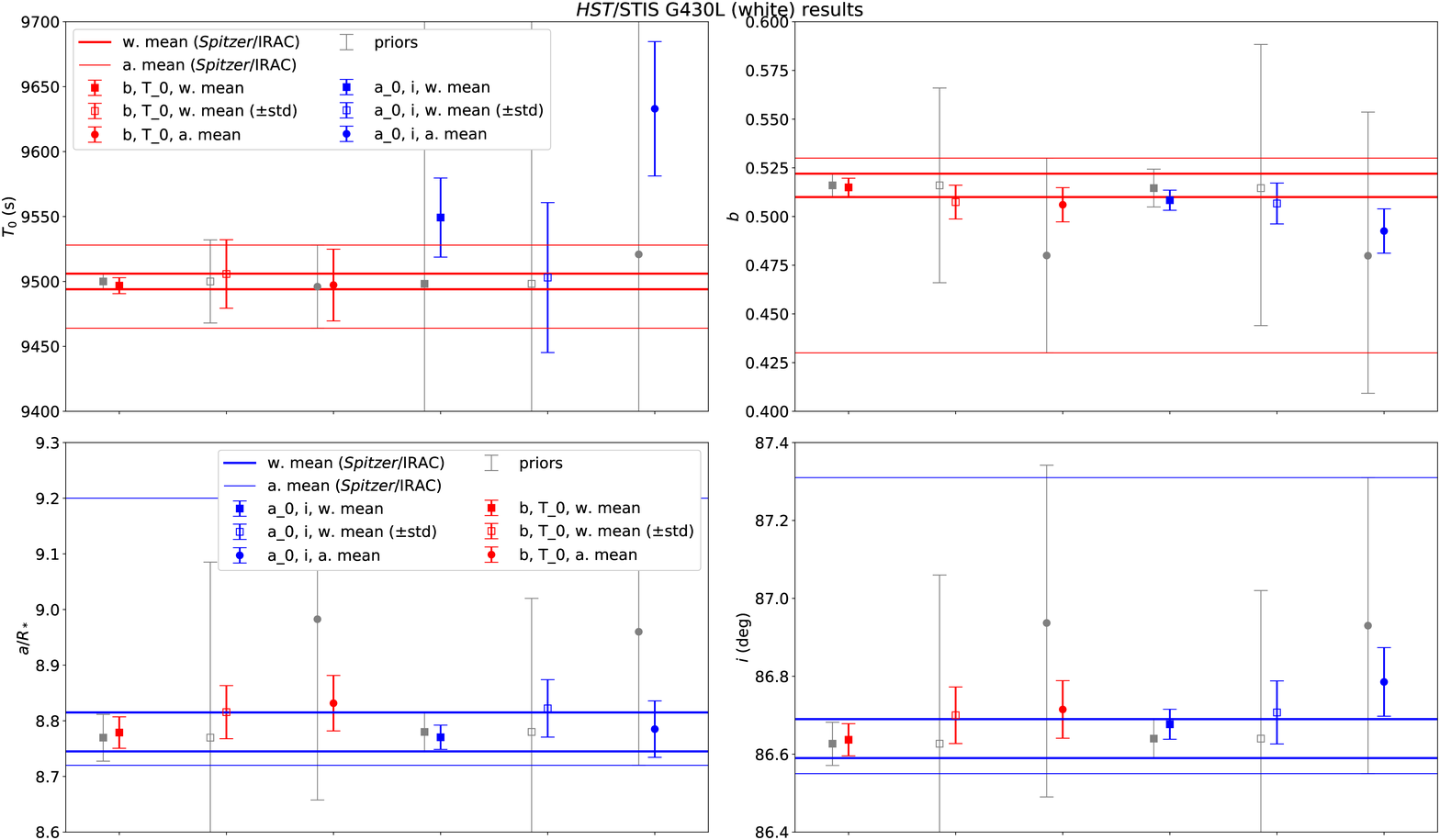}
\caption{\textit{HST}/STIS G430L parameter results for different choices of informative priors on $a/R_*$ and $i$ (blue) or $b$ and $T_0$ (red), and relevant priors (gray). The horizontal lines delimits the 1~$\sigma$ ranges associated with the arithmetic (thinner lines) and weighted (thicker lines) means.
\label{fig12}}
\end{figure*}

\begin{figure*}[!ht]
\epsscale{2.0}
\plotone{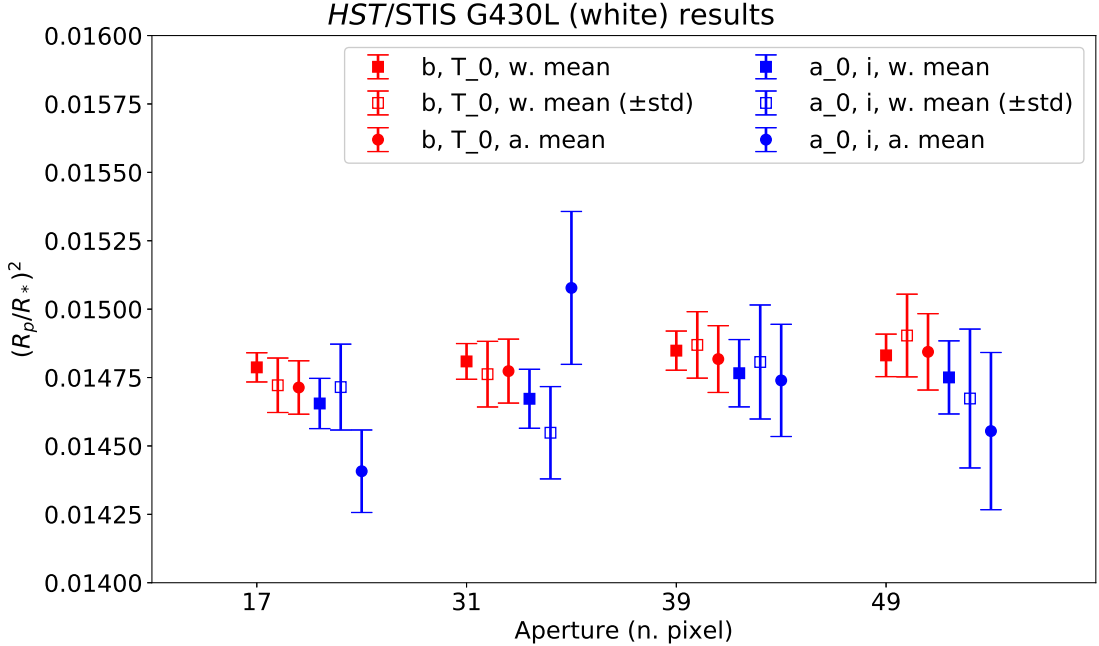}
\caption{\textit{HST}/STIS G430L white transit depth for different choices of informative priors on $a/R_*$ and $i$ (blue) or $b$ and $T_0$ (red), and for different aperture widths.
\label{fig13}}
\end{figure*}

Figure~\ref{fig10} and \ref{fig11} show the two white light-curves integrated over the STIS/G430L passband before and after data detrending, the best-fit transit model and residuals. The $\sim$1.7$\times$10$^{-4}$ rms amplitude of residuals is equal to the photon noise value. Figure~\ref{fig12} reports the best parameter results for $a/R_*$, $i$, $b$ and $T_0$ when adopting different priors (see Section~\ref{ssec:IRAC_priors}). All parameter estimates are mutually consistent within 2~$\sigma$. The parameters obtained with gaussian priors on $b$ and $T_0$ are mutually consistent well within 1~$\sigma$, and also have smaller error bars than those obtained with gaussian priors on $a/R_*$ and $i$. Overall, we note that the results are closer to the weighted rather than arithmetic mean values of the \textit{Spitzer}/IRAC results. In particular, when the arithmetic means of $b$ and $T_0$ are taken as prior (cf., third points of each panel in Figure~\ref{fig12}), the mean values of the posterior distribution are driven more than halfway toward the corresponding weighted means, while the widths are reduced by a factor of up to 6. Figure~\ref{fig13} reports the analogous results in transit depth and for different apertures (widths of 17, 31, 39, and 49 pixels in the cross-dispersion direction). They are all consistent within $\sim$1 $\sigma$, and the smallest error bars are $\sim$50 ppm. From now on, we refer to the results obtained with the 17-pixel aperture and \textit{Spitzer}/IRAC weighted mean priors on $b$ and $T_0$.

\begin{figure*}[!ht]
\epsscale{2.0}
\plotone{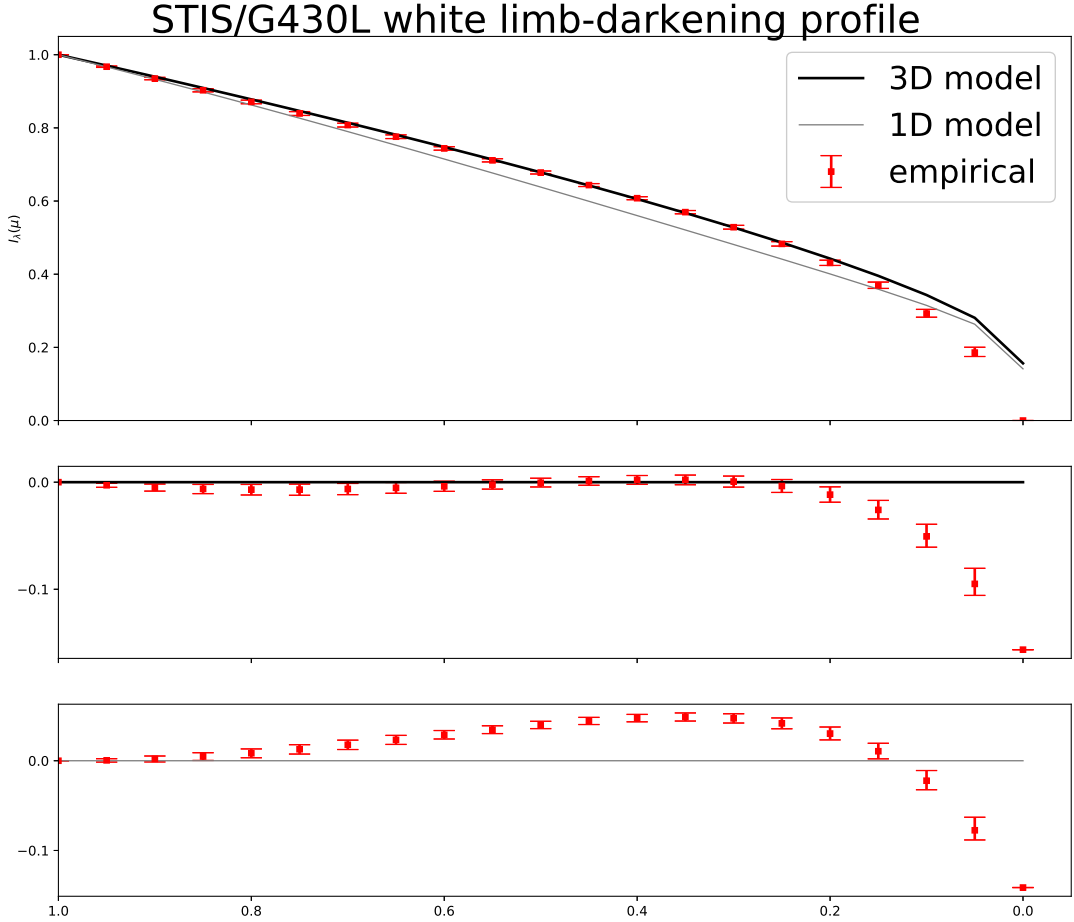}
\caption{Top panel: Empirical four-coefficient limb-darkening profile obtained from the white light-curve fit (red squares) and theoretical profiles computed by \cite{hayek12} using 3D (black, thicker line) and 1D (gray, thinner line) stellar-atmosphere models. Middle panel: Residuals between empirical and 3D profile. Bottom panel: Residuals between empirical and 1D profile.
\label{fig14}}
\end{figure*}
\begin{figure*}[!ht]
\epsscale{1.0}
\plotone{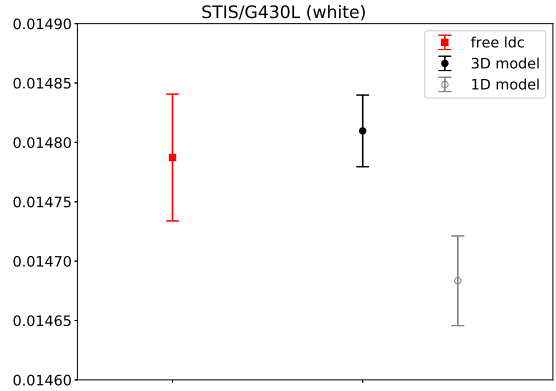}
\caption{Best-fit white-light transit depth with free four-coefficient limb-darkening (red square), and with the fixed limb-darkening coefficients reported by \cite{hayek12} using 3D (black, full circle) and 1D (gray, empty circle) stellar-atmosphere models.
\label{fig15}}
\end{figure*}

\begin{figure*}[!t]
\epsscale{2.0}
\plotone{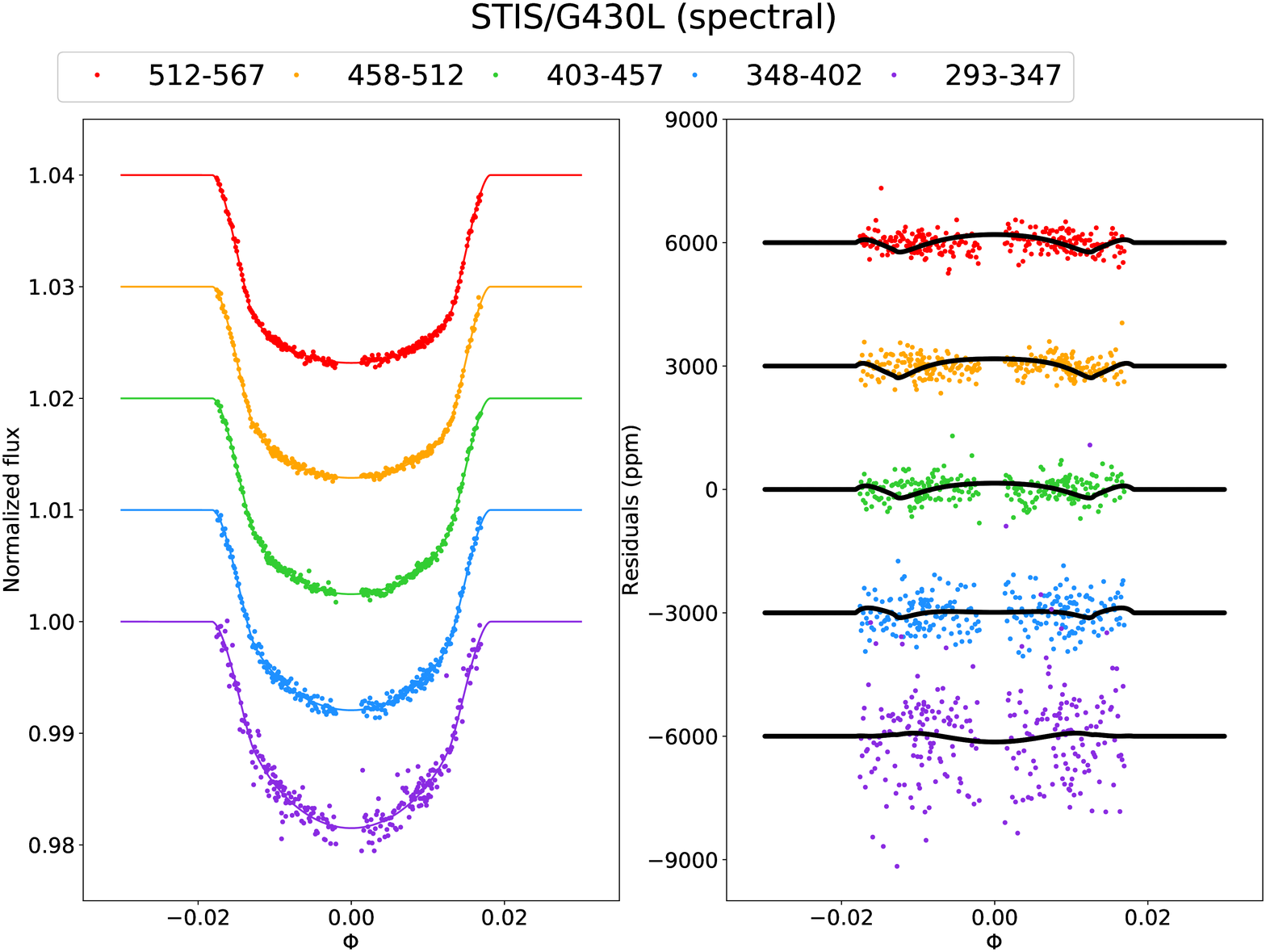}
\caption{Left panel: Detrended spectral light-curves (differently colored points) and best-fit models (same color lines), using  free four-coefficient limb-darkening, with vertical offsets for visualization purposes only. Right panel: Residuals with vertical offsets, and differences between the best-fit transit models and those obtained with limb-darkening coefficients fixed (per passband) to the values reported by \cite{knutson07} (black lines).
\label{fig16}}
\end{figure*}
\begin{figure*}[!ht]
\epsscale{2.0}
\plotone{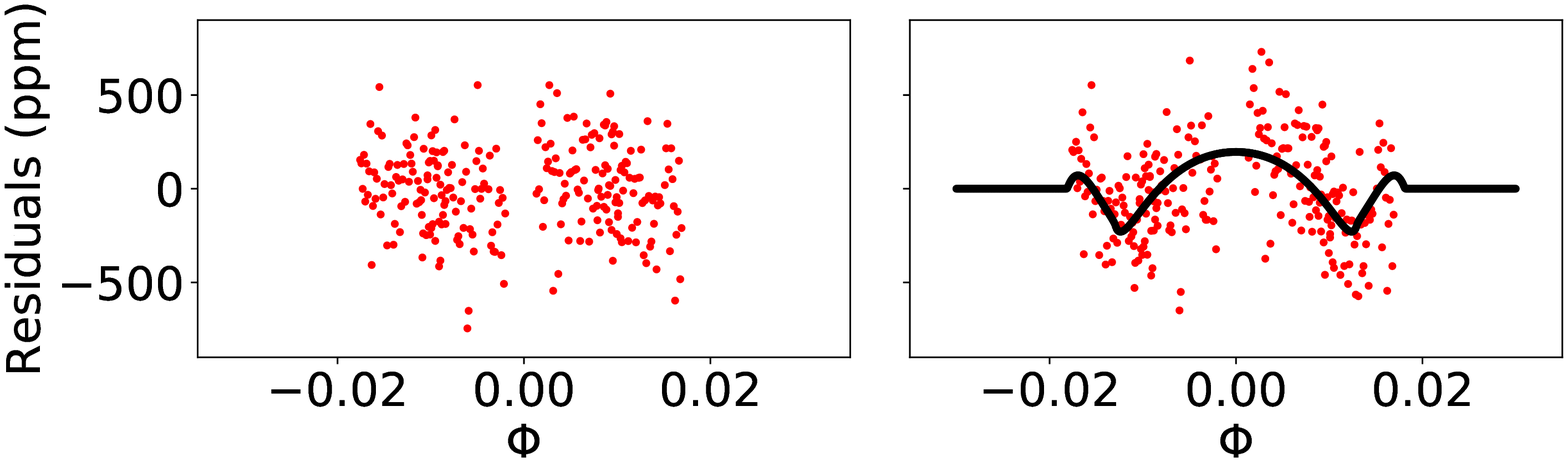}
\caption{Left panel: Zoom on the residuals for the spectral bin 512--567 nm. Right panel: Residuals obtained with limb-darkening coefficients fixed to the values reported by \cite{knutson07}, and relevant model difference (black line).
\label{fig17}}
\end{figure*}

\begin{figure*}[!t]
\epsscale{2.0}
\plotone{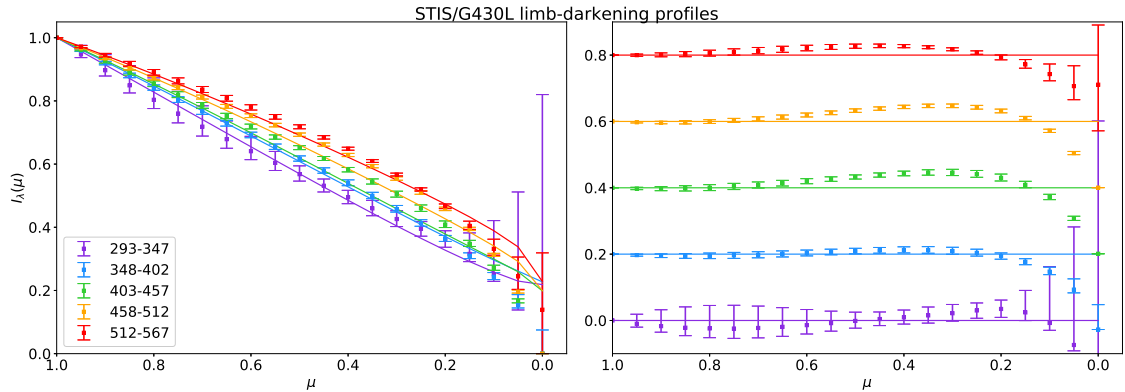}
\caption{Left panel: Empirical four-coefficient limb-darkening profiles obtained from the spectral light-curve fits (differently colored squares) and theoretical models computed by \cite{knutson07} (same color lines). Right panels: Residuals between the empirical and theoretical profiles with vertical offsets.
\label{fig18}}
\end{figure*}

\begin{figure*}[!ht]
\epsscale{2.0}
\plotone{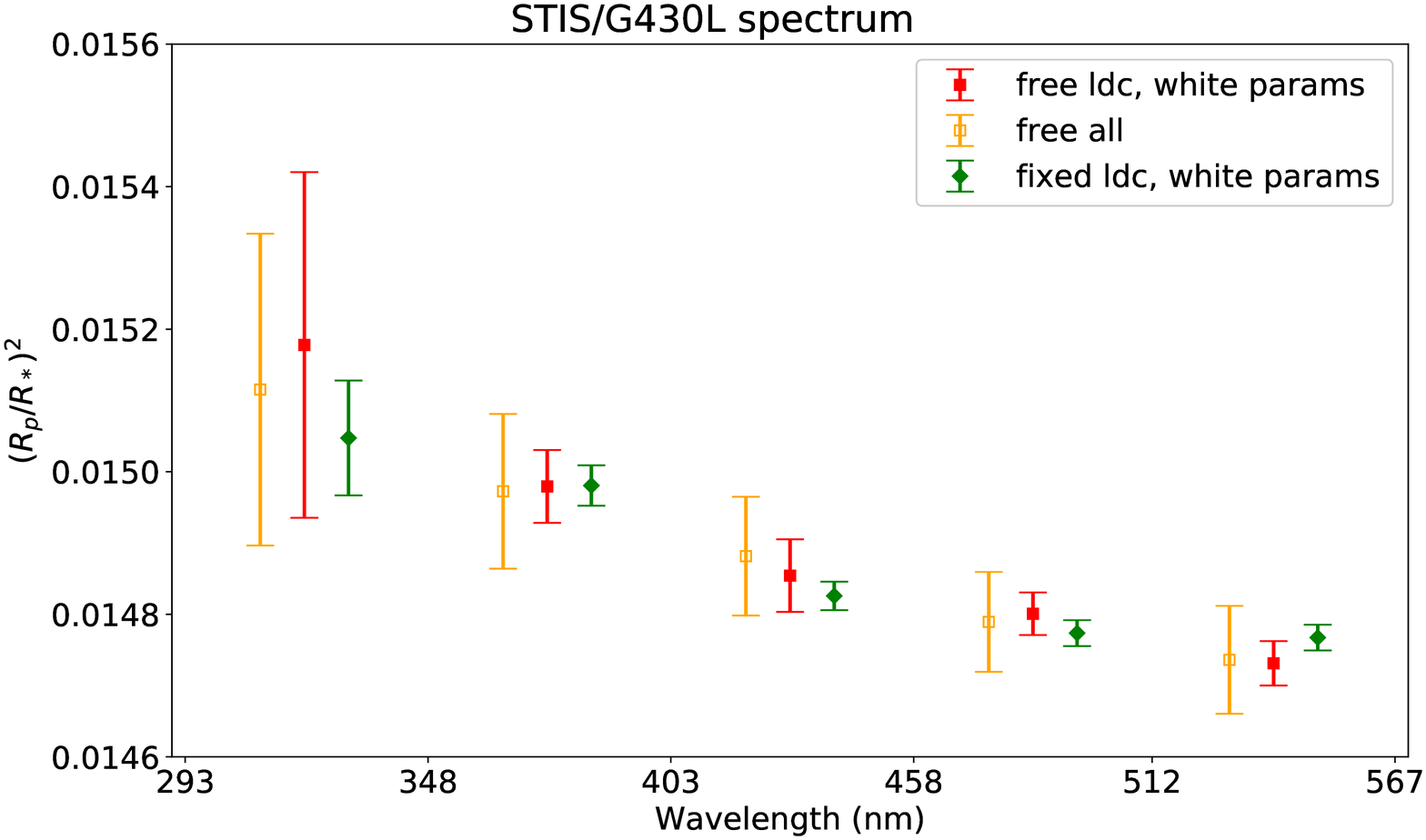}
\caption{Transit depths for the five spectral bins, using: $b$ and $T_0$ fixed to the \textit{Spitzer}/IRAC weighted mean values, phase shifts obtained from the white light-curve fit, and free limb-darkening coefficients (red, full squares), or fixed limb-darkening coefficients reported by \cite{knutson07} (green diamonds), all free parameters with \textit{Spitzer}/IRAC weighted mean priors on $b$ and $T_0$ (orange, empty squares).
\label{fig19}}
\end{figure*}

The empirical limb-darkening profile is shown in Figure~\ref{fig14}, with two theoretical profiles overplotted, derived from \cite{hayek12}. The theoretical profiles were computed ad hoc using 3D hydrodynamical stellar-atmosphere models (\texttt{stagger code}, \citealp{nordlund95}) or interpolated from a precalculated grid of 1D plane-parallel models (MARCS, \citealp{gustafsson08}). The intensities predicted by the 3D model are consistent  within 1 $\sigma$ with the empirical values inferred from our light-curve fit for $\mu>0.2$, i.e., up to 98$\%$ of the projected stellar radius, and significantly overestimated near the edge of the projected disk. In particular, the limb-to-center intensity ratios are $\gtrsim$0.15 and 0 for the theoretical and fitted models, respectively. This discrepancy has a small effect on the light-curve residuals, mainly at the ingress and egress, as shown in Figure~\ref{fig11}. The impact on the inferred transit depth is negligible, as can be noted in Figure~\ref{fig15}. The intensities predicted by the 1D model deviate significantly from the empirical results, leading to much higher systematic residuals and a greater difference in transit depth (just above 1 $\sigma$, however), as also shown in Figures~\ref{fig11} and \ref{fig15}.

The discrepancy between 3D and 1D models for HD209458 has been described before in the literature \citep{hayek12}. The novelty in this paper is the precise measurement of the stellar limb-darkening profile from the data, without using any prior information about the star, stellar-atmosphere models, or precalculated tables of coefficients. The trade-off for this admission of ignorance is an $\sim$1.8 times larger error bar in transit depth.

\subsection{\textit{HST}/STIS G430L - transmission spectrum}
We fixed the impact parameter and transit duration to their \textit{Spitzer}/IRAC weighted mean values (also supported by the white light-curve analysis), and the phase shift for each \textit{HST} visit to those estimated from the white light-curve fit. The only free parameters in the spectral fits are the transit depth and the four stellar limb-darkening coefficients. Fixing the values of the wavelength-independent parameters is common practice in exoplanet spectroscopy, where the focus is on the differences in transit depth over multiple passbands, rather than on their individual transit depth values (e.g., \citealp{sing16, tsiaras17}). The error bars do not account for potential offsets, uniform in wavelength, that may be caused by the adopted parameters.

Figure~\ref{fig16} shows the detrended spectral light-curves, best-fit transit models, residuals, and differences between these models and those obtained with fixed limb-darkening coefficients reported by \cite{knutson07}. The empirical fits are able to eliminate time-correlated noise in the residuals, which appear if the limb-darkening coefficients are fixed to those adopted by \cite{knutson07} (see also Figure~\ref{fig17}). The improvements are statistically significant for the three passbands with the highest S/Ns, leading to smaller rms residuals by 16--34$\%$ (see also Appendix~\ref{app:ld_laws}).

Figure~\ref{fig18} compares the empirical limb-darkening profiles with the reference models, showing, as expected, significant discrepancies over the three passbands with the highest S/Ns. Note that the reference limb-darkening models for the spectral bins are 1D, and the discrepancies are similar to those observed for the white light-curve analysis.
Figure~\ref{fig19} reports the final transmission spectrum and those calculated under different assumptions. Numerical results for the selected configuration are reported in Table~\ref{tab6}.  The spectrum does not change if the impact parameter, transit duration, and two-visit phase shifts are free to vary individually in all passbands, rather than fixed to the same values, but the error bars are typically twice larger. This test shows no evidence of systematic errors introduced by our choice of fixed parameters in the spectral fits. Some biases of tens of ppm are expected when using the fixed limb-darkening coefficients, but they are overcome by the error bars. Similar discrepancies will be statistically significant in the near future, thanks to the smaller error bars that are expected with future instruments onboard the \textit{James Webb Space Telescope} (JWST; \citealp{beichman14}) and other missions.

\begin{table}[h]
\begin{center}
\caption{Final transit depth estimates for \textit{HST}/STIS G430L passbands (white and spectral bins). \label{tab6}}
\begin{tabular}{cc}
\tableline\tableline
Passband & $(R_p/R_*)^2$ \\
(nm) & percent \\
\tableline
290--570 & 1.479$\pm$0.005 \\
\tableline
512--567 & 1.4706$\pm$0.0024 \\
458--512 & 1.4736$\pm$0.0023 \\
403--457 & 1.4772$\pm$0.0027 \\
348--402 & 1.494$\pm$0.004 \\
293--347 & 1.510$\pm$0.012 \\
\tableline
\end{tabular}
\\
\textbf{Note.} The spectral error bars are not marginalized over the orbital and timing parameters, as discussed in the text.
\end{center}
\end{table}

\section{CONCLUSIONS}
We performed a self-consistent reanalysis of infrared-to-visible transits of the exoplanet HD209458b, applying an upgrade of the Bayesian scheme proposed by \cite{morello17}, here named SEA BASS. The SEA BASS procedure enables simultaneous investigation of the stellar and exoplanetary atmospheres by spectroscopic measurements of the stellar limb-darkening profile and the optically thick area of the transiting exoplanet. This paper reports the first application of the SEA BASS approach on real datasets. The upgrade on the previous version, tested on simulated datasets by \cite{morello17}, consists of using more than one observation over multiple infrared passbands to assess the geometric parameters with higher confidence, and the use of a set of less degenerate parameters (impact parameter and transit duration, rather than semimajor axis and inclination). 

The improved data-detrending techniques lead to consistent parameter results for all the \textit{Spitzer}/IRAC observations, including those that appear as outliers in the literature.

We measured the stellar limb-darkening profiles of the HD209458 star over multiple passbands in the wavelength range 290--570 nm. The retrieved profiles are in good agreement with the 3D \texttt{stagger} model computed by \cite{hayek12} for the whole passband, and present non-negligible deviations from the corresponding 1D MARCS model and the 1D ATLAS models reported by \cite{knutson07} for the spectral bins. The potential biases in transit depth that are due to the inaccurate limb-darkening models are not larger than a few tens of ppm, hence they are comparable to the 1~$\sigma$ error bars.

The analysis presented in this paper confirms the reliability of the SEA BASS approach with the current datasets. Future missions, such as the JWST, are expected to provide higher quality datasets that will lead to smaller error bars, which in turn will increase the importance of SEA BASS for avoiding the potential biases of theoretical models. We also note that HD209458 is relatively well known, because it is a Sun-like star. The uncertainties in the theoretical models are larger for other stellar types, especially for cooler stars.

\acknowledgments
G.M. thanks G. Tinetti, A. Tsiaras, I. P. Waldmann, I. D. Howarth, D. Dicken, C. Danielski, M. Martin-Lagarde, and P.-O. Lagage for useful discussions.
This work was supported by the LabEx P2IO, the French ANR contract 05-BLAN-NT09-573739 and the ERC project ExoLights (617119).
This research has made use of the NASA/IPAC Infrared Science Archive, which is operated by the Jet Propulsion Laboratory, California Institute of Technology, under contract with the National Aeronautics and Space Administration.
Some of the data presented in this paper were obtained from the Mikulski Archive for Space Telescopes (MAST). STScI is operated by the Association of Universities for Research in Astronomy, Inc., under NASA contract NAS5-26555.

\clearpage

\appendix

\section{\textit{Spitzer}/IRAC background}
\label{app:background}

\begin{figure}[!h]
\epsscale{0.95}
\plotone{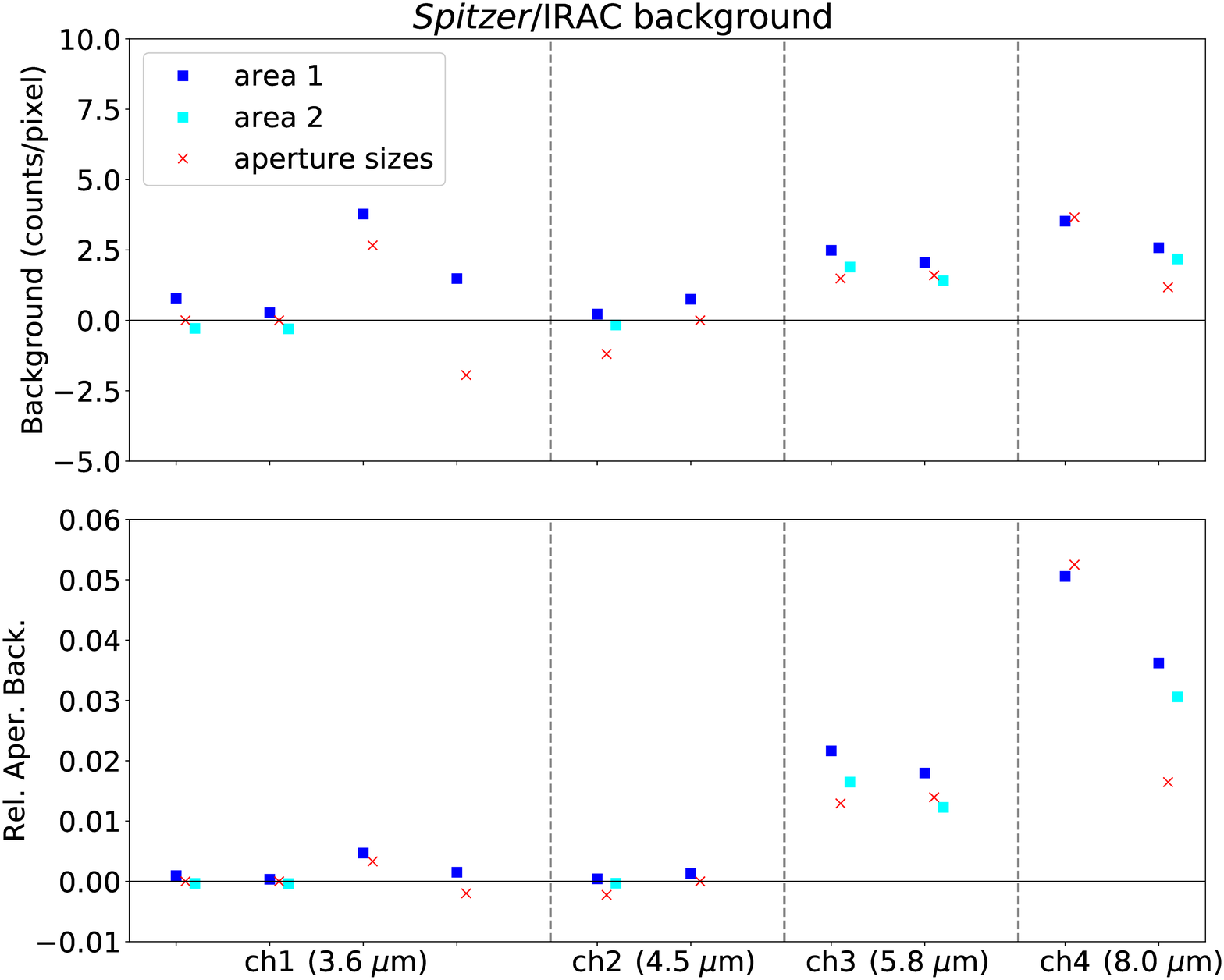}
\caption{Top panel: Background estimates in counts per pixel (observations in the same channel are ordered by date, as in Table~\ref{tab2}). Mean values over 5$\times$5 arrays in the corners of the sub-array area (area 1, blue squares), and of the full-array (area 2, cyan squares), and best-fit results from the tests of the aperture sizes (Section~\ref{ssec:test_aperture}, red crosses). Bottom panel: Array-integrated background divided by the out-of-transit flux. It shows the relatively higher background contribution in the longer wavelength channels. The gaussian noise error bars ($\sim$10$^{-3}$--10$^{-2}$ counts/pixel) cannot be visualized at the plot scale.
\label{fig20}}
\end{figure}

Figure~\ref{fig20} compares the background estimates obtained with the test of aperture sizes with classical estimates from two detector areas with no apparent astrophysical sources. The first area consists of four 5$\times$5 arrays in the corners of the sub-array area, excluding the edges. For the full-array observations, this area was defined to have the same position relative to the centroid. Analogously, the second area consists of four 5$\times$5 arrays at the edges of the full-array area (not applicable to sub-array observations). The three measurements are correlated over the multiple datasets, but the background measured within the sub-array area is systematically higher than measured at the corners of the full array, denoting some light contamination from the source. Owing to the limited size of the sub-array, this small contamination cannot be eliminated. The results of the test of aperture sizes are generally closer to the full-array estimates, in particular for the observations at 5.8~$\mu$m and the first two observations at 3.6 $\mu$m during the cold \textit{Spitzer} era. In a few cases, the test suggests significantly lower background than the classical estimates.

\begin{figure}[!t]
\epsscale{0.49}
\plotone{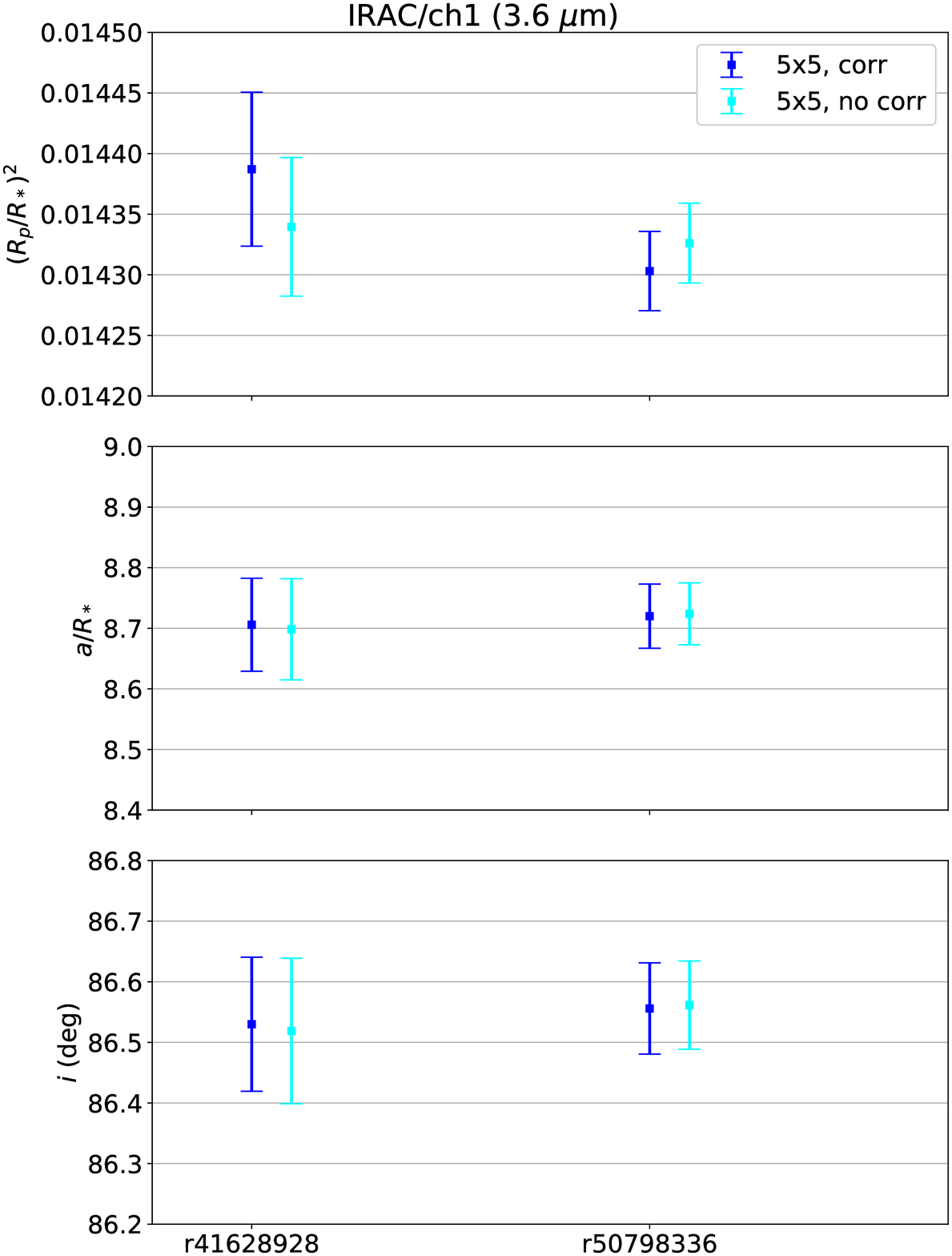}
\plotone{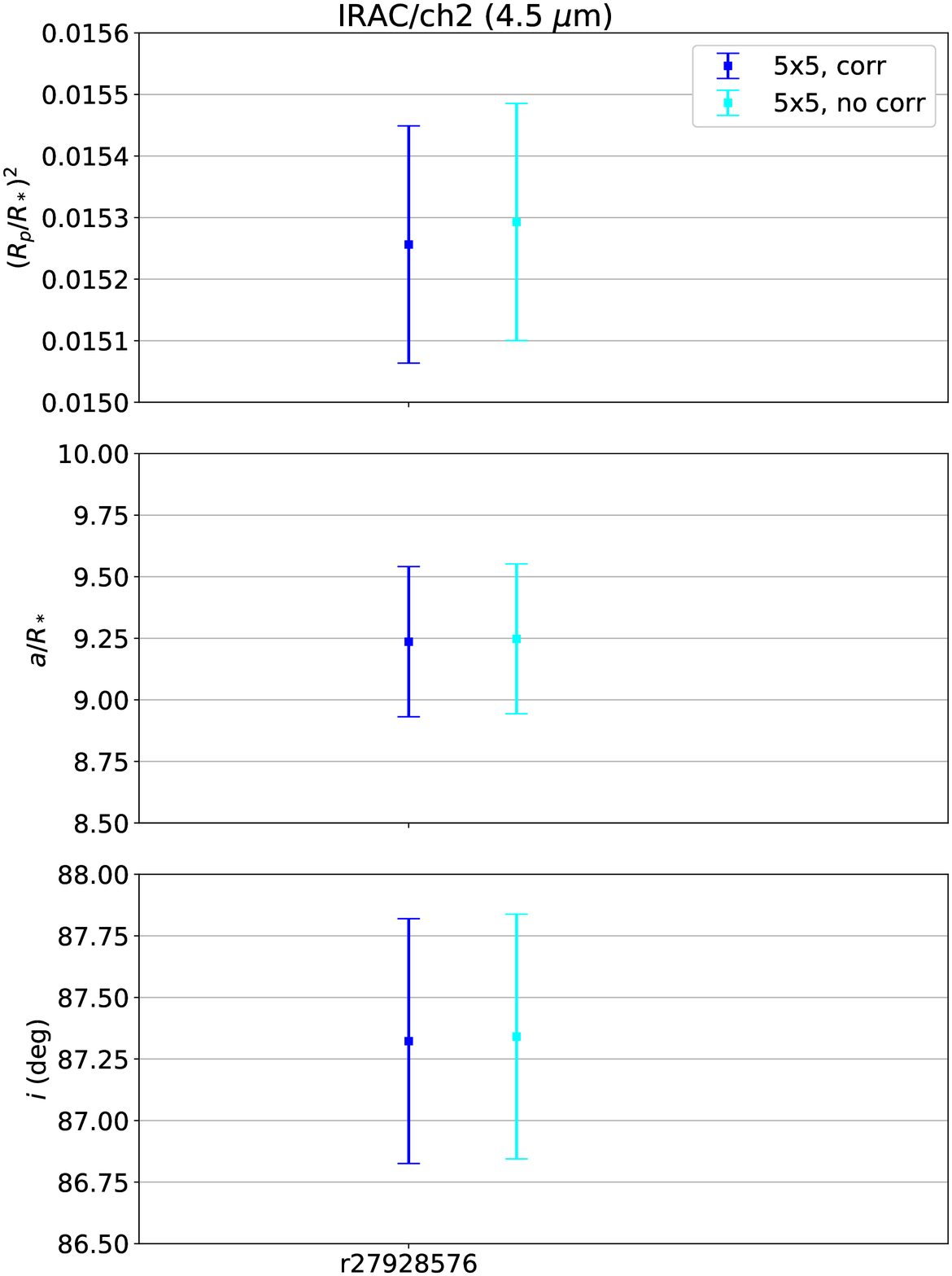}
\caption{Top, left panel:  Best-fit transit depths for two observations at 3.6 $\mu$m using 5$\times$5 arrays as photometric apertures, background corrected through the test of aperture sizes (blue) or not corrected (cyan). Top, right panel: Analogous transit depths for one observation at 4.5 $\mu$m. Middle and bottom panels: Corresponding best-fit values for $a/R_*$ and $i$. Note that different scales have been adopted on the plot axes for the two channels.
\label{fig21}}
\end{figure}
\begin{figure}[!ht]
\epsscale{0.49}
\plotone{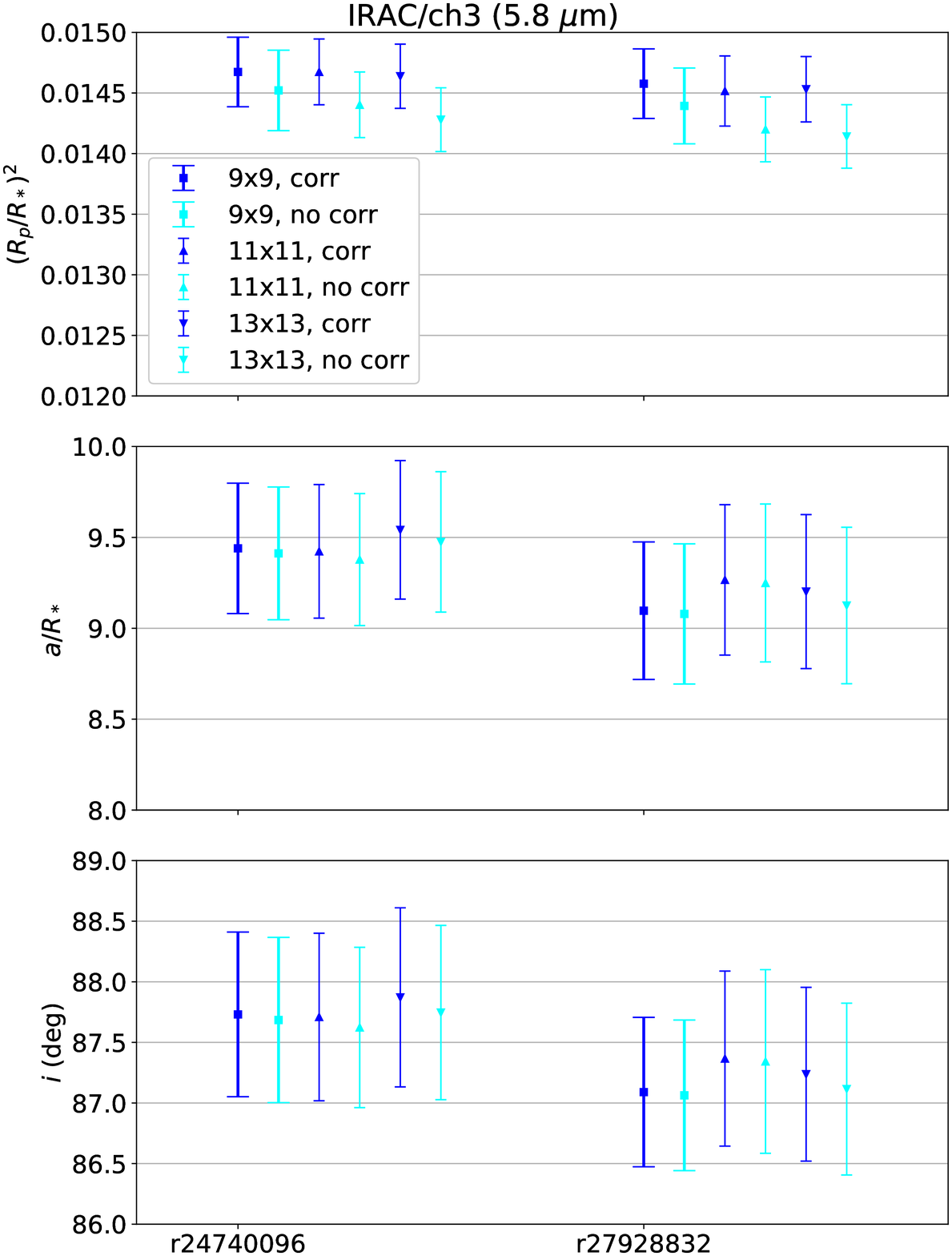}
\plotone{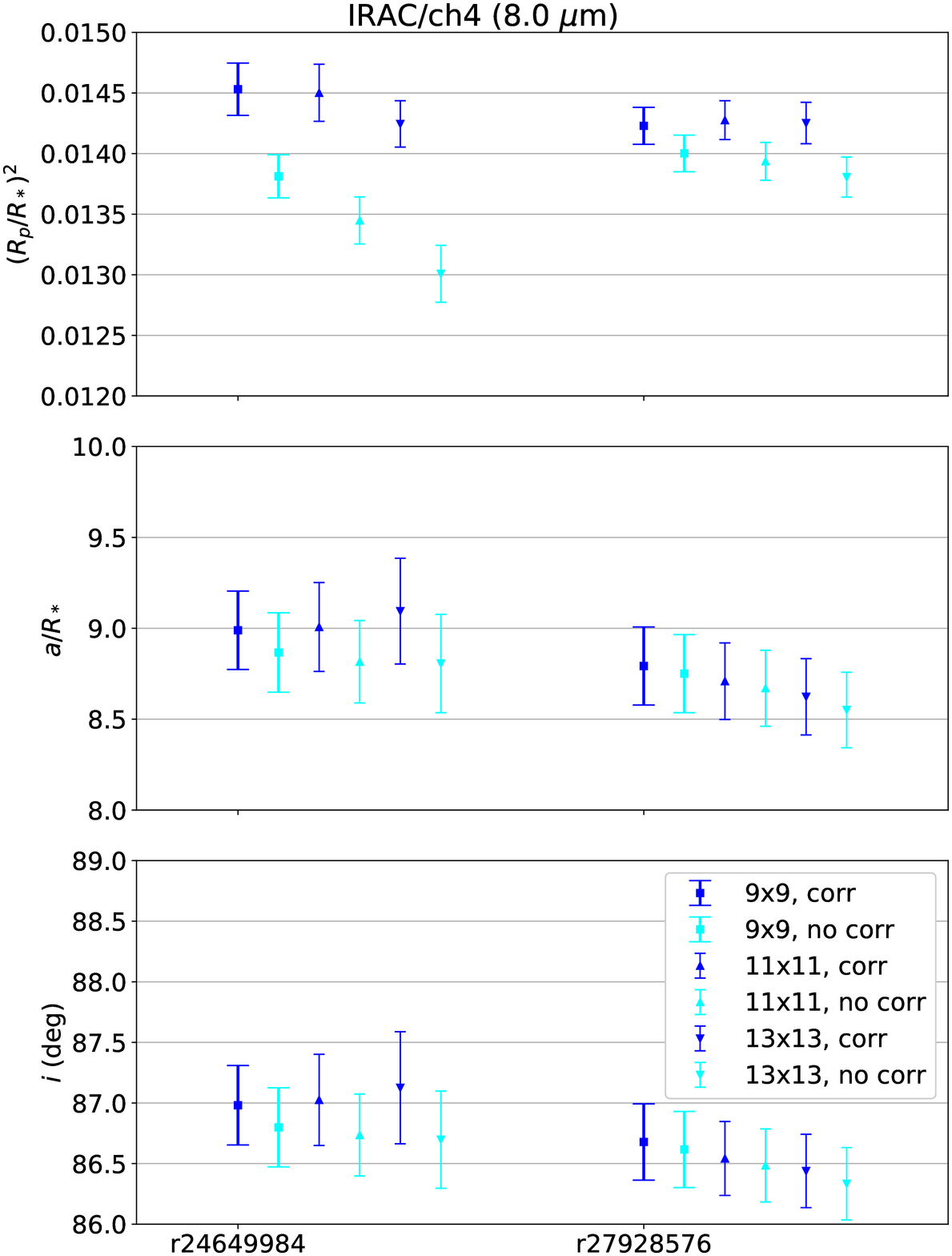}
\caption{Top, left panel:  Best-fit transit depths for the observations at 5.8 $\mu$m using 9$\times$9 (squares), 11$\times$11 (upward-pointing triangles), and 13$\times$13 (downward-pointing triangles) arrays as photometric apertures; background corrected through the test of aperture sizes (blue) or not corrected (cyan). Top, right panel: Analogous transit depths for the observations at 8.0~$\mu$m. Middle and bottom panels: Corresponding best-fit values for $a/R_*$ and $i$.
\label{fig22}}
\end{figure}

Figures~\ref{fig21} and \ref{fig22} report the transit parameters obtained using different aperture sizes with and without background corrections. Except for the earlier observation at 8.0~$\mu$m, all parameter results are consistent within 1 $\sigma$, as expected for observations that are not background limited. The correction at 8.0 $\mu$m is necessary to obtain consistent transit depth values over the two observations, especially when using larger apertures. The orbital parameters, $a/R_*$ and $i$, are less strongly affected by the choice of the array and background correction.

\clearpage

\section{The cold-warm \textit{Spitzer} offset}
\label{app:offset}

\begin{figure}[!t]
\epsscale{0.49}
\plotone{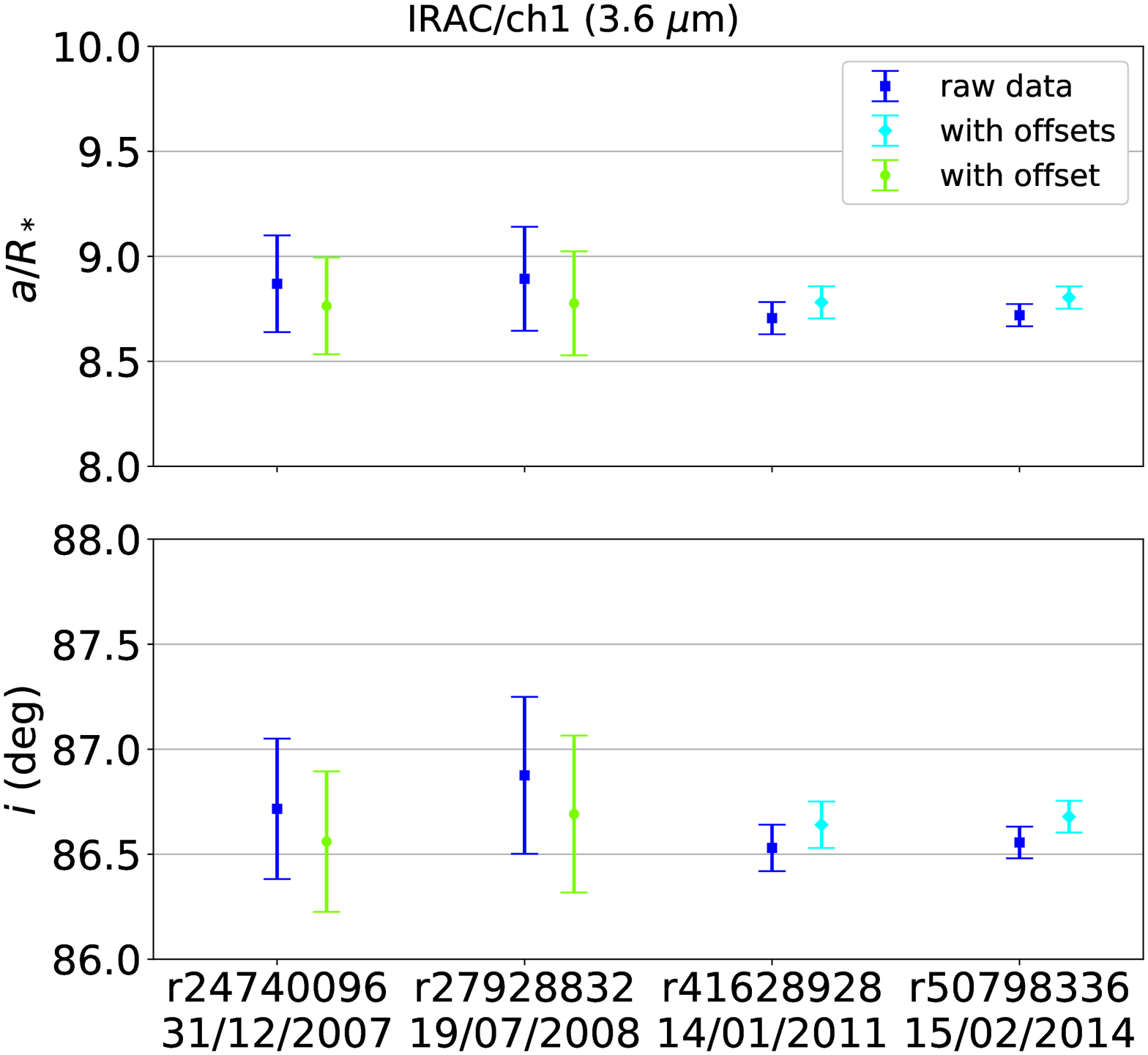}
\plotone{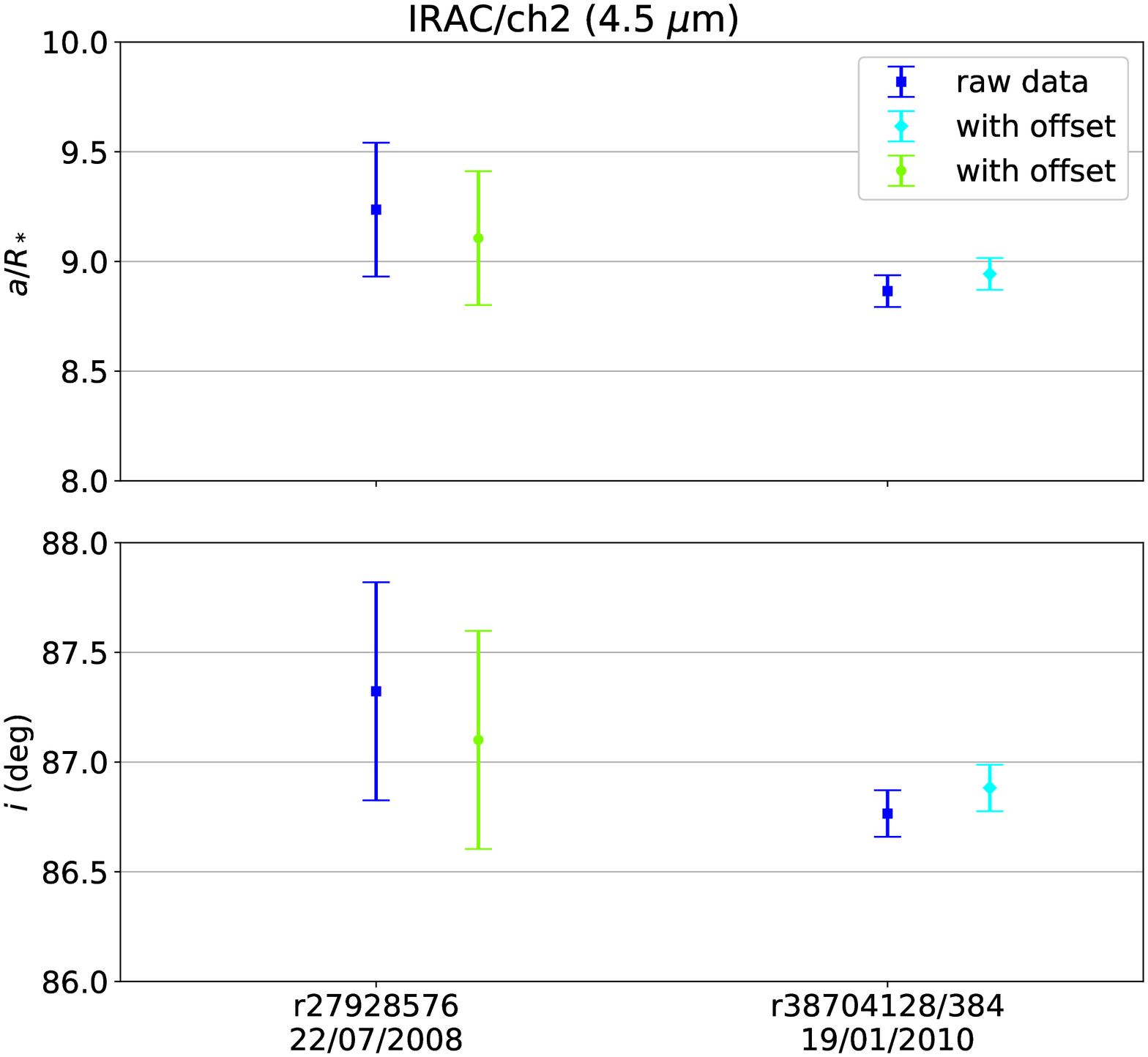}
\caption{Best-fit $a/R_*$ and $i$ for the observations at 3.6 and 4.5 $\mu$m (blue squares), with an offset in flux to force the warm transit depths to the cold value (cyan diamonds), and with an offset in flux to force the cold transit depths to the warm value (green circles).
\label{fig23}}
\end{figure}

We previously showed that warm \textit{Spitzer} transit depth measurements are systematically smaller than the cold ones in the same passband, with discrepancies at the 2--5 $\sigma$ level (see Section~\ref{ssec:results_irac1e2}). \cite{evans15} observed a similar trend, but they attributed the discrepancy to the presence of high-frequency noise of unknown origin during part of the warm \textit{Spitzer} observations at 3.6 $\mu$m. The same explanation cannot be applied to the observations at 4.5 $\mu$m as they themselves stated, although the discrepancy is also less significant ($\lesssim$2 $\sigma$). It is reasonable to assume that there is an offset in the calibration of the two \textit{Spitzer} eras. \cite{evans15} obtained inconsistent results for the orbital parameters in the 3.6 $\mu$m warm measurements. This problem is minimized by our pixel-ICA reanalysis, i.e., the new orbital parameter estimates are mutually consistent within 1~$\sigma$ (see Section~\ref{ssec:results_irac1e2}); the only inconsistency remains between the warm and cold transit depth measurements. We also noted that if we add appropriate offsets to the flux measurements in order to obtain identical transit depths over the cold and warm eras, the orbital parameters also become more similar (but they are always within 1~$\sigma$, see Figure~\ref{fig23}).

Such offsets are more than eight times greater than the highest background estimate in the 3.6 and 4.5 $\mu$m channels. Any potential offset equal for all pixels can be discarded, as it would be revealed by the test of aperture sizes (see Section~\ref{ssec:test_aperture}). We show that a possible explanation is a differential error in the nonlinearity correction between the cold and warm \textit{Spitzer} eras, which can be mimicked to first order by an offset in the total flux.

If the instrument response is linear, the flux integrated over the photometric aperture is proportional to the source flux:
\begin{equation}
\label{eqn:Flin}
F_{0}(t) = \alpha F_* f(t) = \alpha F_* (1-\lambda(t)) ,
\end{equation}
where $F_*$ is the unperturbed stellar flux, and $\lambda(t)$ is the occulting factor as a function of time ($\lambda(t)=0$ out-of-transit, $0<\lambda(t) \ll 1$ during the transit). We refer to $f(t) = 1-\lambda(t)$ as the ``transit model'' (cf., \citealp{mandel02}). Note that in the simple case defined by Equation~\ref{eqn:Flin}, the transit model is coincident with the observed flux normalized to the out-of-transit value:
\begin{equation}
\label{eqn:Flin_norm}
F_{0,\mbox{norm}}(t) = \frac{F_{0}(t)}{\alpha F_*} = f(t) .
\end{equation}
If the instrument response is nonlinear, the proportionality constant in Equation~\ref{eqn:Flin}, $\alpha$, must be replaced with a non-trivial function of the source flux. We assume a first-order nonlinearity:
\begin{equation}
\label{eqn:Fnonlin1}
F_{1}(t) = \alpha (1-\beta F_* f(t))  F_* f(t) .
\end{equation}
The $F_{1}$ time series normalized to the out-of-transit value is
\begin{equation}
\label{eqn:Fnonlin1_norm}
F_{1,\mbox{norm}}(t) = \frac{F_{1}(t)}{\alpha F_* (1-\beta F_*)} = \frac{1-\beta F_* f(t)}{1-\beta F_*} f(t) .
\end{equation}
In the limit $\beta F_* \ll 1$, using the second-order Taylor expansion for the denominator, and expliciting $f(t) = 1-\lambda (t)$:
\begin{equation}
\label{eqn:Fnonlin1_norm_approx_inter}
F_{1,\mbox{norm}}(t) \approx (1+\beta F_* + (\beta F_*)^2) \times (1-\beta F_* (1-\lambda (t))) \times (1-\lambda (t)) .
\end{equation}
By expanding the products, and retaining up to second-order terms in $\beta F_*$ and $\lambda$:
\begin{equation}
\label{eqn:Fnonlin1_norm_approx}
F_{1,\mbox{norm}}(t) \approx 1 - \lambda (t) + \beta F_*  \lambda (t) .
\end{equation}

Now, we consider the case with linear instrument response and an additive offset, $C$:
\begin{equation}
\label{eqn:Foffset}
F_{\mbox{C}}(t) = \alpha F_* (1-\lambda(t)) + C,
\end{equation}
The $F_{\mbox{C}}$ time series normalized to the out-of-transit value is:
\begin{equation}
\label{eqn:Foffset_norm}
F_{\mbox{C},\mbox{norm}}(t) = \frac{F_{\mbox{C}}(t)}{\alpha F_* + C} = \frac{ 1-\lambda(t) + \frac{C}{\alpha F_*}}{1 + \frac{C}{\alpha F_*}} .
\end{equation}
In the limit $C / (\alpha F_*) \ll 1$, using the second-order Taylor expansion for the denominator:
\begin{equation}
\label{eqn:Foffset_norm_approx_inter}
F_{\mbox{C},\mbox{norm}}(t) \approx \left ( 1 - \frac{C}{\alpha F_*} + \left ( \frac{C}{\alpha F_*} \right )^2 \right ) \times \left ( 1-\lambda(t) + \frac{C}{\alpha F_*} \right ) .
\end{equation}
By expanding the products, and retaining up to second-order terms in $C / (\alpha F_*)$ and $\lambda$:
\begin{equation}
\label{eqn:Foffset_norm_approx}
F_{\mbox{C},\mbox{norm}}(t) \approx 1 - \lambda (t) + \frac{C}{\alpha F_*}  \lambda (t)
\end{equation}

Note that Equations~\ref{eqn:Fnonlin1_norm_approx} and \ref{eqn:Foffset_norm_approx} have the same structure, and are equivalent if
\begin{equation}
\beta F_* = \frac{C}{\alpha F_*} \equiv x .
\end{equation}
The third terms in Equation~\ref{eqn:Fnonlin1_norm_approx} and \ref{eqn:Foffset_norm_approx}, with $\lambda (t) = (R_p/R_*)^2$, approximate the bias in transit depth due to the nonlinearity or the additive offset, respectively. As the observed cold-warm discrepancy is $\sim$300 ppm in transit depth (see Section~\ref{ssec:results_irac1e2}), using $(R_p/R_*)^2 \sim 1.5 \times 10^{-2}$, we obtain
\begin{equation}
x \approx 2 \times 10^{-2} .
\end{equation}
This result is consistent with the $\sim$3$\%$ error in the absolute flux calibration reported by \cite{carey12}.

\section{Different ramp models for \textit{Spitzer}/IRAC channels 3 and 4}
\label{app:ramps}
In this Appendix, we compare the performances of the following ramp models in statistical terms:
\begin{itemize}
\item single exponential,
\begin{equation}
\label{eqn:ramp_exp_appendix}
F(t) = a_0 - a_1 e^{-t/\tau_1} \ ,
\end{equation}
\item double exponential,
\begin{equation}
\label{eqn:ramp_exp2_appendix}
F(t) = a_0 - a_1 e^{-t/\tau_1} - a_2 e^{-t/\tau_2} \ ,
\end{equation}
\item linear + quadratic log,
\begin{equation}
\label{eqn:ramp_linlog2_appendix}
F(t) = a_0 + a_1 t + a_2 \log{(t-t_0)} + a_3 [ \log{(t-t_0)} ]^2 \ ,
\end{equation}
\item linear + logarithmic,
\begin{equation}
\label{eqn:ramp_linlog_appendix}
F(t) = a_0 + a_1 t + a_2 \log{(t-t_0)} \ ,
\end{equation}
\item quadratic log,
\begin{equation}
\label{eqn:ramp_linlog2_appendix}
F(t) = a_0 + a_1 \log{(t-t_0)} + a_2 [ \log{(t-t_0)} ]^2 \ ,
\end{equation}
\item fifth-order polynomial,
\begin{equation}
\label{eqn:ramp_poly5_appendix}
F(t) = a_0 + a_1 t + a_2 t^2 + a_3 t^3 + a_4 t^4 + a_5 t^5 \ .
\end{equation}
\end{itemize}
Following the prescription by \cite{beaulieu10}, we set $t_0$ to be 30~s before the observations begin to prevent the divergence of the logarithm at $t=0$.

We focus on the two observations at 5.8 and 8.0 $\mu$m  taken in 2008 July. 
\begin{figure}[!h]
\epsscale{0.49}
\plotone{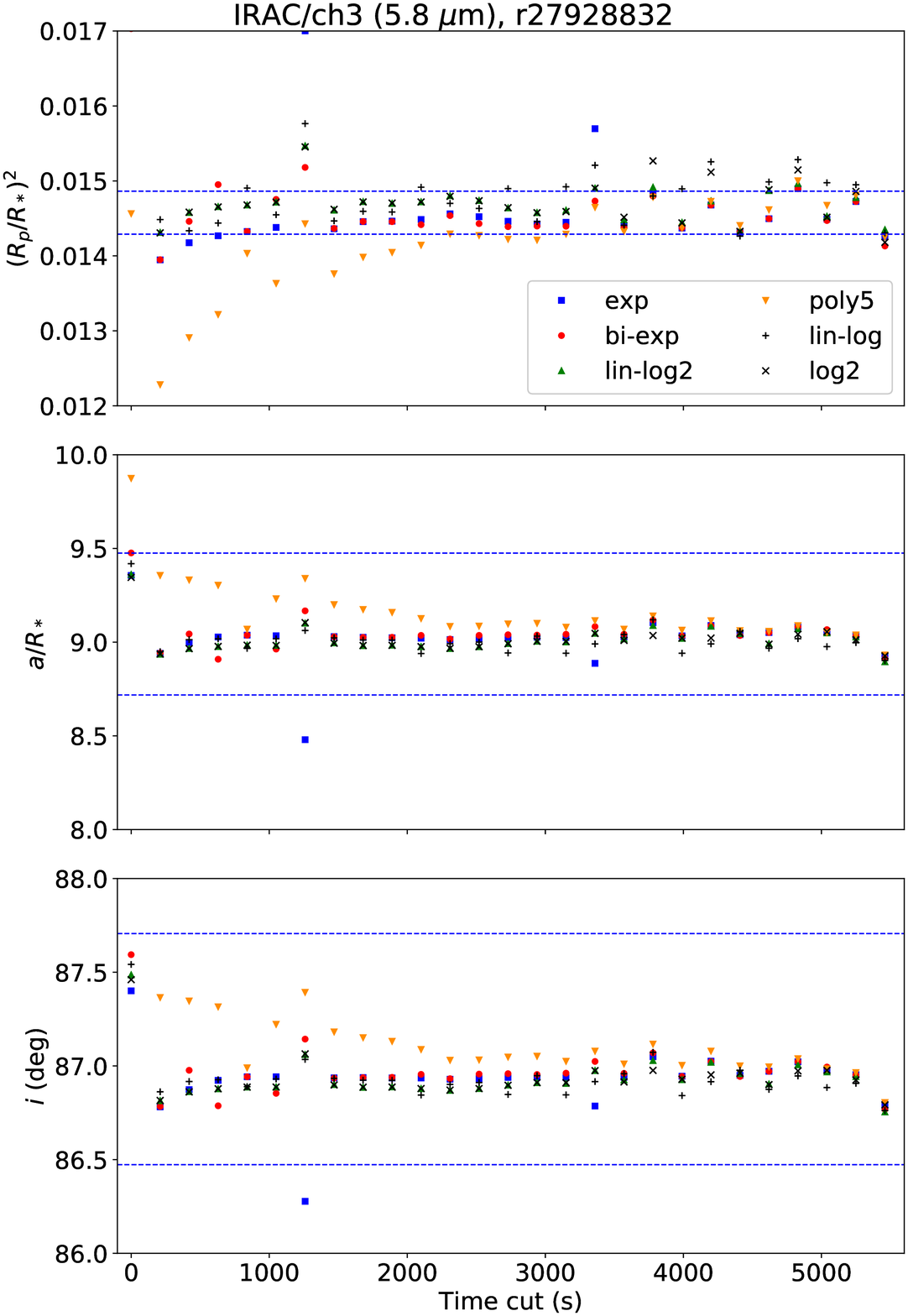}
\plotone{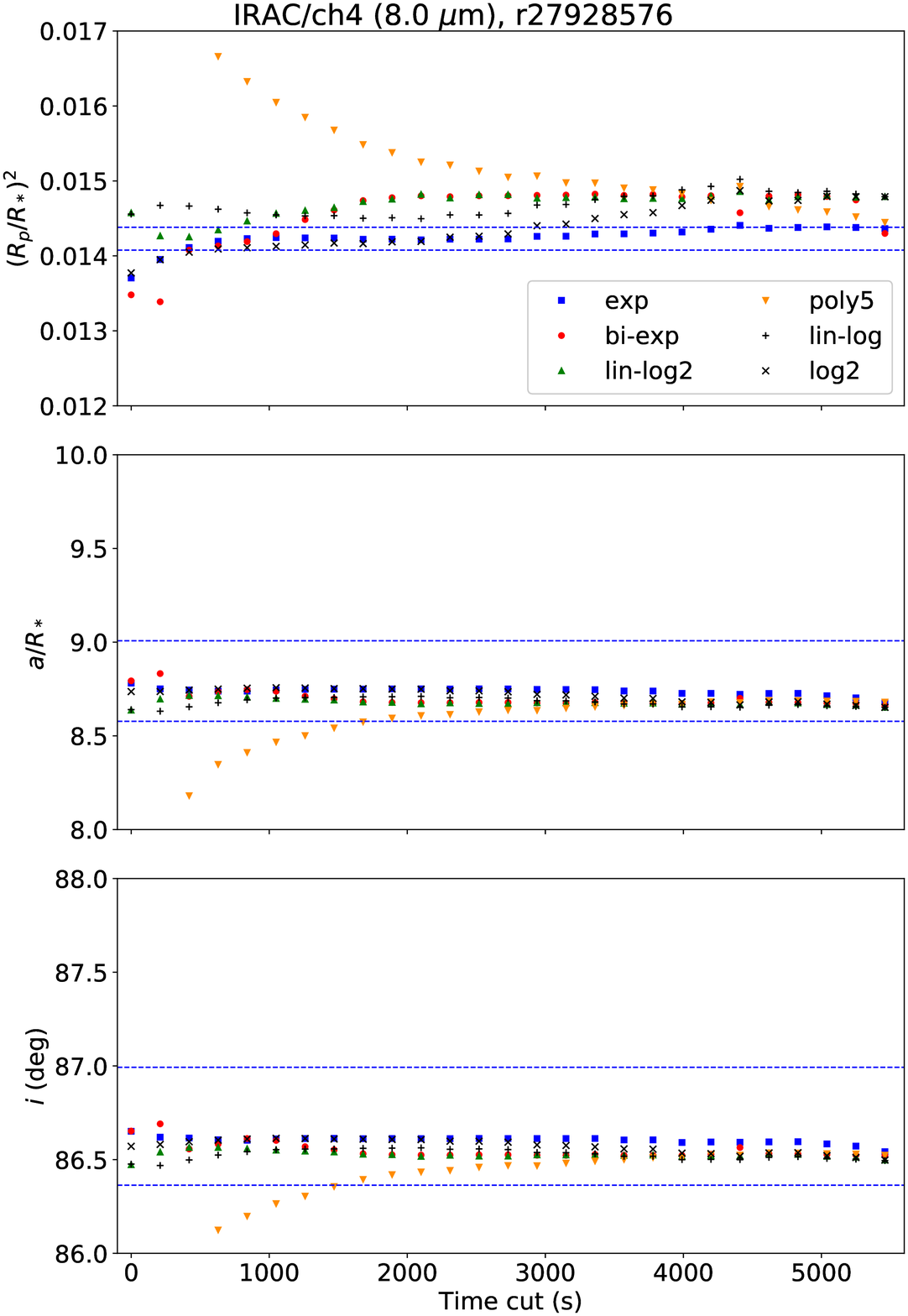}
\caption{Top, left panel:  Best-fit transit depths for one observation at 5.8 $\mu$m as a function of the trimmed interval from the beginning, adopting a single exponential (blue squares), double exponential (red circles), linear + quadratic log (green upward-pointing triangles), fifth-order polynomial (yellow downward-pointing triangles), linear + logarithmic (black plus), and quadratic log (black cross) ramp parameterizations. The blue dashed lines delimit the 1 $\sigma$ intervals for the final parameter estimates, reported in Table~\ref{tab5}. Top, right panel: Analogous transit depths for one observation at 8.0 $\mu$m. Middle and bottom panels: Corresponding best-fit values for $a/R_*$ and $i$.
\label{fig24}}
\end{figure}
Regardless of the choice of the ramp parameterization, the root mean square (rms) of residuals decreases monotonically while discarding more data points from the beginning until random fluctuations become dominant.  The best-fit parameters also follow a trend before stabilizing.
Figure~\ref{fig24} shows the best transit parameters obtained with six different ramp models (Equations~\ref{eqn:ramp_exp_appendix}--\ref{eqn:ramp_poly5_appendix}), discarding different time intervals from the beginning. When the single exponential formula (Equation~\ref{eqn:ramp_exp_appendix}) is used, the estimated transit depth becomes stable after the first 10--20 minutes are discarded, i.e., it varies by less than 1 $\sigma$ if more data are discarded. The orbital parameters are even more robust, as the amplitudes of the observed trends across 0--91 minute cuts are much smaller than 1 $\sigma$. The results obtained with the other ramp parameterizations show similar or larger variations as a function of the discarded interval, the worst case is the fifth-order polynomial (Equation~\ref{eqn:ramp_poly5_appendix}).  Overall, the orbital parameters, $a/R_*$ and $i$, are very stable with respect to the choice of the ramp model and the length of the discarded interval. Only for the case of the polynomial ramp at 8.0 $\mu$m and discarded interval shorter than 30 minutes do they fall outside the accepted ranges. The linear + quadratic log and the double exponential formulas lead to systematically larger transit depths at 8.0 $\mu$m than those obtained with the single exponential. \cite{beaulieu10} adopted the linear + quadratic log formula to correct for the ramp at 8.0 $\mu$m. We conclude that the most likely cause of discrepancy with our analysis in this passband is the choice of the ramp parameterization. The statistical tests discussed in this Appendix are moderately in favor of the single exponential form.

We consider two quantitative model selection techniques: the Bayesian Information Criterion (BIC, \citealp{schwarz78}) and the Akaike Information Criterion (AIC/AIC$_c$, \citealp{akaike74, cavanaugh97}). They have been calculated under the assumptions of uniform uncertainties per data point and gaussian distributed residuals \footnote{It is equivalent to rescaling the \textit{Spitzer}-provided uncertainties, a common, well-motivated practice in the analysis of transits \citep{blecic13}.}:
\begin{equation}
\label{eqn:bic}
BIC = n \log{\sigma_{res}^2} + k \log{n}
\end{equation}
\begin{equation}
\label{eqn:aic}
AIC_c = ( n \log{\sigma_{res}^2} + 2k ) + \frac{2k(k+1)}{n-k-1} = AIC + \frac{2k(k+1)}{n-k-1}
\end{equation}
Here $\sigma_{res}$ is the rms of residuals, $n$ is the number of data points, and $k$ is the number of ramp parameters. The first term in Equations~\ref{eqn:bic} and \ref{eqn:aic} is a measure of the goodness of fit, while the second terms are penalties proportional to the number of free parameters, and the third term in Equation~\ref{eqn:aic} is a correction for small sample sizes (negligible for $n \gg k$). In our cases, the differences between AIC and AIC$_c$  are $\sim$0.01 (negligible), therefore we can here equivalently refer to AIC instead of AIC$_c$. The best models according to these selection techniques are those with the lowest BIC or AIC \citep{schwarz78, akaike74}, respectively. BIC tends to select simpler models than AIC as a result of the different penalties for the number of parameters (greater for BIC). The relative strength of evidence between two models, $i$ and $j$, is defined by their probability ratio \citep{burnham02}:
\begin{eqnarray}
\label{eqn:bic_prob_ratio}
\left ( \frac{p_i}{p_j} \right )_{BIC} = \exp{ \left ( - \frac{1}{2} (BIC_i - BIC_j) \right ) } = \exp{ \left ( - \frac{\Delta BIC}{2}  \right ) }\\
\label{eqn:aic_prob_ratio}
\left ( \frac{p_i}{p_j} \right )_{AIC} = \exp{ \left ( - \frac{1}{2} (AIC_i - AIC_j) \right ) } = \exp{ \left ( - \frac{\Delta AIC}{2}  \right ) }\\
\end{eqnarray}
According to the interpretation provided by \cite{raftery95}, absolute differences smaller than 2 in BIC/AIC\footnote{\cite{raftery95} only refers to $\Delta$BIC, but Equations~\ref{eqn:bic_prob_ratio}--\ref{eqn:aic_prob_ratio} allow to extend the same interpretation to $\Delta$AIC.} are not statistically significant and only differences greater than 6 provide ``strong evidence'' against the model with higher BIC/AIC.

Table~\ref{tab_bic_ramp} reports the differences in BIC and AIC obtained for different ramp models relative to the single exponential. Positive differences denote a preference for the single exponential model, negative differences show a preference for the alternative model. The single exponential is the favorite model for the ramp at 5.8 $\mu$m according to both selection criteria, but the linear + logarithmic and the quadratic log models have nearly the same likelihoods. The situation is reversed for the ramp at 8.0 $\mu$m with a slight preference for the linear + logarithmic model, but there is no strong evidence against the single exponential. The double exponential formula at 8.0 $\mu$m is strongly disfavored by BIC, but only marginally disfavored by AIC. The fifth-order polynomial is by far the worst parameterization for all ramps (the same is valid for lower-order polynomials).
\begin{table}[h]
\begin{center}
\caption{$\Delta$BIC and $\Delta$AIC$_c$ using different ramp parameterizations for two observations at 5.8 and 8.0 $\mu$ relative to the single exponential formula. \label{tab_bic_ramp}}
\begin{tabular}{ccc}
\tableline\tableline
IRAC/ch3 & \multirow{2}{*}{$\Delta$BIC} & \multirow{2}{*}{$\Delta$AIC$_c$} \\
r27928832 & & \\
\tableline
poly5 & +20.2 & +8.6 \\
lin-log2 & +8.1 & +2.3 \\
lin-log & +0.2 & +0.2 \\
log2 & +0.3 & +0.3 \\
bi-exp & +15.6 & +4.0 \\
\tableline
\end{tabular}
\qquad \qquad
\begin{tabular}{ccc}
\tableline\tableline
IRAC/ch4 & \multirow{2}{*}{$\Delta$BIC} & \multirow{2}{*}{$\Delta$AIC$_c$} \\
r27928576 & & \\
\tableline
poly5 & +30.3 & +18.7 \\
lin-log2 & +4.8 & -1.0 \\
lin-log & -2.8 & -2.8 \\
log2 & -0.7 & -0.7 \\
bi-exp & +12.4 & +0.8 \\
\tableline
\end{tabular}
\end{center}
\end{table}

\begin{figure}[!h]
\epsscale{0.49}
\plotone{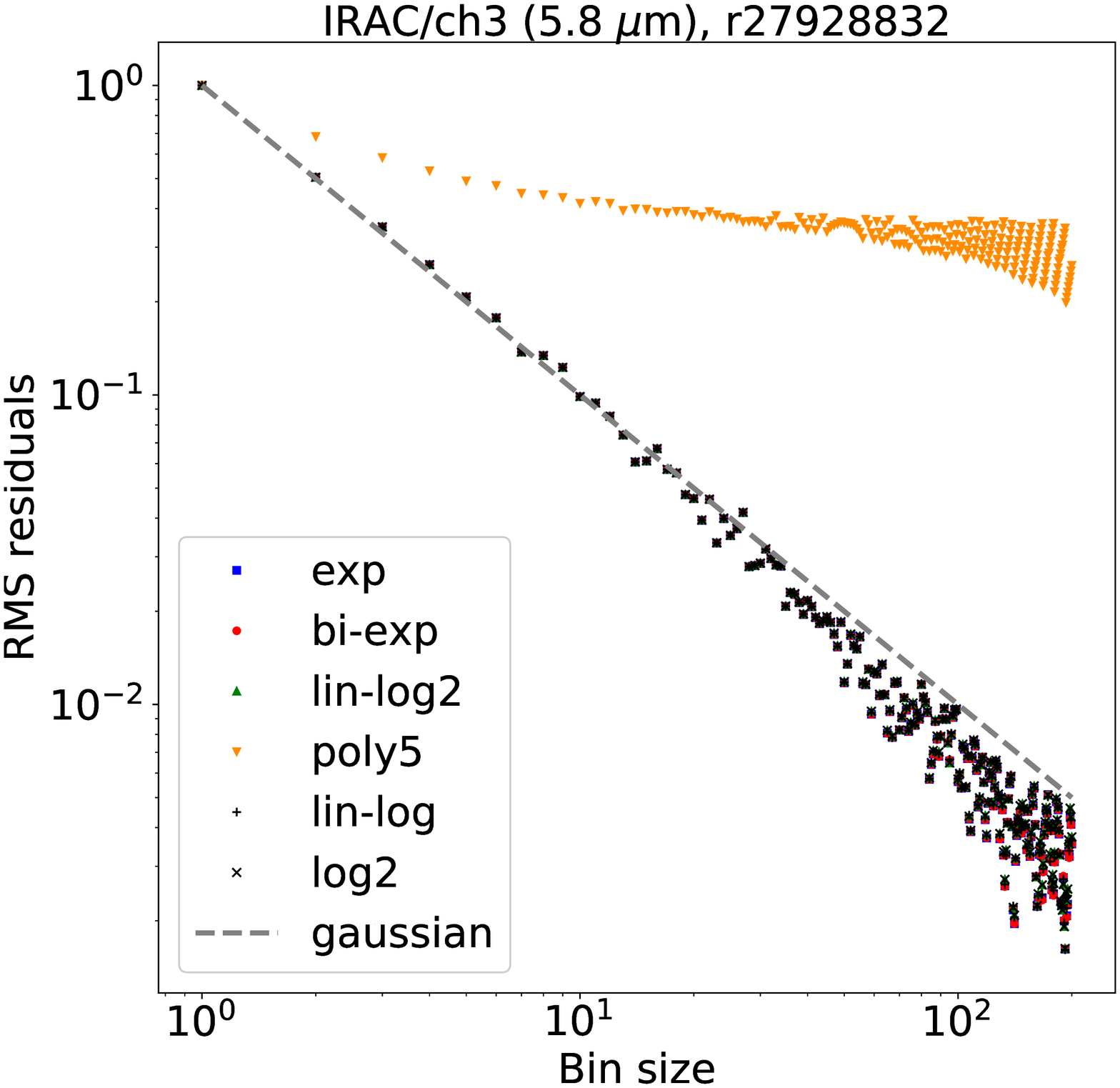}
\plotone{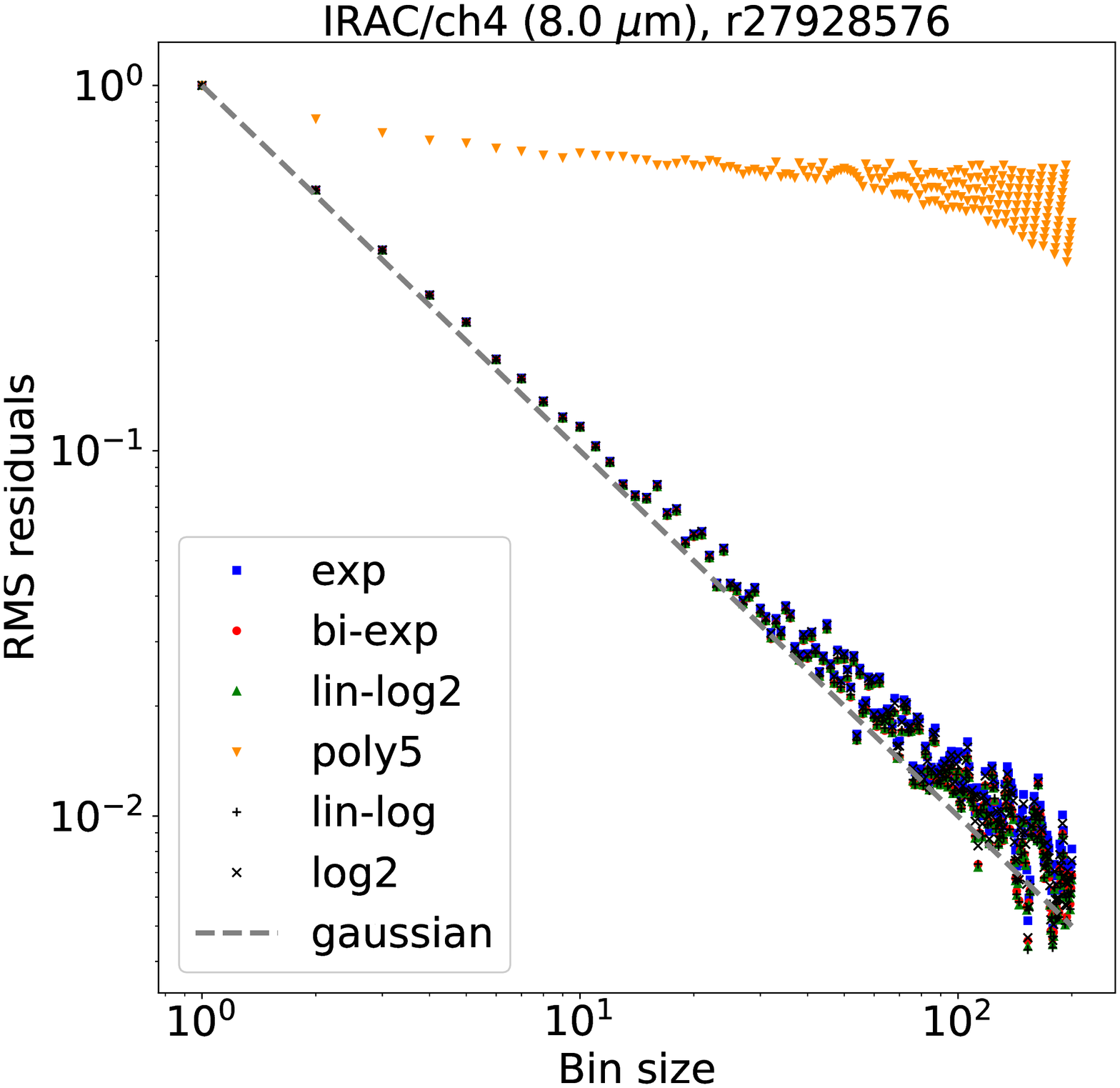}
\caption{Left panel: Root mean square of residuals for one observation at 5.8~$\mu$m as a function of the bin size, scaled to their non-binned values, adopting a single exponential (blue squares), double exponential (red circles), linear + quadratic log (green upward-pointing triangles), fifth-order polynomial (yellow downward-pointing triangles), linear + logarithmic (black plus), and quadratic log (black cross) ramp parameterizations; the gray dashed line shows the theoretical behavior for gaussian residuals. Right panel: Analogous plot for one observation at 8.0~$\mu$m.
\label{fig25}}
\end{figure}

We also checked for correlated noise in the residuals, but all the autocorrelation coefficients turned out to be smaller than 0.07, regardless of the formula adopted to remove the ramp. Figure~\ref{fig25} shows that the residuals scale with the bin size according to the expected behavior for gaussian noise, except when the polynomial ramp is adopted.

\clearpage

\section{Accuracy of two-coefficient limb-darkening laws}
\label{app:ld_laws}
\begin{figure}[!h]
\epsscale{1.0}
\plotone{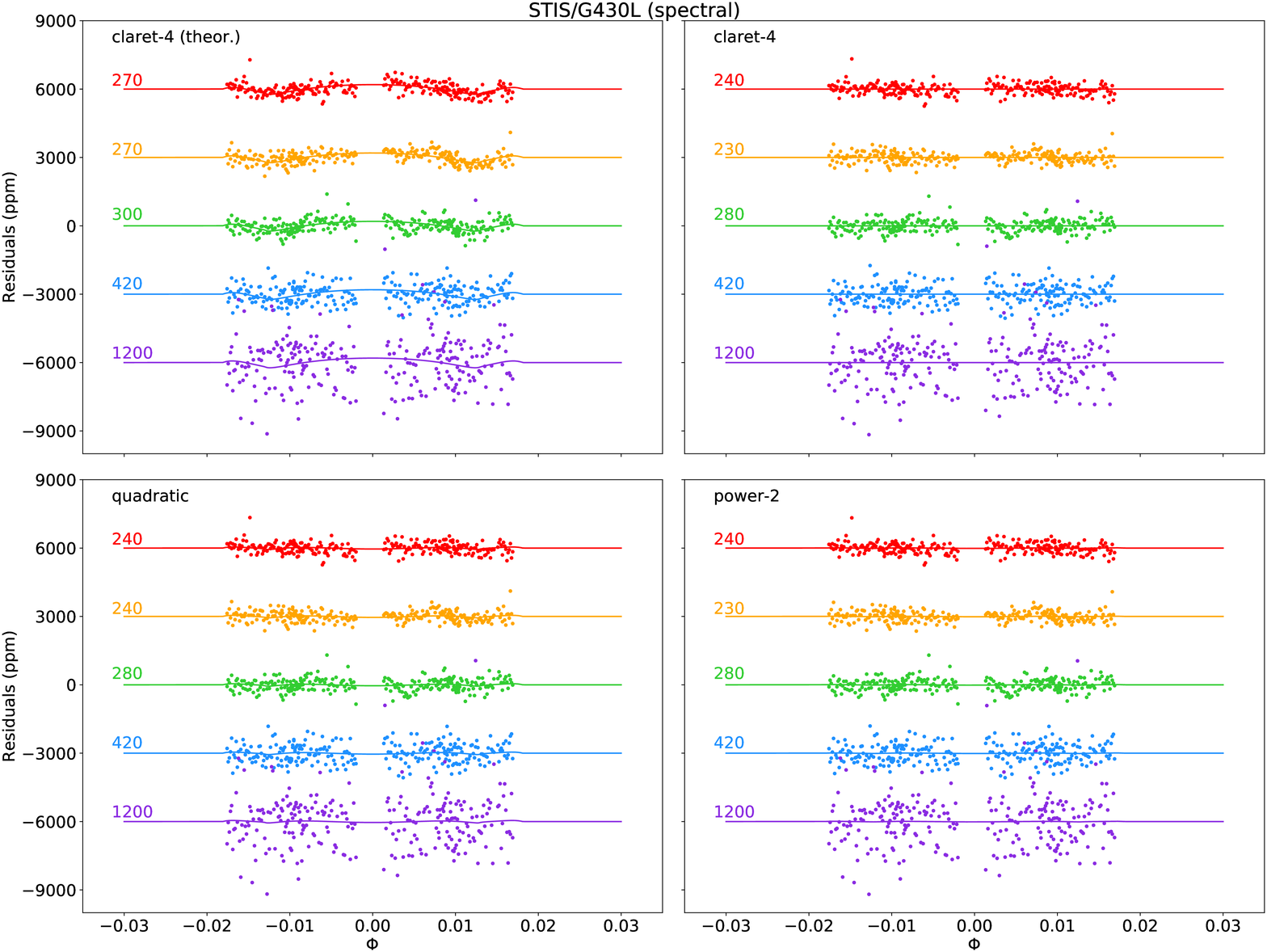}
\caption{Top, left panel:  Spectral light-curve residuals obtained with fixed claret-4 limb-darkening coefficients taken from \cite{knutson07}, and overplotted difference between the relevant best-fit transit models and those obtained with free claret-4 coefficients. The numbers on the left indicate the rms of residuals. Top, right panel: Analogous plot with free claret-4 coefficients. Bottom, left panel: Analogous plot with free quadratic coefficients. Bottom, right panel: Analogous plot with free power-2 coefficients.
\label{fig26}}
\end{figure}

\begin{figure}[!h]
\epsscale{1.0}
\plotone{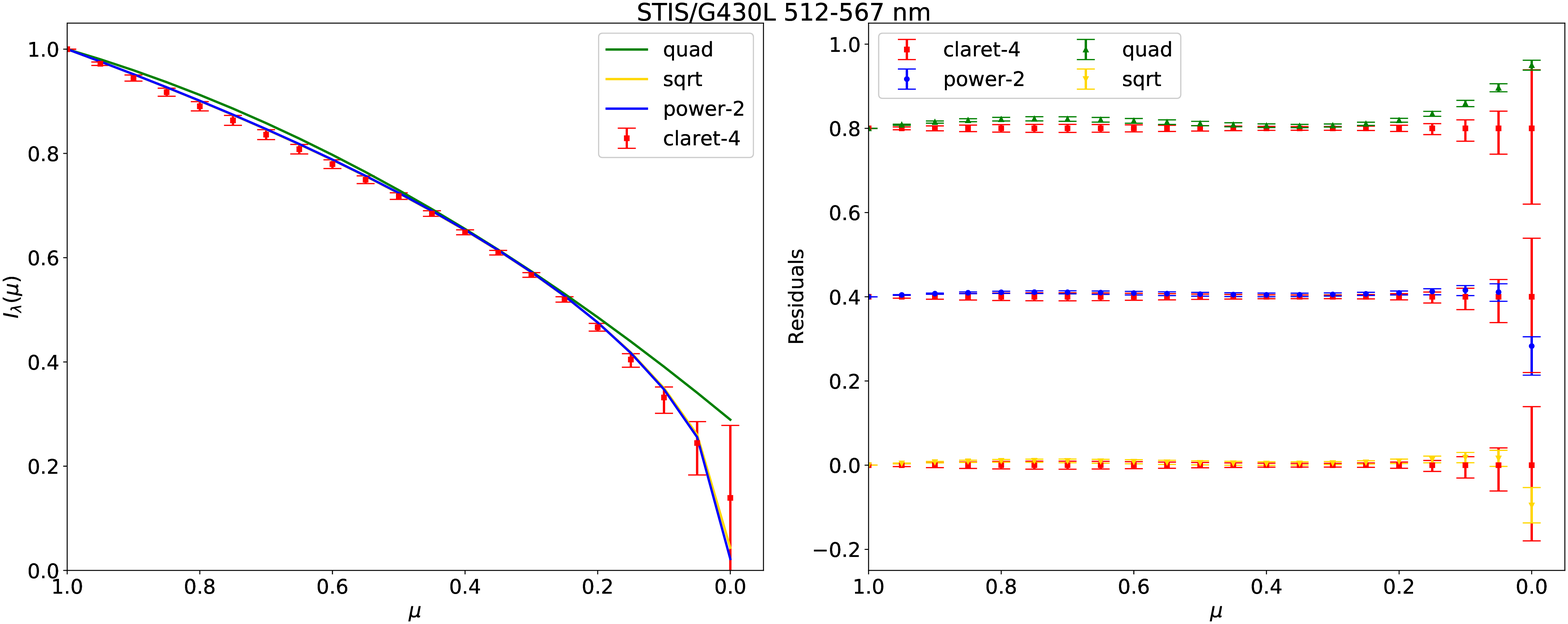}
\plotone{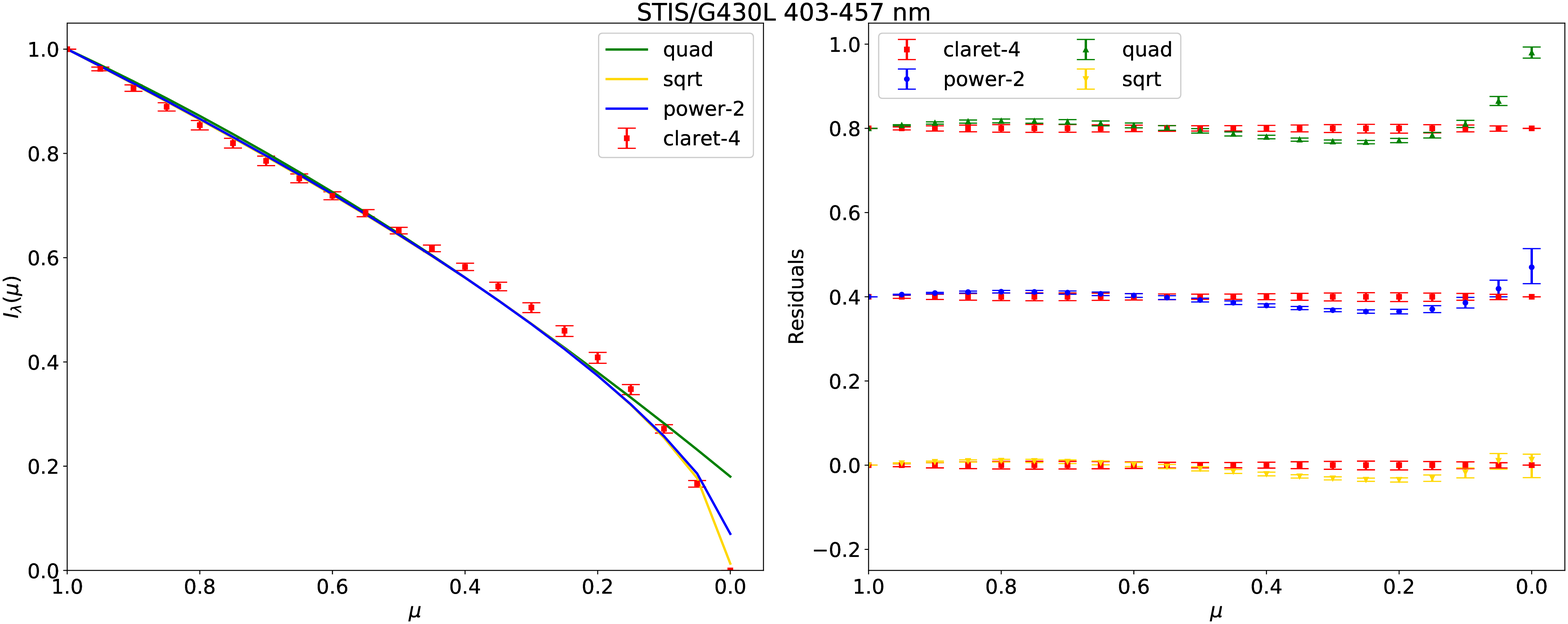}
\caption{Top, left panel: Empirical limb-darkening profile for the 512--567 nm passband using the claret-4 (red squares), quadratic (green), square-root (yellow) and power-2 (blue) formulas. Top, right panel: Differences between each model and the claret-4 with vertical shifts to improve their visualization. Bottom panels: Analogous plots for the 403--457 nm passband.
\label{fig27}}
\end{figure}

\begin{figure}[!h]
\epsscale{1.0}
\plotone{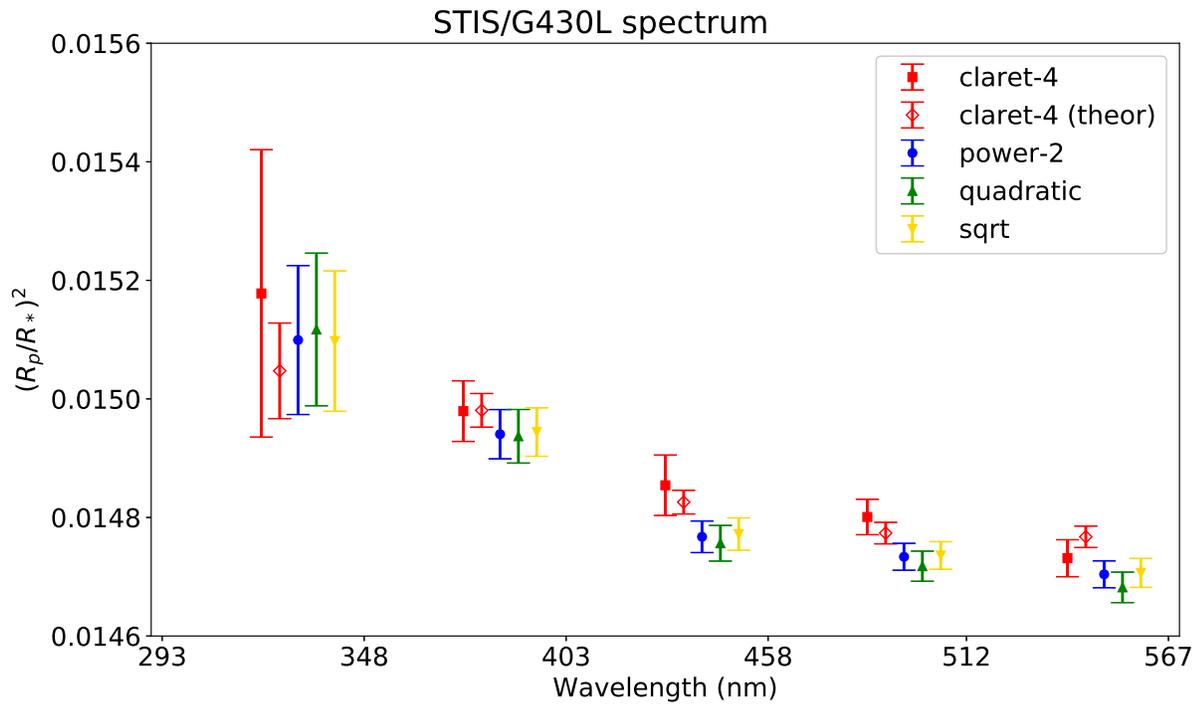}
\caption{Transit depth for the five spectral bins in the \textit{HST}/STIS G430L passband using different limb-darkening parameterizations: Free claret-4 (red full squares), fixed claret-4 from \cite{knutson07} (red empty diamonds), free power-2 (blue circles), free quadratic (green upward-pointing triangles), and free square-root (yellow downward-pointing triangles) coefficients.
\label{fig28}}
\end{figure}

\cite{morello17} claimed that two-coefficient limb-darkening laws are inaccurate in the 290-570 nm wavelength range covered by \textit{HST}/STIS G430L, even if the corresponding light-curve fit residuals appear to be limited by photon noise. This result was based on simulated data using PHOENIX stellar-atmosphere models \citep{allard12}.
Figures~\ref{fig26} and \ref{fig27} confirm the same behavior for real HD209458 datasets. In particular, we obtained nearly identical light-curve residuals when using free quadratic \citep{kopal50}, power-2 \citep{morello17}, or claret-4 \citep{claret00} coefficients, despite that the corresponding stellar intensity profiles are, in some cases, significantly different. The bias in transit depth can be $\sim$100 ppm when adopting two-coefficient limb-darkening laws, and in particular, it is slightly above 1 $\sigma$ for the 403--457 and 458--512 nm passbands for the quadratic, power-2, or square-root law. We note that for the case analyzed in this paper, even the 1D claret-4 theoretical models outperform the empirical two-coefficient parameterizations (see also Figure~\ref{fig28}). 

\clearpage

\clearpage

\end{document}